\documentclass{IEEE-Con-Sys-mag}
\usepackage{hhline}
\usepackage{soul}
\usepackage{graphicx}
\usepackage{fixmath}
\usepackage[T1]{fontenc}
\usepackage{siunitx}
\usepackage[version=4]{mhchem}
\usepackage{amsmath,amssymb,amsfonts}
\usepackage{float}
\usepackage{algorithmic}
\usepackage{graphicx}
\usepackage{adjustbox}
\usepackage{textcomp}
\usepackage{hyperref}
\usepackage{gensymb}
\usepackage{xcolor}
\usepackage[backend=biber, sorting=none]{biblatex}
\newcounter{sidebarref}
\renewcommand*{\thesidebarref}{S\arabic{sidebarref}}
\usepackage{cancel}
\usepackage{subcaption}
\usepackage{comment}
\usepackage{hyperref}

\newcommand{\sidecite}[1]{%
  \stepcounter{sidebarref}%
  [\thesidebarref]}
\jname{Preprint for IEEE CONTROL SYSTEMS publication}
\pubyear{2026}
\begin{document}

\title{The essential role of ribosomal feedback in bacterial cell growth and metabolic load\stitle{A systems biology approach for unveiling  shared resources regulation within synthetic genetic circuits}}

\author{{C}HIARA CIMOLATO, ELISA GAETAN, LORENZO PASOTTI, LUCA SCHENATO and MASSIMO BELLATO\\\textcolor{red}{The final peer-reviewed version of this paper is available on IEEE Control System Magazine.}}
\affil{}

\maketitle

\begin{figure*}[!htp]
    \centering
    \includegraphics[width=\linewidth]{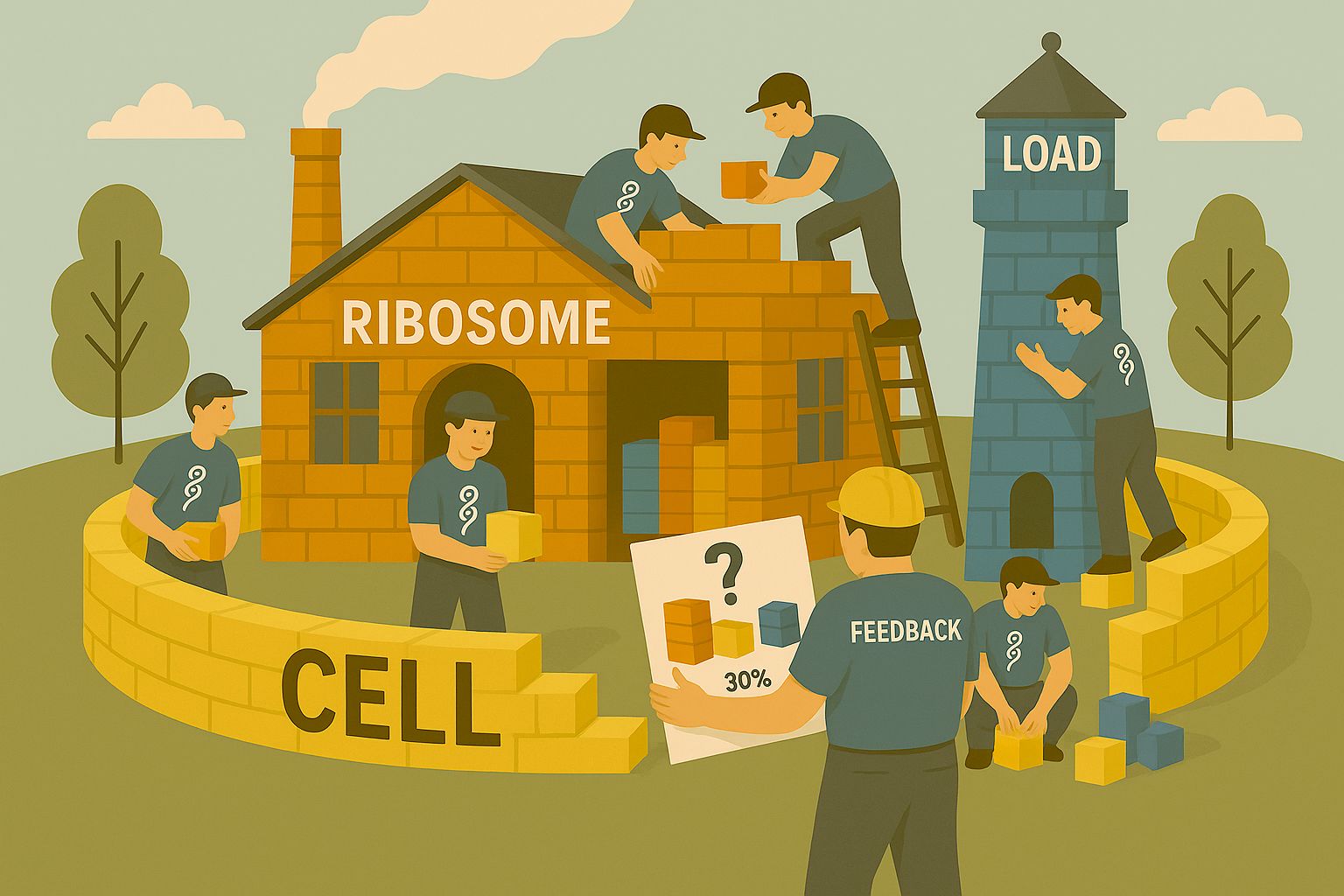}
    \caption{Ribosomes serve as the central builders of the cell, producing proteins that fulfill three interconnected demands: (i) maintaining cellular structure and function, (ii) synthesizing recombinant proteins, which can impose a metabolic load, and (iii) generating new ribosomes to expand translational capacity. Achieving the right balance among these competing tasks is inherently complex: overcommitment to one pathway can compromise others, leading to reduced cell viability or impaired performance. Feedback mechanisms are therefore essential to dynamically regulate ribosomal allocation, enabling robust protein production without destabilizing the broader cellular system. Image generated with AI (ChatGPT 5.4).}
    \label{fig:graphabstract}
\end{figure*}

\chapterinitial{S}{ynthetic} biologists engineer living cells to create novel biological systems, intending to implement new functionalities leading to biotechnologically relevant outcomes.
Rationally designed genetic circuits are therefore introduced into host cells to implement novel functions, enabling the production of additional proteins beyond those typically synthesized by the biological chassis. However, due to the finite nature of shared resources within cells and the intricate regulation that hinders the entire engineering process, it often leads to unpredictable outcomes in the behavior of engineered cells~\cite{burden_maks,burden_ddvb, Pasotti2017,resource_competition_Qian}. This can even affect cell growth when there is increased demand for protein synthesis. As a result, a control system's understanding of these feedback mechanisms is crucial for the effective design of synthetic cells. An allegorical representation of the described context is shown in Fig.~\ref{fig:graphabstract}.

\begin{summary}
\summaryinitial{M}odeling growth in bacterial cells is a major issue in systems and synthetic biology. Despite several growth rate functions proposed in the literature, most focus on nutrient composition without explicitly accounting for the possible perturbation provided by the expression of recombinant genes, an effect known as cell load or burden. 

On the other hand, mathematical models that attempt to provide mechanistic details on the phenomena, leveraging ribosome partitioning and nutrient availability, are generally too detailed and complex to be easily applied to the rational design of synthetic genetic circuits. A bottom-up approach is adopted herein to identify and analyze the minimal model structure, thereby unveiling the fundamental role of negative feedback in ribosomal synthesis in predicting the effects of cell load on both gene expression and growth rate. Indeed, to ensure cellular efficiency, ribosome synthesis must be finely regulated. While an increased number of ribosomes generally enhances protein production and cellular performance, their synthesis incurs a high energetic cost. For this reason, cells have evolved mechanisms to tightly control ribosome synthesis, avoiding unnecessary accumulation. One of the key regulatory strategies, usually neglected in previous cell models, involves a negative feedback loop that modulates the production of ribosomal components. This feedback ensures that ribosomes are produced only in the amount strictly needed, balancing functionality and energy expenditure.

This work evaluates the individual contribution of this feedback under heterologous expression conditions using minimal gene-circuit models, explicitly linking ribosome allocation, hidden couplings between protein synthesis levels, and growth rate. The paper is structured as follows:
\begin{itemize}
    \item \textbf{Biological context -} The article begins by explaining the role of ribosomes in bacterial cells and the feedback mechanisms that regulate ribosome abundance.
    \item \textbf{Experimental motivation -} It then reviews experimental evidence showing how burdensome gene expression affects bacterial growth rate and recombinant protein production.
    \item \textbf{Mathematical framework -} A minimal mechanistic model is introduced, including the relevant molecular reactions, differential equations, the growth-rate function, and the selected parameter values.
    \item \textbf{Model hierarchy -} Three models of increasing complexity are considered to isolate and evaluate the effect of ribosomal feedback.
    \item \textbf{Key outputs -} The results focus on three indicators of cell-load effects: ribosome levels, growth rate, and protein expression, with emphasis on the biological plausibility of their trends.
    \item \textbf{Robustness analysis -} Sensitivity analysis is used to test whether the main conclusions remain valid under parameter variations.
    \item \textbf{Supplementary Sidebars material -} Additional material summarizes bacterial gene-expression mechanisms, experimental procedures for characterizing cell-load effects, reaction-network modeling and simplification, and growth laws reported in the literature.
\end{itemize}

\end{summary}

\section{Introduction}\label{sec:introduction}

While the \textit{growth rate} of the bacterium \textit{Escherichia coli} in its exponential growth phase is typically assumed to be conserved (cells can double every $\sim20$ min at $37^{\circ}C$), it has been observed that this time can be reduced~\cite{growth_rate_of_Ecoli} or increased~\cite{hans_bremer} by several factors, such as the amount of available nutrients in the surrounding environment. Despite several examples of growth rate functions, including those influenced by temperature or nutrient starvation, found in the literature~\cite{ratkowsky1982relationship,prince04,arana10}, most of them do not include cell load perturbation, which is the burden imposed by recombinant protein expression due to the competition of endogenous and heterologous genes for shared cellular resources. Empirical models linking gene expression levels and growth rate have subsequently been proposed~\cite{LP_Scott2010,LP_Scott2011} and adopted in various models to investigate the sources of bacterial growth variation, such as antibiotic exposure or heterologous expression. All the models agree that the growth rate function should depend on the ribosomal component, which is the main limiting resource of bacterial gene expression machinery; however, a consensus on the model structure is still an open question. Controversially, for example, in~\cite{growth_rate_of_Ecoli} and~\cite{A_numbers_game_ribosome_density}, the proposed model for cell growth is a function of the total number of ribosomes; however, while in the first case the relationship is linear, a Michaelis-Menten function is adopted in the latter. On the other hand, in other models~\cite{Cellular_perception_of_growth_rate,proof_load_effect,weisse}, the proposed growth rate is a linear function, as in~\cite{growth_rate_of_Ecoli}, but it depends on the active fraction of ribosomes (i.e., ribosomes bound with mRNAs), and also on available energy and total protein mass, as perturbations in gene expression or nutrients may all have an effect on cell growth.

To make the scenario even more fuzzy, one more canonical example of ribosomes usage in synthetic biology applications can be found in~\cite{del_vecchio} where, while a growth rate function was not explicitly defined, the authors provided the ratios of free ribosomes over their total amount; based on this information, a growth rate relationship with ribosomes must have at least a region of load where variation in ribosomes distributions among mRNAs do not imply significant growth defects. Similar considerations were also made by Bartoli et al., who showed that coupled transcription–translation control reveals how ribosome limitation nonlinearly links gene expression to growth rate~\cite{Bartoli2020}.

The use of a specific growth function depends on the level of detail included in a model in terms of molecules or processes that are included. Consequently, growth functions must be evaluated for their biological plausibility.

\begin{sidebar}{Biological scenario} \label{biological scenario}
\setcounter{sequation}{0}
\renewcommand{\thesequation}{S\arabic{sequation}}
\setcounter{stable}{0}
\renewcommand{\thestable}{S\arabic{stable}}
\setcounter{sfigure}{0}
\renewcommand{\thesfigure}{S\arabic{sfigure}}
DNA is an extraordinarily versatile molecule, serving as the genetic blueprint for all living organisms.
DNA is a molecule composed of nucleotides that encodes the information required for the development and functioning of all living cells.
The process by which DNA is decoded and used is described by the central dogma of molecular biology~\cite{crick1970central}. According to this framework, the coding sequences in DNA, known as genes, are transcribed into RNA molecules. These RNA molecules then serve as templates for the translation into proteins by ribosomes, which are responsible for carrying out cellular functions. Proteins, as the functional end-products of the genetic program, play a crucial role in regulating all aspects of cellular life, including metabolism, structure, and signal transduction. Additionally, DNA undergoes replication, ensuring the faithful transmission of genetic information to progeny cells during cell division (Fig.~\ref{fig:centralDogma}).
\sdbarfig{\includegraphics[width =.7\columnwidth]{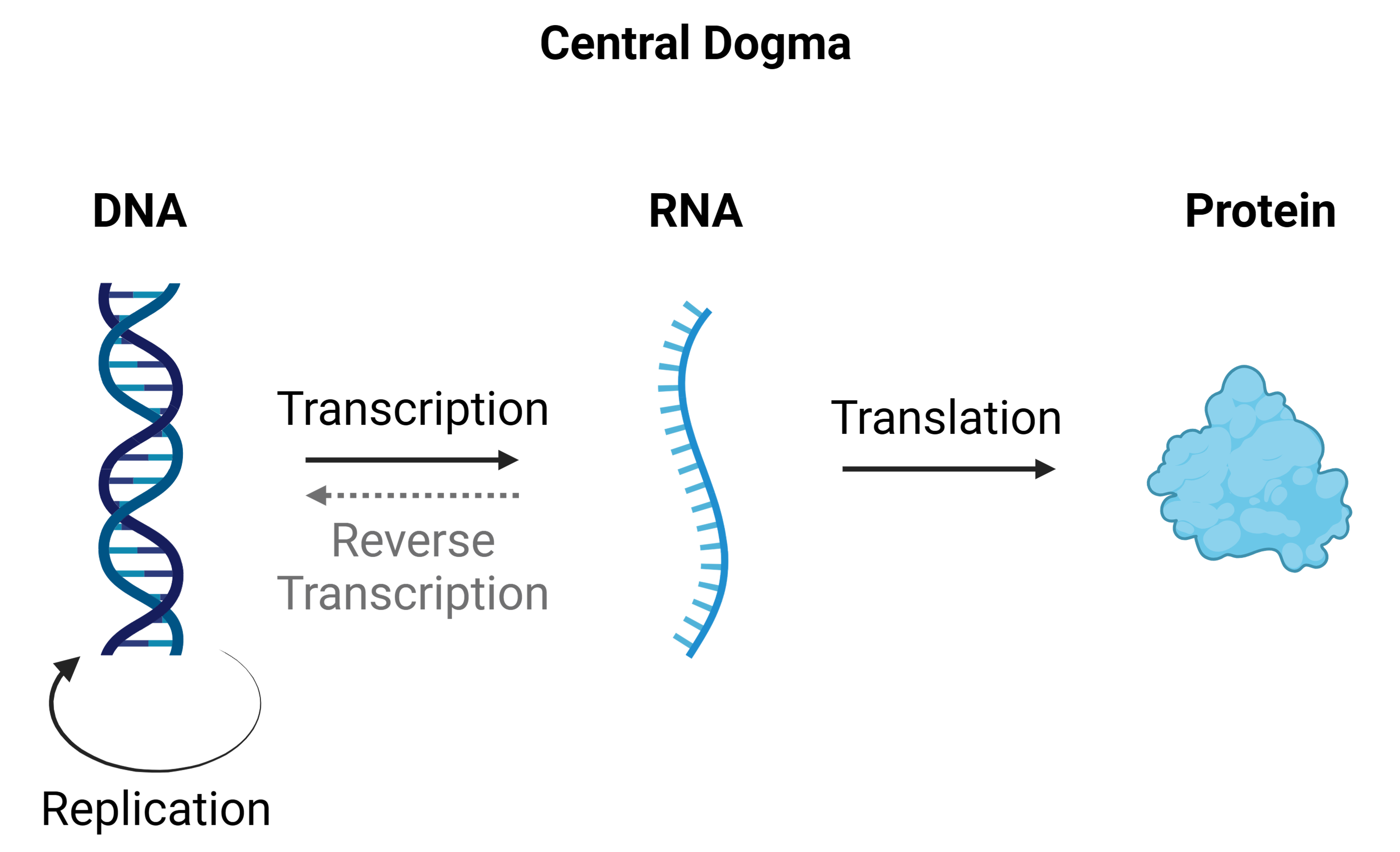}}{\textbf{The Central dogma.} DNA coding sequences are transcribed into RNA molecules by RNA polymerase, which are in turn translated into proteins by ribosomes. Moreover, replication of DNA molecules enables the transmission of genetic information to subsequent generations.\label{fig:centralDogma}}	
In this study, the growth of an \textit{Escherichia coli} cell is described, starting with an explanation of the chemical reactions and the components involved.
The expression dynamics of a gene can be modeled starting from three main processes, described by the following reactions: (a) transcription, (b) translation, and (c) molecule loss. In addition, cellular functions often involve (d) the formation of new complexes or molecules through association reactions. These chemical reactions are illustrated in Fig.~\ref{fig:resume}; mathematical representations of these processes are extensively examined in the literature \cite{del_vecchio,alon2019introduction,Ingallis2013} and can be summarized as follows:
\begin{itemize}
\item\textit{Transcription}: its standard form is:
		\begin{equation}\label{transcription}
		\ce{D + RNAP <=>[p^+][p^-] CTR ->[$\omega$] D + m + RNAP}	\end{equation} 
		During this phase, the \textit{DNA} ($D$) is used as a template for the synthesis of \textit{mRNA} ($m$). More specifically, this happens with the support of the \textit{RNA} polymerase ($RNAP$), which can "open" the double helix of the \textit{DNA}. This initial process forms an open complex ($CTR$). Afterward, \textit{RNAP} can proceed with the formation of the \textit{mRNA} sequence, which also contains a region called \textit{Ribosome Binding Site} (\textit{RBS}), needed for the downstream process.
		\\In the following, the transcription process will be approximated, under the assumption that $RNAP$ is constant, as:
		\begin{equation}
		\ce{D ->[$\omega$] D + m}
		\end{equation}
\item\textit{Translation}:  its standard form is:
		\begin{equation}\label{translation}
		\ce{m + R <=>[a][d] c ->[$\beta$]m + R + P}
		\end{equation}
		It occurs downstream of the \textit{Transcription} process. Once the \textit{mRNA} is produced, the ribosome $R$ binds it in the Ribosome binding site (\textit{RBS}) region, forming a complex, $c$, enabling the actual translation process to begin (reaction in Eq.~\eqref{translation}). The final product is the protein $P$.
		
\item\textit{Molecule loss}: all the intracellular components are subject to a decrease, due to molecule decay or dilution during cell division. This is pointed out by a molecule loss reaction, which is: 
		\begin{equation}\label{degradation}
		\ce{A ->[$\psi_A$] 0}
		\end{equation}
		where $A$ is a generic component and $\psi_A$ is the rate of loss that depends on the component, and represents the actual summation of the spontaneous decay of $A$ and the cell growth rate.
		
\item\textit{Molecular interactions}: The final reactions considered are auxiliary interactions that are essential for cellular dynamics, leading to the formation of new molecular complexes.
		Its general form contains two reactants $A$ and $B$ and a final product $C$ (usually one reactant is a protein):
		\begin{equation}
		\ce{A + B ->[$\epsilon$] C}
		\end{equation}
		where $\epsilon$ is the rate of association between the reactants.
	\end{itemize}
	\sdbarfig{\includegraphics[width=.9\linewidth]{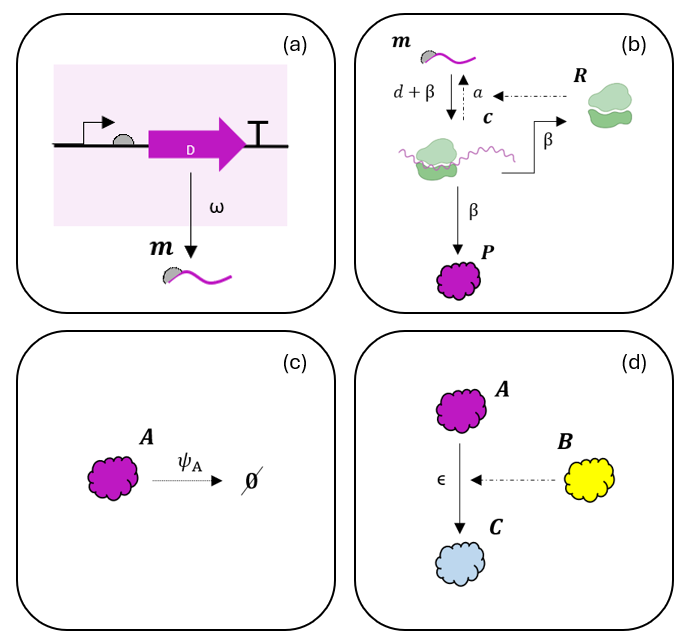}}{\textbf{Sketches of chemical reactions.} (a) Transcription of gene $D$ into mRNA $m$ with rate $\omega$. (b) Traslation of mRNA $m$: the ribosome $R$ binds $m$ with  association constant $a$ and dissociation constant $d$ forming a ribosome-mRNA complex that produces protein $P$ at rate $\beta$. (c) Loss of protein $A$ at rate $\psi_A$. (d) Molecular interactions of reactants $A$ and $B$, producing the final product $C$. \label{fig:resume}}
        \end{sidebar}
\setcounter{equation}{0}
\setcounter{figure}{1}

The main challenge in modeling growth rate and expression burden primarily resides in the fact that ribosomes, which are the common, shared factory for producing all proteins in a cell, are also responsible for producing proteins necessary to create the ribosomes themselves, thus giving rise to a complex feedback interaction. Such a mechanism also includes ribosomal regulations at multiple levels (transcriptional and post-transcriptional), detailed below, which inevitably complicate the model definition and identification steps.

Models describing resource partitioning have been proposed to investigate the hidden interactions that emerge at the gene expression level in the presence of a significant cell load, imposed by the incorporation of additional expression units~\cite{LP_Qian2015,LP_CarbonellBallestero2016,LP_Gorochowski2016,LP_Darlington2018,LP_Darlington2023}. These models have a relevance for the study of specific phenomena observed in experimental data, such as couplings between gene expression or growth feedback, and include a small number of parameters. Nonetheless, such models often lack an explicit representation of ribosomal species or do not link the burden effects to growth rate perturbation. To provide a more comprehensive description of cell physiology, cell models have been recently proposed~\cite{proof_load_effect,weisse,sechkar24,santosnavarro21,liao2017,Resource_aware_whole_cell_model,LP_AhnHorst2022,LP_Bosdriesz2015,LP_Erickson2017,LP_Droghetti2025}, integrating several biological processes that involve limiting resources, encompassing ribosomes and nutrients (providing ATP to the cell). Despite being coarse-grained models, their granularity is too detailed to be efficiently generalized and exploited in the rational design of synthetic genetic circuits, requiring notable efforts in retrieving the values of all the involved parameters.

Noteworthy complementary efforts in synthetic biology aim to overcome the detrimental effects of resource sharing by control theory-based designs in individual populations or multicellular consortia to define efficient production and division of labor strategies~\cite{Martinelli2025,Fiore2017,Ruolo2021,Annunziata2017,Campanile2025,Gorochowski2012,CeroniFeedback}.

Coarse-grained cell models efficiently capture naturally occurring autocatalytic loops and resource limitations in ribosomes (which serve as the translation machinery needed for their own production) and in energy (which is necessary for protein production, including proteins needed to import nutrients and convert them into ATP). Nonetheless, such models often lack additional regulatory mechanisms on the ribosomal species that occur in bacteria. One of them is the post-transcriptional negative feedback of ribosomal proteins and their mRNAs~\cite{Nomura_feedback}, a key topic of this work, which is illustrated in detail in the next section.

Given this scenario, the contribution of this paper is to propose a minimal mathematical model that extends the preliminary work presented in~\cite{GNB} to describe a bacterial cellular system  with this usually neglected feedback loop, yielding results consistent with biological evidence. Rather than enhancing the earlier cell models with the feedback mechanism, a minimal, specific model was developed to isolate the effects of this endogenous control strategy and to determine whether it could account for experimental observations of cell load and growth rate.

This work is organized as follows. First, the role of ribosomes in bacterial cells will be elucidated, along with the feedback mechanisms that regulate their levels. Second, an overview of experimental data demonstrating the effect of burdensome gene expression on bacterial growth rate and protein expression is provided. Third, the mathematical model is illustrated, defining the minimal set of molecular reactions occurring in a cell, their differential equations, the growth-rate function, and the chosen parameter values. Three increasingly complex models were considered to effectively investigate the effects of the ribosomal feedback, as detailed in the next section. Fourth, a results section includes the analysis of three key model outputs that will be discussed considering the biological plausibility of their trends, namely ribosome levels, growth rate, and protein expression, which define the cell load effects. Finally, a sensitivity analysis is provided to discuss whether the conclusions are general upon parameter value variations. A set of sidebars is also reported to illustrate: the biological mechanisms underlying bacterial gene expression; experimental procedures for characterizing cell load effects in recombinant bacteria; how to define gene reaction networks, the related differential equations, and how to simplify them under specific assumptions; details on bacterial growth and its mathematical modeling.

\section{The role of ribosomes}
The transcription and translation processes work together to lead to protein synthesis, the primary biochemical process that occurs within cells. Indeed, proteins are responsible for many aspects of cellular life, including cell shape and internal organization, product synthesis and waste disposal, and routine maintenance; furthermore, they receive signals from outside the cell and trigger intracellular responses~\cite{importance_of_proteins}.

Even though different proteins, translated from different \textit{mRNAs}, are assigned to different tasks, all of them share one component: the ribosome $R$. From this statement, it is straightforward to understand the importance of the ribosomes; they are constituted by some of the 52 ribosomal proteins contained in the cell (\textit{R-Proteins}, $P$) and three ribosomal \textit{RNAs} (\textit{rRNA}, $r$) \cite{Nomura_feedback},\cite{num_r_proteins}, namely:
	\begin{equation}\label{R_production}
	\ce{P + r ->[$\sigma_p$]R}
	\end{equation} 

Since the cellular system is highly complex and contains many different species, some proteins have been grouped together based on their components and/or final products.  
First, \textit{ribosomal} species serve as a general category that includes all molecular species that involve ribosomes $R$, either as free ribosomes or as ribosome-mRNA complexes, regardless of the specific mRNA species being translated.
Then, the existence of only three species of proteins will be assumed:
\begin{itemize}
\item\textit{Basal} species $B$: it gathers together all the proteins that are fundamental for the survival of the cell (i.e., proteins that are needed for the basic cell life).
\item\textit{R-Protein} species $P$: it is known from the literature that the bacterial cell contains 52 types of \textit{R-Proteins}~\cite{num_r_proteins}, which participate in the ribosome formation. Since it is not worth considering them singularly (if every type of \textit{R-Protein} were considered, there would need to be $52$ equations just for describing the \textit{R-Proteins} dynamics), they have been gathered into the more general \textit{R-Protein} species.
\item\textit{Heterologous} species, further split into $L$ and $F$: both are additional proteins added to the cell via synthetic genetic circuits. $L$ requires a non-negligible amount of energy and consequently influences the system by eventually causing cell \textit{load}. On the other hand, $F$ does not require significant cellular resources to be expressed and is used as a proxy for resource availability only in the \textit{Analysis of Load effect} section, for the sake of clarity. The most common proteins used to study cellular load are constitutively expressed reporter proteins - such as the red or the green fluorescent protein (\textit{RFP} and \textit{GFP}) - used in laboratory experiments, as they can be quantified using instruments like fluorimeters, plate readers, fluorescence microscopy, or flow cytometers~\cite{LP_Ceroni2015,LP_Gyorgy2014}. Roughly, the device measures the intensity of the emitted light in response to excitation, as a function of wavelength. In biochemical experiments, this spectrum enables indirect quantification of protein expression levels in cell populations or single cells.
\end{itemize}	

\begin{sidebar}{From kinetic reactions to dynamic systems}
\setcounter{equation}{5}
\renewcommand{\thesequation}{S\arabic{sequation}}
\setcounter{stable}{0}
\renewcommand{\thestable}{S\arabic{stable}}
\renewcommand{\thesfigure}{S\arabic{sfigure}}
\sdbarinitial{B}iological systems can be modeled in different ways, depending on the level of description and resolution wanted.

Depending on the system size, the impact of noise, and the type of questions being addressed, several mathematical frameworks can be applied. At a fine-grained, stochastic level, the Chemical Master Equation (CME) captures the probabilistic nature of molecular interactions. CME becomes analytically intractable for systems with high complexity, i.e., with different molecular species that are produced at large molecule counts. For these systems, approximations such as the Chemical Langevin Equation or the Fokker–Planck Equation can be used to reduce computational complexity. In cases where fluctuations are negligible, deterministic approaches such as reaction rate equations are sufficient. The choice of the appropriate modeling framework requires balancing accuracy, computational feasibility, and the biological insights sought \cite{del_vecchio}. Starting from the chemical reactions that occur in it, the aim is to derive a set of differential equations.\\
The chemical reactions have been modeled using the \textit{Reaction Rate Equations} method. According to this method, the system is considered as a set of species $S_i$ that interact with each other, and in particular, it takes into account the concentration $x_i$:
\begin{equation}
		x_i = [S_i] = \frac{n_{S_i}}{\Omega}
\end{equation}
where $n_{S_i}$ is the number of molecules of the species $S_i$ and $\Omega$ is the given volume.\\
Furthermore, the application of this method is based on a strong assumption: all the reactions occur in a well-stirred volume, which means that the rate of interaction between two species is uniform and does not have any spatial influence.\\
The final aim is, as already declared, to describe the system with a set of differential equations 
\begin{equation}
		\dot{x} = f(x,\theta)
\end{equation}
where $x \in \mathbb{R}^n$ is the vector that contains all the species of the system, $\theta \in \mathbb{R}^p$ is the vector of the parameters, and $f: \mathbb{R}^n \times \mathbb{R}^p \longrightarrow \mathbb{R}^n$ rules the change in the concentrations.\\ 
To describe the process of derivation of the differential equations, a basic biomolecular reaction is proposed as an example, namely:
\begin{equation}\label{general reaction}
		\ce{A + B <=>[k_f][k_r] AB}
\end{equation}
The reaction in Eq.~\eqref{general reaction} can be interpreted as follows: every time the forward response occurs, the number of molecules of \textit{A} ($n_A$) and \textit{B} ($n_B$) must be decreased by one, while the number of molecules of \textit{AB} ($n_{AB}$) must be increased by one. The reasoning is the same with the reverse reaction ($n_{AB}$ must be decreased while $n_{A}$ and $n_{B}$ must be increased by one). However, only one reaction can occur at a time, and it is regulated by the likelihood. In particular, the likelihood of the forward reaction (in a time interval $dt$) is:
\begin{equation}\label{likelihood forward}
		a_f(q)dt = (k_f/\Omega)n_An_Bdt
\end{equation}
where $q$ is the configuration of the system and $k_f$ is the parameter that depends on the reaction (rate of association). For the reverse reaction, the likelihood is:
\begin{equation}\label{likelihood reverse}
		a_r(q) = k_r n_{AB}
\end{equation}
where $k_r$ still depends on the reaction (dissociation rate).\\
Now the equation that expresses the variation of $n_{AB}$ can be written also considering the contribution of the likelihoods:
\begin{equation}\label{eq in number of mol}
		n_{AB}(t + dt) = n_{AB}(t) + a_f(q- \xi_f)dt - a_r(q)dt.
\end{equation}
The number of molecules after a time interval $dt$ is the amount at the previous instant $n_{AB}$ plus the contribution of the forward reaction ($\xi_f$ represents the change in the configuration) and minus the contribution of the reverse reaction. Both likelihood contributions can be roughly interpreted as the probabilities that the reaction occurs.
    
To convert Eq.~\eqref{eq in number of mol} into an equation that involves the concentrations, it is enough to divide each term by the given volume $\Omega$ and to replace the expression of the likelihoods with Eq.~\eqref{likelihood forward} and Eq.~\eqref{likelihood reverse}:
	\begin{equation}
		[AB](t+dt) = [AB](t) + (k_f/\Omega^2)n_An_Bdt - k_r n_{AB}/\Omega dt
	\end{equation}
	Furthermore, since $n_A / \Omega = [A]$, $n_B / \Omega = [B]$ and $n_{AB} / \Omega = [AB]$:
	\begin{equation}
		[AB](t+dt) - [AB](t) = (k_f[A][B] - k_r[AB])dt
	\end{equation}
	If $dt \longrightarrow 0$:
	\begin{equation}
		\frac{d}{dt}[AB] = k_f[A][B] - k_r[AB]
	\end{equation}
which is the differential equation that regulates the changes in concentration of species \textit{AB}.
    
Analogously, the equations of \textit{A} and \textit{B} can be derived:
	\begin{equation}
		\frac{d}{dt}[A] =  k_r[AB] - k_f[A][B]  \qquad \text{and} \qquad \frac{d}{dt}[B] =  k_r[AB] - k_f[A][B]
	\end{equation}
	For the sake of simplicity, in the next chapter, the concentrations of the species will not be indicated with the squared brackets (i.e, $AB$ instead of $[AB]$).

    \end{sidebar}
\setcounter{figure}{1} 
\setcounter{equation}{1}

\subsection{Ribosomal feedback}
Given the leading role of ribosomes, the cell tends to control their production to ensure the right amount to fulfill its metabolic requirements. 

It is known from the literature that an inner feedback mechanism regulates the synthesis of ribosomes, aiming to prevent their accumulation in the cell~\cite{Nomura_feedback,yates1981feedback, robert2001ribosomal,LP_Shen2025}. It may sound like a counterintuitive strategy, as the more ribosomes are available, the more proteins are synthesized, and the more tasks are fulfilled. However, the reason the cell seeks to limit ribosome concentration lies in the energy required. Indeed, like all cellular processes, ribosome synthesis requires energy (i.e., \textit{ATP}); therefore, it is not worth producing more ribosomes than needed.
Because of the components involved in the synthesis (as shown in Eq.~\eqref{R_production}), the feedback should concern at least one between \textit{R-Protein} and \textit{rRNA}.

Feedback mechanisms occurring at the ribosomal level rely on transcriptional and post-transcriptional regulations, aimed at preventing their accumulation and optimizing the allocation of translational resources~\cite{LP_Burgos2017,LP_Shea2023}. In the post-transcriptional regulation mechanism, certain \textit{R-Proteins} work as inhibitors of protein synthesis from their own \textit{mRNAs}, limiting their unnecessary translation~\cite{Nomura_feedback,LP_Zengel1994}. At the transcriptional regulation level, one of the ribosome feedback models proposed in the literature~\cite{Control_of_ribosomes} states that the cell can prevent their accumulation by using a feedback regulation on \textit{rRNA}. In particular, this is an auto-regulatory process through a negative feedback loop~\cite{mechanism_regulation_rRNA} mediated by a class of molecules, called alarmones (e.g., ppGpp), produced in stressful conditions when amino acids are limiting~\cite{Snoeck2024}. Other authors~\cite{control_of_ribosome_synthesis} reported that there may be a mechanism that regulates the parallel production of \textit{R-Proteins} and \textit{rRNAs}. This should guarantee similar concentrations for both reactants, since the insufficient amount of one component is detrimental to ribosome synthesis, as observed in Eq.~\eqref{R_production}. In other words, it does not make sense to produce a massive amount of $rRNA$ if the \textit{R-Proteins} are missing and vice versa. Overall, the alarmone-mediated transcriptional feedback acts as a switch to reallocate the cellular resources under stressful conditions. While cell models typically do not include these feedback mechanisms, a few of them explicitly consider ppGpp as a resource-dependent regulator of ribosome biogenesis ~\cite{sechkar24,liao2017,LP_AhnHorst2022,LP_Bosdriesz2015,LP_Erickson2017,LP_Droghetti2025}. Only one cell model of \textit{E. coli} was found to consider both feedback mechanisms, even though their individual contribution was not investigated~\cite{roy2021}.

In this work, we focus on the \textit{R-Proteins}- mediated post-transcriptional feedback, which is usually neglected in other works on ribosome allocation models. This post-transcriptional feedback loop could be predominant in bacteria with compromised alarmone biosynthesis, which is required for efficient triggering of the transcriptional response~\cite{LP_Lee2018}.

Three different models with incremental complexity have been evaluated, namely: 
\begin{itemize}
    \item[\textbf{\textit{M0:}}] a simplified model of gene expression lacking the autocatalytic production of ribosomes;
    \item[\textbf{\textit{M1:}}] a model decoupling ribosomal components into ribosomal RNA (rRNA) and ribosomal proteins (R-protein) that are individually expressed, restoring the autocatalytic production of ribosomes;
    \item[\textbf{\textit{M2:}}] a model including the negative feedback in R-protein translation by the inhibition of the relative mRNA species. 
\end{itemize}
These models assume that ribosomes are the bottleneck of burdensome protein expression, as it is assumed in other works~\cite{sechkar24}, while transcriptional resources are not limiting in the scenarios of interest of this study.

We demonstrate by mathematical modeling that the minimal level of description requires the existence of this post-transcriptional feedback to generate three different regions in which increasing cell load can lead to: (i) no effects on growth and protein synthesis, (ii) burden effects on protein production, maintaining a relatively conserved growth rate, and, lastly, (iii) defects in both protein production and growth rate. It is worth noting that the last scenario is underrepresented in the literature regarding experimental data on synthetic circuits. In fact, circuit designs with no impairment on cell growth are typically desired, and a part of the studies on cell load also rely on simplified biological systems in which hidden interactions in gene expression occur before growth rate defects become apparent (i.e., the region (ii), corresponding to an intermediate burden regime, where protein synthesis is impaired while the growth rate remains relatively stable), to isolate the contribution of resource limitation on gene expression~\cite{LP_Gyorgy2014,LP_McBride2021}. Experimental results demonstrating the possible effect of metabolic burden on the growth rate of bacterial cells are presented in the \textit{Experimental evidence of metabolic burden} section. 

A schematic representation of the feedback mechanism mediated by \textit{R-Proteins} is shown in Fig.~\ref{fig:es1}c, where the negative feedback loop is highlighted in blue. As already specified in this section, for the sake of simplicity, all the different species of \textit{R-Proteins} are grouped together in a unique species $P$, and the \textit{rRNAs} are grouped together into the generic component $r$.

Moreover, the regulation of ribosomal protein synthesis can be seen as a competition between \textit{rRNA} and \textit{mRNA} for the same pool of R-proteins ($P$). 
Indeed, the inhibitory feedback can be described by the following reactions:
	\begin{equation}
		\ce{P + m_p ->[$\alpha_p$] {P:m_p}} \label{feedback_eqn}
	\end{equation}
	\begin{equation}
		\ce{{P:m_p} ->[$\gamma$] P }
	\end{equation}
	\begin{equation}
				\ce{P ->[$\mu$] 0}
	\end{equation}
where $\alpha_p$ denotes the binding rate of the ribosomal protein to its own mRNA, and therefore quantifies the strength of the negative autoregulatory feedback. In particular, the reaction in Eq.~\eqref{feedback_eqn} competes with the binding of ribosomal proteins to \textit{rRNA} described in Eq.~\eqref{R_production}. However, it is reasonable to assume that ribosomal proteins have a higher affinity for \textit{rRNA} than for \textit{mRNA}~\cite{control_of_ribosome_synthesis,LP_Burgos2017}. Otherwise, the binding for mRNA inhibition would be stronger than the binding to rRNAs, impairing ribosome assembly, which is required for the translation of R-proteins and any other cellular protein species, which would represent an inefficient scenario for cellular economy.

\section{Experimental evidence of metabolic burden}
\begin{sidebar}{Details on the experimental burden proof}
\renewcommand{\thesequation}{S\arabic{sequation}}
\renewcommand{\thesfigure}{S\arabic{sfigure}}
\section{Genetic circuit} \sdbarinitial{T}he recombinant NR34 strain is based on the TOP10F1 bacterial host. A medium-copy plasmid (pSB3K3 from the MIT Registry of Standard Biological Parts; \url{https://registry.igem.org/)} contains a reporter gene coding for the red fluorescent protein (RFP; BBa\_E1010) downstream of the $P_{lux}$ promoter from \textit{Vibrio fischeri} and a strong ribosome binding site (RBS) (see \cite{DeMarchi2024} for the full sequences of the plasmids and details on the used strain). RFP expression is triggered in response to an exogenously added small and diffusible molecule (HSL) via the LuxR/$P_{lux}$ system. In particular, HSL binds LuxR, which activates the cognate inducible promoter. A high expression of LuxR is guaranteed by a constitutive cassette placed on a low-copy plasmid (pSB4C5).

RFP causes cellular load due to its high expression level, not due to protein-specific toxicity. In fact, its copy number, transcription rate and translation efficiency were the main drivers of cell load in this circuit, as previously reported \cite{DeMarchi2024}. The NR34 strain also includes a constitutive expression cassette for the green fluorescent protein (GFP; BBa\_E0040), used as a proxy of cell load \cite{LP_Ceroni2015} in the low-copy plasmid.

\section{Experimental setup} Bacteria were grown in 50-ml tubes at 37$\degree$C, 200 rpm, in 4 ml of M9 medium (11.28 g/L M9 salts M6030 - Sigma Aldrich, 2 mM MgSO4, 0.1 mM CaCl2) with 0.2\% casamino acids, 0.4\% glycerol, 100 mg/l ampicillin, 25 mg/l kanamycin, and 12.5 mg/l chloramphenicol. Pre-cultures from single colonies (N = 3) were grown overnight, then diluted 70-fold in fresh medium containing HSL (K3007, Sigma-Aldrich) at the indicated concentrations, and incubated for 6 h. The optical density at 600 nm ($OD_{600}$) of 200 $\mu L$ samples taken from the culture was measured every hour in a microplate reader (Infinite F200, Tecan).

Bacterial samples were also taken for flow-cytometry analysis (N$\geq3$) from exponentially growing cultures prepared as above, or incubated in a deep-well plate (0.5 mL, grown at 37$\degree$C, 750 rpm for 5 h). The two procedures gave highly reproducible data. Flow-cytometric analysis was performed as previously described \sidecite{Bertaux2022}\cite{DeMarchi2024}. Briefly, 200 $\mu$l of cultures were transferred into a 96-well microplate, 1 mg/ml gentamicin was added to stop translation, and samples were incubated at 37 $\degree$C for 1 h to guarantee fluorescent protein maturation. Samples were finally analyzed using a benchtop cytometer (Guava EasyCyte 14HT, Luminex), measuring 5,000 events.

\section{Data analysis} Raw $OD_{600}$ data were background-subtracted and log-transformed, using the absorbance of M9 as the blank. Growth rate ($\mu$) was computed as the slope of log($OD_{600}$) vs time. The GRN-B and ORG-G channels were used for detecting GFP and RFP, respectively. Data were gated with forward scatter (FSC, $10^{2}$/$10^{3}$) and side scatter (SSC, $10^{0}$/$10^{4}$) lower/upper thresholds. Fluorescence values were normalized by FSC \sidecite{Bertaux2022}, and the per-cell fluorescent protein synthesis rate ($S_{cell}$ was computed by multiplying the normalized fluorescence value by the growth rate $\mu$ \sidecite{Kelly2008}. The $S_{cell}$ values at each HSL concentration were finally normalized by the highest mean value.

Statistical tests were performed using GraphPad Prism 10.6.1. One-way ANOVA was performed to compare growth rate and GFP values across the HSL conditions. Multiple comparisons were performed to detect statistical differences compared with the no-induction condition using the BKY false discovery rate (FDR). A p-value cutoff of 0.05 was always used for statistical significance.

\end{sidebar}

\begin{figure}[H]
    \centering
\includegraphics[width =1\columnwidth]{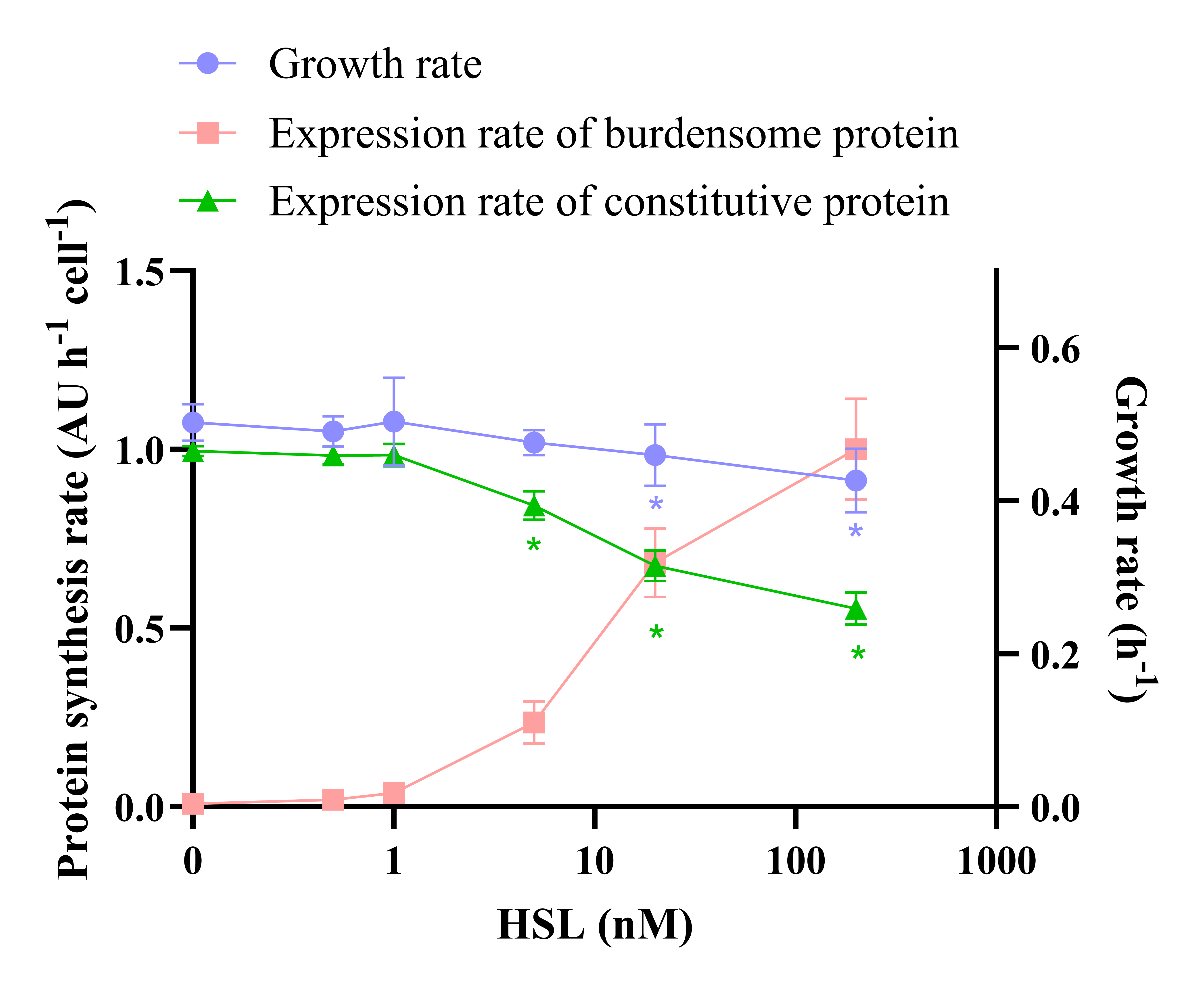}
\caption{Growth rate and expression rates of GFP (cell load proxy) and RFP (burdensome gene) in recombinant \textit{E. coli} strain. Data are reported at steady state and during the exponential growth phase as a function of HSL, which drives RFP expression. Asterisks indicate conditions in which a significant decrease of GFP and $\mu$ values was detected (p-value<0.05, ANOVA).}
\label{fig:spaso}
\end{figure}

An example of variation in cellular load and growth rate of a burdened bacterial culture is reported in Fig.~\ref{fig:spaso} and details are provided in the \textit{Experimental characterization of cell load} sidebar.
In this case, a recombinant strain (NR34, previously described and characterized \cite{DeMarchi2024}), expresses the red fluorescent protein (RFP) in response to an exogenously added small and diffusible molecule (N-3-oxo-C6 homoserine lactone - HSL) via the inducible LuxR/$P_{lux}$ system. LuxR is produced by a constitutive expression cassette within the circuit. RFP causes cell load due to its high expression level, driven by a strong promoter ($P_{lux}$) and highly efficient ribosome binding site (RBS). The strain also includes a constitutive expression cassette for the green fluorescent protein (GFP), used as a proxy of cell load: the per-cell GFP synthesis rate is expected to be constant, and its decrease indicates cell load in the context of limited translational resource availability \cite{LP_Ceroni2015}. With the copy number, promoter, and RBS used in this circuit, GFP expression does not exert any detectable metabolic load on the strain \cite{Pasotti2017}.

The expression rate of the burdensome protein (RFP) spanned a 130-fold range upon induction by HSL, showing a half-induction concentration of  $\sim$10 nM. Both GFP and growth rate were affected by cell load. GFP showed a 15\% decrease at 5 nM HSL, and larger effects at 20 and 200 nM HSL (32\% and 44\% decrease, respectively). On the other hand, the growth rate showed a more modest decline, not significant at HSL concentrations below 20 nM (less than a 5\% decrease). The maximum reduction in growth rate (15\%) was observed at the 200 nM inducer level.

This example highlights that the growth rate can remain stable even when the cell load is increasing and affecting the production rate of other proteins. This trend is consistent with previous reports \cite{Pasotti2017,LP_Ceroni2015,LP_Gyorgy2014} in which burdensome expression affects the synthesis of other uncoupled proteins, but bacterial growth rate shows a lower fold-reduction or even no reduction at intermediate levels of heterologous protein expression.

\section{Mathematical modeling}
\begin{sidebar}{Strategies for simplifying gene expression models}
\setcounter{equation}{15}
\renewcommand{\thesequation}{S\arabic{sequation}}
\setcounter{stable}{0}
\renewcommand{\thestable}{S\arabic{stable}}
\renewcommand{\thesfigure}{S\arabic{sfigure}}
\section{Time-scale separation: Rapid Equilibrium versus Steady State}\label{type of equilibria}
\sdbarinitial{O}nce the differential equations have been derived, the dynamics can be investigated. The following derivations are standard in the literature; see the textbook by Uri Alon \cite{alon2019introduction} for further details.\\
For a generic system in the form:
    \begin{equation}
        \frac{d x}{dt} = f(x,\theta)
    \end{equation}
the study of the equilibrium points is usually performed. The configuration that the system reaches at the equilibrium is called \textit{Steady State}, and it can be obtained by setting all the differential equations to zero:
	\begin{equation}
		\frac{d x}{dt} = 0
	\end{equation}
However, a different analysis can be executed with biological systems, which takes into account the velocity of the reactions and is called \textit{Rapid Equilibrium}~\cite{michaelis_menten}.\\
Consider a generic enzymatic reaction:
\begin{equation}\label{enzymatic reaction}
		\ce{E + S <=>[a][d] C ->[k] E + P}
\end{equation}
where $E$ is the enzyme, $S$ is the substrate, $C$ the complex and $P$ the final product. Furthermore, $a$ and $d$ are the association and dissociation constants, respectively, while $k$ is the catalytic rate constant.
    
The conservation of species is at the basis of reaction rate models, as species are typically transformed during reactions but neither created from nothing nor annihilated. Therefore, the following conservation of mass laws can be derived: 
\begin{eqnarray}
    &E^{TOT} = E + C \\
    &S^{TOT}= S + C + P
\end{eqnarray}
The corresponding set of differential equations (without considering any degradation) is:
	\begin{eqnarray}
		\dot{S} &=& - aES + dC \label{substrate}\\
		\dot{E} &=& - aES + dC + kC \label{enzyme}\\
		\dot{C} &=& aES - (d+k)C \\
		\dot{P} &=& kC  \label{product}
	\end{eqnarray}
Nevertheless, it can be assumed that the first reaction in Eq.~\eqref{enzymatic reaction}, namely the formation of complex $C$, is faster than the synthesis of the final product $P$, and consequently, it reaches the equilibrium first.\\
Setting  Eq.~\eqref{substrate} equal to zero, one obtains:
\begin{equation}\label{rapid_eq}
		\frac{E S}{C } = K_d
\end{equation}
where $K_d = \frac{d}{a}$ is called \textit{dissociation constant}. In addition, the total amount of the enzyme $E^{TOT} = E + C$ is usually constant at the equilibrium; thus, it can be substituted into the equation at the equilibrium Eq.~\eqref{rapid_eq}:
\begin{equation}
		\frac{(E^{TOT} - C)S}{C} = K_d \qquad \longrightarrow \qquad  C = \frac{S}{K_d + S }E^{TOT }
\end{equation}
Indeed, due to the mass conservation law, the total amount of a species $A^{TOT}$ is defined as the sum of the free species itself $A$ and the complexes that are directly derived from it.\\
Going back to the starting enzymatic reaction, now the formation rate of product $P$ at the \textit{Rapid Equilibrium} can be computed as follows:
\begin{equation}\label{rapid eq}
		\frac{d P}{dt} = k C = k E^{TOT } {\dfrac{S}{K_d + S} } = P_{max}{\dfrac{S}{K_d + S}}
\end{equation}
which is called \textit{Michaelis-Menten kinetics} and it describes the evolution of the final product $P$ as function of the substrate $S$ (Fig.~\ref{fig:michaelis-mentenaxis}).
    
\sdbarfig{\includegraphics[width =\columnwidth]{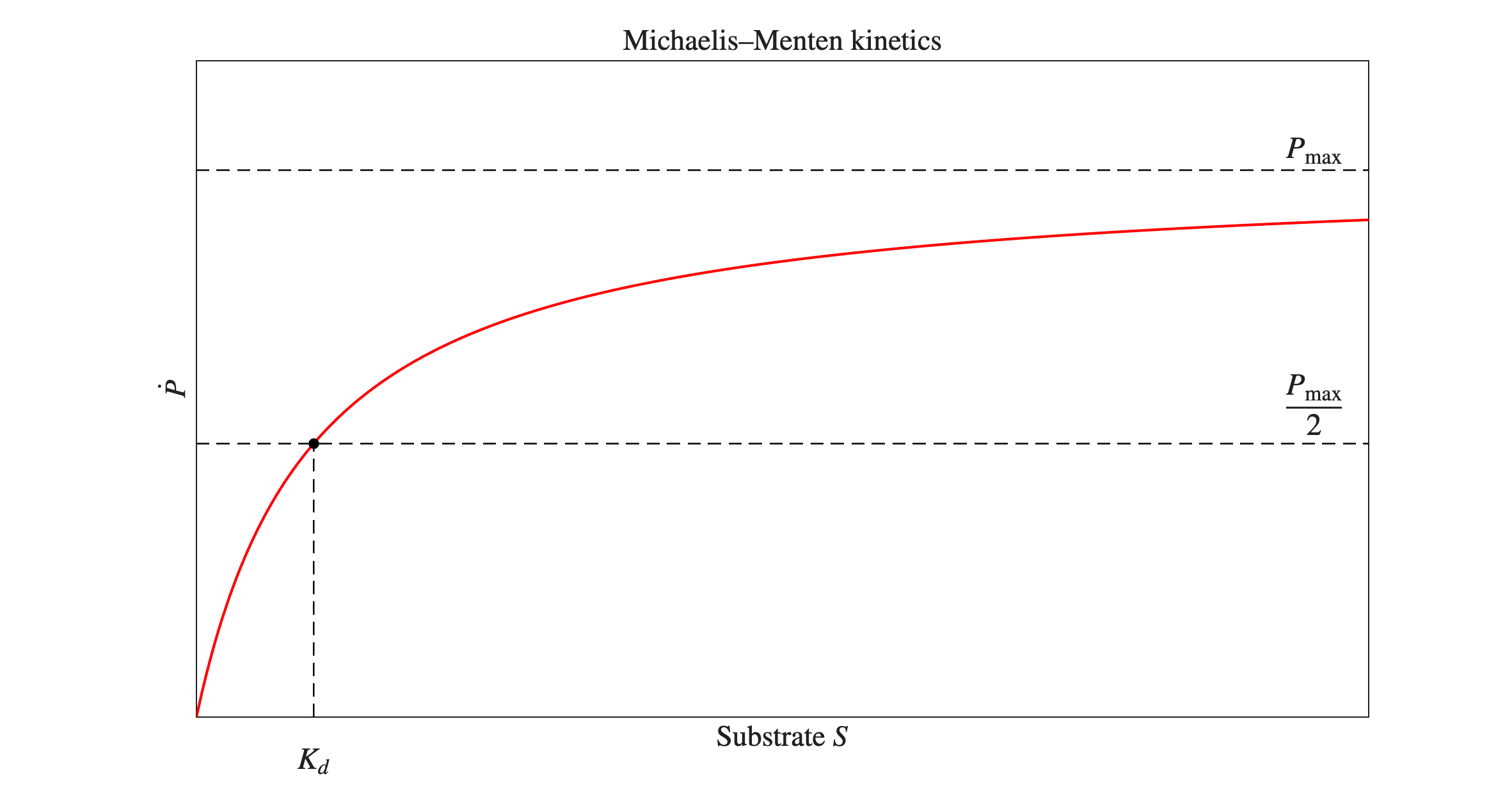}}{\textbf{Michaelis-Menten function.} $K_d = d/a$ is the dissociation constant.\label{fig:michaelis-mentenaxis}}
It may be beneficial to compare this result with the \textit{Steady State} solution. From  Eq.~\eqref{enzyme}:
\begin{equation}
		\dot{E} = - aES + dC + kC = - a(E^{TOT}-C)S + dC + kC = 0  
\end{equation}
And then:
\begin{equation}
		C = \frac{E^{TOT}}{1 + \frac{d + k}{aS}} = \frac{E^{TOT}}{ 1 + \frac{ K_m }{ S}}
\end{equation}
where $K_m = \frac{d+k}{a}$\text{ is called \textit{half saturation constant}}

Hence, the product $P$ at the \textit{Steady State} is:
\begin{equation}\label{ss}
		\dot P = kC = k E^{TOT} {\dfrac{S}{K_m+S}} = P_{max} {\dfrac{S}{K_m+S}}
\end{equation}
which is instead called \textit{Briggs-Haldane kinetics}.\\
In conclusion, the \textit{Rapid Equilibrium} study is based on the assumption $k << d$, which means that the association and dissociation of the substrate are faster than the product formation. Instead, the \textit{Steady State} does not consider any relation between the rates of the reaction. Due to these considerations, it is possible to assert that the Michaelis-Menten equation is a special case of the Briggs-Haldane equation. It is worth recalling that the set of ODEs (from Eq.~\eqref{substrate} to Eq.~\eqref{product}) does not account for any loss reaction (i.e., decay or dilution) of the components. However, if it were considered, the equation of $P$ (\eqref{rapid eq}) at \textit{Rapid Equilibrium} would not change, since it is based on the velocity of the reactions. Instead, the computation of the equilibria at the \textit{Steady State} would lead to a different formulation compared to Eq.~\eqref{ss}. 
\end{sidebar}

\begin{sidebar}{\continuesidebar}
\section{Overabundances: Negligibility of RNAP}\label{rnap_neg}
Gene expression in a circuit depends on shared transcriptional and translational resources. While both RNAP and ribosome availability can limit expression, the literature identifies ribosome limitation as the primary bottleneck \cite{resource_competition_Qian}. A generic chemical reaction that describes the transcription can be written as follows:
\begin{equation}
	\ce{D + RNAP <=>[p^+][p^-]CTR ->[$\omega$]D + m + RNAP} 
\end{equation}
where $D$ is the \textit{DNA}, $p^+$ and $p^-$ are the association and dissociation rate respectively between \textit{DNA} and \textit{RNAP}, \textit{CTR} is pre-\textit{mRNA} molecule, $\omega$ is the transcription rate, $m$ is the \textit{mRNA}.\\
The relative mass law kinetics are (molecule loss due to decay or dilution is not considered):
\begin{eqnarray}
		\dot{D} &=& p^- CTR - p^+ D \cdot RNAP + \omega CTR \\
		\dot{m} &=& \omega CTR \label{m in rnap}\\
		\dot{CTR} &=&  - p^- CTR + p^+ D \cdot RNAP - \omega CTR \\
		\dot{RNAP} &=& 0 \label{rnap cons}
\end{eqnarray}
where Eq.~\eqref{rnap cons} is equal to zero under the hypothesis of constant \textit{RNAP}.\\
Studying the differential equations at the steady state, \textit{CTR} is:
\begin{equation}\label{ctr}
		CTR_{ss} = {\dfrac{RNAP}{K}} D \qquad \text{ with } K = {\dfrac{p^-+ \omega}{p^+}}
\end{equation}
Assuming the concentration of RNAP constant, $CTR_{ss}$ depends uniquely on $D$. 

This consideration can lead to a possible reformulation of the chemical reaction:
\begin{equation}\label{simplified rnap_S}
	\ce{D ->[$\omega^{'}$]D + m}
\end{equation} 
Now, the relation between $\omega$ and $\omega^{'}$ must be found.\\
By the substitution of Eq.~\eqref{ctr} in Eq.~\eqref{m in rnap}, the dynamics of \textit{mRNA} becomes:
\begin{equation}
		\dot{m} = \omega {\dfrac{RNAP}{K}} D
\end{equation}
while the mass action law derived from Eq.~\eqref{simplified rnap_S} is:
\begin{equation}
		\dot{m} = \omega^{'}D
\end{equation}
Now it can be easily seen that the relation between the transcription rates is:
\begin{equation}
		\omega^{'} = \omega {\dfrac{RNAP}{K}}
\end{equation}
From this point on, the concentration of \textit{RNAP} can be considered as constant, leading to the reformulation of all transcription chemical reactions as in Eq.~\eqref{simplified rnap_S}.\\
For the sake of simplicity, the new transcription rate $\omega^{'}$ will be indicated with the symbol $\omega$ as well.
\end{sidebar}
\setcounter{figure}{2}
\setcounter{equation}{4}
\subsection{Biological hypothesis and mathematical assumptions}
The protein synthesis process consists of two phases: transcription and translation. The following chemical reaction generally describes the transcription phase:
\begin{equation}
	\ce{D + RNAP <=>[a][d]CTR ->[$\omega$]D + m + RNAP} \label{transcription rnap}
\end{equation}
Where $a$ and $d$ are the association and dissociation rates, respectively, between DNA (D) and RNA polymerase (RNAP), CTR is the pre-mRNA molecule, $\omega$ is the transcription rate, and $m$ is the mRNA.
As a first assumption, RNAP concentration can be assumed to be constant~\cite{del_vecchio}. Under this assumption, the chemical reaction Eq.~\eqref{transcription rnap} can be rewritten as:
\begin{equation}\label{simplified rnap}
    \ce{D ->[$\omega$^']D + m}
\end{equation}
The relation between the transcription rates can be found by considering the steady state condition of the mass law kinetics derived by Eq.~\eqref{transcription rnap}, which is:
\begin{equation}
    \omega^{'} = \omega {RNAP \dfrac{a}{d+\omega}}.
\end{equation}
The detailed mathematical derivation of this simplification is provided in the \textit{Overabundamces: negligibility of RNAP} sidebar. 
For the sake of simplicity, the new transcription rate $\omega^{'}$ will be indicated with the symbol $\omega$ as well.
To limit the number of parameters and assuming equal order of magnitude over time, the decay is considered equal for proteins, ribosomes, and ribosomal complexes (regardless of the species) and indicated with $\delta$; analogously, mRNA and rRNA decay is indicated with $\gamma_{R}$.\\
Another assumption concerns the loss rate of mRNA and rRNA ($\mu_{mRNA,rRNA}$), defined as the sum of the spontaneous decay rate ($\gamma_{R}$) and the growth rate ($\lambda$). From~\cite{del_vecchio} and ~\cite{LP_Page2018}, it is known that $\gamma_{R}$ is faster than $\lambda$. This leads to the approximation: $\mu_{mRNA,rRNA} = \gamma_{R} + \lambda \approx \gamma_{R}$.
Thus, the rate of decrease of mRNA and rRNA can be considered independent of the growth rate.

The last aspect concerns protein synthesis. For ribosomes $R$, as a first simplifying assumption (used in the \textit{M0} model, Fig. \ref{fig:es1}a), the process of ribosome synthesis can be described as a direct first-order production dynamic of rate $\beta_r$ from the DNA encoding rRNA ($D_r$):
\begin{equation}
    \ce{D_r ->[$\beta$_r] D_r + R}.\label{react_modA_R_first}
\end{equation} On the other hand, at a mechanistic level, ribosome production can be modeled  by decoupling the synthesis of ribosomal rRNA $r$ and R-proteins $P$ to form a mature ribosome $R$ (Fig.~\ref{fig:es1}b) as follows:
\begin{align}
	\ce{D_r& ->[$\omega_r$] D_r + r} \label{eqn rna}\\
        \ce{D_p& ->[$\omega$_p] D_p + m_p} \label{eqn mp}\\
        \ce{m_p + R& <=>[a_p][d_p] c_p ->[$\beta$_p] m_p + R + P} \label{eqn P}\\
        \ce{r + P& ->[$\sigma$_p]R}
\end{align}
where $\omega_r$ denotes the transcription rate of the DNA sequence D$_r$ which encodes for $r$, while $\omega_p$ represents the transcription rate of D$_p$, encoding the R-protein P, into its corresponding mRNA $m_p$. The parameters $a_p$ and $d_p$ are the association and dissociation constants between $m_p$ and $R$, $\beta_p$ denotes the translation rate of m$_p$ into P, and $\sigma_p$ represents the association rate between r and P to produce the ribosome.  This ribosome synthesis model is adopted in models \textit{M1} and \textit{M2} to describe the autocatalytic ribosomal production. Lastly, as reported by Nomura et al.~\cite{Nomura_feedback}, an additional level of complexity can be incorporated, as certain R-Proteins have been shown to act as inhibitors of protein synthesis from their own mRNAs. This negative feedback has been incorporated into model \textit{M2} (Fig.~\ref{fig:es1}c).

Accordingly, it is possible to identify up to four different ribosomal species (i.e., components involving free ribosomes in their characteristic chemical reactions): (i) free ribosomes R, (ii) the basal species c$_b$ representing all ribosomes bound to mRNA coming from the host genome or other constitutive expression cassettes, (iii) the load species c$_\ell$ which corresponds to the tunable extra demand of protein to be synthesized by exogenous genetic circuits and thus possibly leading to burden, and (iv) the ribosomal-protein species c$_p$, which are those ribosomes involved in the production of R-proteins P.

In all the models, the synthesis of endogenous (basal proteins $B$) and heterologous (load proteins $L$) is assumed to follow the same two-step process as in ribosomal proteins in \textit{M1} and \textit{M2} (Eq.~\eqref{eqn P}), including the binding of mRNA and ribosome, and then translation of the active complex:
\begin{equation}
    \ce{m_x + R <=>[a_x][d_x] c_x ->[$\beta$_x] m_x + R + x} \label{eqn Pgeneral}
\end{equation}
where $x$ is a protein species, $a_x$ depends on the ribosome binding site on the mRNA of the gene encoding $x$ (i.e., it defines the efficiency of ribosome recruitment by each mRNA); $d_x$ is the rate of unbinding and $\beta_x$ is the translational elongation rate, both assumed to be constant and equal for all the genes of the same species.

The last assumption regards the dissociation constants $K_x = \frac{d_x}{a_x}$. Basal transcripts $m_b$ are assumed to exhibit the highest effective affinity for ribosomes, followed by ribosomal protein mRNAs $m_p$, whereas external load transcripts are translated with lower efficiency. This reflects the observed prioritization of ribosome allocation \cite{weisse} toward essential cellular functions (like $m_b$ and $m_p$) under limited translational capacity and imply that: 
\begin{equation}
    K_b < K_p <K_{\ell}. 
\end{equation}

\setcounter{figure}{2}
\begin{figure*}[ht]
    \centering
    \includegraphics[width = \textwidth]{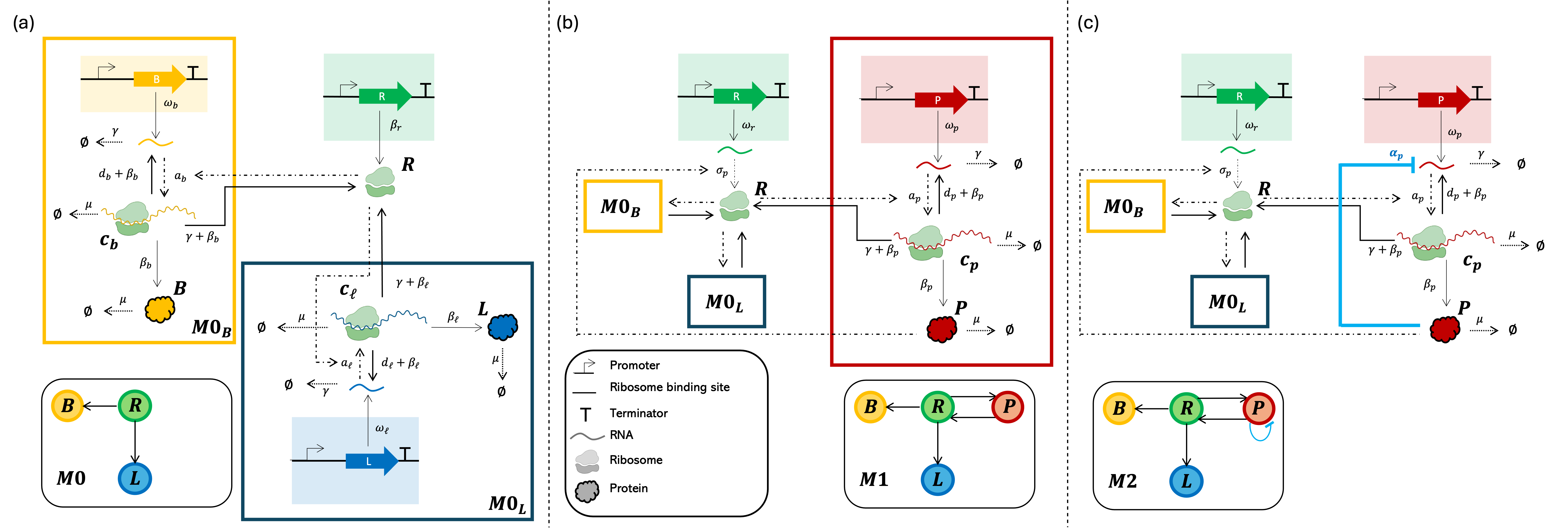}
    \caption{\label{fig:es1}  \textbf{Models of the cell system with kinetic reactions}. In (a):  \textit{M0} with direct ribosomal synthesis without R-proteins and distribution over basal and load mRNAs. In (b): \textit{M1}, augmentation of \textit{M0} with ribosomal distribution over basal, load mRNAs and R-protein \textit{P}, necessary for the formation of mature ribosomes. In (c): \textit{M2}, augmentation of \textit{M1} involving negative feedback on R-protein \textit{P} on its own synthesis.}
\end{figure*}

\subsection{ODE Model implementation}

The simpler model, \textit{M0}, is characterized by the following set of chemical reactions (with $i = \{b,\ell\}$):
\begin{align}
    \ce{D_i& ->[$\omega$_i] D_i + m_i} \label{react_in_M0}\\
	\ce{D_r& ->[$\beta$_r] D_r + R}\label{react_modA_R}\\
	\ce{m_b + R& <=>[a_b][d_b] c_b ->[$\beta$_b] m_b + R + B}\\
	\ce{m_{\ell} + R& <=>[a_{\ell}][d_{\ell}] c_{\ell} ->[$\beta$_{\ell}] m_{\ell} + R + L}\\
   \ce{m_i& ->[$\gamma_{R}$]0}\\
    \ce{c_i& ->[$\gamma_{R}$]R}\\
    \ce{R,c_i,B,L& ->[$\mu$]0} \label{react_fin_M0}
\end{align}
and mass conservation laws:
\begin{align}
	R^{T} =& R + \sum_i{c_i}\label{Rtot_m1_M0}\\
        m_i^{T} =& m_i + c_i \label{mb_tot1_M0}
\end{align}
The derived model, with relative kinetic reactions, is represented as in Fig.~\ref{fig:es1}a. 

The model \textit{M1} in Fig.~\ref{fig:es1}b is characterized by an extended set of chemical reactions counting for the formation of rRNA $r$ and R-protein $P$ (hence, with $i = \{b,\ell,p\}$):
\begin{align}
    \ce{D_i& ->[$\omega$_i] D_i + m_i} \label{react_in}\\
	\ce{m_b + R& <=>[a_b][d_b] c_b ->[$\beta$_b] m_b + R + B}\\
	\ce{m_{\ell} + R& <=>[a_{\ell}][d_{\ell}] c_{\ell} ->[$\beta$_{\ell}] m_{\ell} + R + L}
    \end{align}
\begin{align}
\ce{D_r& ->[$\omega_r$]D_r +r} \\
    \ce{m_{p} + R& <=>[a_{p}][d_{p}] c_{p} ->[$\beta$_{p}] m_{p} + R + P}\\
    \ce{r +P& ->[$\sigma_p$]R} \\
   \ce{m_i,r& ->[$\gamma_{R}$]0}\\
    \ce{c_i& ->[$\gamma_{R}$]R}\\
    \ce{R,c_i,B,L,P& ->[$\mu$]0} \label{react_fin}
\end{align}
and mass conservation laws:
\begin{align}
	R^{T} =& R + \sum_i{c_i}\label{Rtot_m1}\\
        m_i^{T} =& m_i + c_i \label{mb_tot1}\\
        r^{T} =& r + R^{T}\\
        P^{T} =& P + R^{T} 
\end{align}

As previously mentioned, in both models the structure involves up to four different types of ribosomal species: the free ribosomes $R$, and the complexes $c_b=R\!:\!m_b$, $c_{\ell}=R\!:\!m_\ell$ and $c_{p}=R\!:\!m_p$ (only in models \textit{M1} and \textit{M2}) translating basal \textit{B},  load \textit{L} and R-proteins \textit{P}, respectively; together, they form the total amount of ribosomes of the cell $R^{T}$. Ribosomes bind RNA with rate $a_i$ (where \textit{i} indicates the species, namely $ i=\{b,\ell,p\}$) and dissociate with rate $d_i$. In the building phase of the model, an additional hypothesis is assumed: the mRNA m$_i$ within the complex $c_i$ degrades at a rate $\gamma_{R}$, thereby releasing the free ribosome. Furthermore, the translation process synthesizes the protein with rate $\beta_i$ from the ribosomal complex $c_i$, but also returns the mRNA and the free ribosome used in the binding process; this justifies the arrows with rate $d_i + \beta_i$ from the complex to the mRNA and the one with rate $\gamma_{R} + \beta_i$ from the complex to the ribosome in Fig. \ref{fig:es1}.
The model structure also accounts for loss rates ($\gamma_{R}$ for mRNA and $\mu$ for ribosomes, ribosomal complexes, and protein).

Lastly, model \textit{M2} in Fig.~\ref{fig:es1}c can be derived from \textit{M1} by adding the reactions associated with the negative feedback, namely:
\begin{align}
    \ce{P + m_p& ->[$\alpha$_p] $Z$}\label{react_modC_r}\\
    \ce{$Z$& ->[$\gamma_{R}$] P}\\
    \ce{$Z$& ->[$\mu$] 0}
\end{align}
where $Z$ is a complex between ribosomal proteins $P$ and their own mRNAs $m_p$, which become translationally repressed, i.e., they are unable to bind ribosomes to generate new \textit{R-proteins}~\cite{Nomura_feedback}. The bound mRNA is assumed to be as sensitive to degradation as the free mRNA species (i.e., the binding with R-proteins does not protect it from the degradation machinery), and the $Z$ complex can be diluted due to cell growth and degraded at a rate comparable to that of proteins. The $\alpha_p$ parameter is the association rate between $P$ and $m_p$.  
The last mass conservation law to be added is:
\begin{equation}
	m_p^{T} = m_p + c_p + Z\label{mp_tot_m3}
\end{equation}
In model \textit{M2}, it is reasonable to expect that the affinity of R-Protein with rRNA is higher than with mRNA~\cite{control_of_ribosome_synthesis}, otherwise the inhibition would be stronger than the synthesis, which is a contradiction.

\paragraph{\textbf{ODE model for M0:}}
The associated kinetic reactions 
can be translated into a system of differential equations, as it follows (with $i=\{b,\ell\}$):
\begin{align}
	\dot{m_i} =& \omega_i D_i - a_i m_i R + (d_i + \beta_i) c_i - \gamma_{R} m_i \label{mb m1_M0} \\
	\dot{c_i} =&  a_i m_i R - (d_i + \beta_i) c_i - \gamma_{R} c_i - \mu c_i \label{cb m1_M0} \\
	\dot{R} =& \beta_r D_r - \sum_i a_i m_i R + \sum_i d_i c_i + \nonumber\\ 
	 &  \sum_i \beta_i c_i + \gamma_{R} \sum_i c_i - \mu R \label{R m1_M0}\\
	\dot{B} =& \beta_b c_b - \mu B \\
	\dot{L} =& \beta_{\ell} c_{\ell} - \mu L \\
	\dot{R}^{T} =& \beta_r D_r - \mu R^{TOT} \label{r_tot m1}\\
    \dot{m}_i^{T} =& \omega_i D_i - \gamma_{R} m_i^{T} - \mu c_i \label{mb_tot_ode_M0} 
\end{align}
A further approximation can be performed: by considering the relation between $\gamma_{R}$ and $\mu$ ($\gamma_{R} >> \mu$, from~\cite{del_vecchio}) and assuming $m_{i}^{TOT} \geq c_i$ (from Eq.~\eqref{mb_tot1_M0}), then Eq.~\eqref{mb_tot_ode_M0} becomes 
\begin{equation}
    \dot{m}_{i}^{T} \approx\omega_{i} D_{i} - \gamma_{R} m_{i}^{T} \label{miTOT_approx}
\end{equation}
with closed-form solution:
\begin{equation}
    m_{i}^{T} = {\frac{\omega_{i} D_{i}}{\gamma_{R}}}  \label{m_tot closed form}
\end{equation} 
The set of Eqs.~\eqref{mb m1_M0}-\eqref{r_tot m1} can be studied in steady state conditions by numerical methods (here via \textit{Matlab} script adopting the solver \textit{ODE15s}).
However, by exploiting time scale separation for the reactions (see the \textit{Strategies for simplifying gene expression models} sidebar for more details on time scale separation), it is possible to find an approximated solution also for the Eqs.~\eqref{mb m1_M0}-\eqref{R m1_M0} (and thus for Eq.~\eqref{r_tot m1}) \cite{briggs1925note, thomson2009unlimited}.
Considering the set of differential equations, the fast dynamics are represented by Eq.~\eqref{mb m1_M0} and Eq.~\eqref{cb m1_M0}, while the slow dynamics are described by Eq.~\eqref{mb_tot_ode_M0}.
This assumption is standard and commonly adopted in the literature (see \cite{alon2019introduction, del_vecchio,Ingallis2013}).
Replacing $a_i$ as $\frac{d_i}{K_i}$ (where $K_i=\frac{d_i}{a_i}$ is called dissociation constant) we have
\begin{align}
    \dot{m_i} =& \omega_i D_i - d_i m_i \frac{R}{K_i} + d_i c_i + \beta_i c_i - \gamma_{R} m_i \label{manifold 1 M0}\\
	\dot{c_i} =&  d_i m_i \dfrac{R}{K_i} - d_i c_i - \beta_i c_i - \gamma_{R} c_i - \mu c_i 
\end{align}
To formalize the time-scale separation, we introduce a small parameter $\epsilon = \frac{\gamma_R}{d_i}>0$
and rewrite $d_i$ as $d_i = \frac{\gamma_R}{\epsilon}$. The resulting equation is:
\begin{equation}
    \dot{m_i} = \frac{1}{\epsilon} (-m_i \frac{R}{K_i} + c_i) + \omega_i D_i + \beta_b c_i - \gamma_{R} m_i 
\end{equation}
Due to the condition of fast dynamics, $\epsilon \longrightarrow 0$. Consequently, the obtained (slow-dynamics) manifold is:
\begin{equation}\label{cb manifold_M0}
    c_i = \frac{R}{K_i} m_i
\end{equation}
Replacing $c_i$ from Eq.~\eqref{cb manifold_M0} in Eq.\eqref{mb_tot1_M0}, one obtains:
\begin{align}
     m_i^{T} =& \frac{R}{K_i}m_i + m_i\label{mtot_in_appendix}\\
     c_i =& \frac{R}{R + K_i}m_i^{T} \label{manifold c}\\ 
     m_i =& \frac{K_i}{R + K_i}m_i^{T} \label{manifold m}
\end{align}
This illustrates how the evolution of $c_i$ and $m_i$ depends uniquely on the total amount of mRNA and the strength of the bound mRNA-ribosome $K_i$. Since it is proved that $m_i^{T}$ depends on the transcription rate $\omega_i$ (by Eq.~\eqref{m_tot closed form}), the only factor that does not influence the mRNA and complex concentration is the translation rate $\beta_i$. 
By~replacing Eq.~\eqref{manifold c} in Eq.~\eqref{Rtot_m1_M0}:
\begin{equation}\label{r tot analysis_M0}
    R^{T} = R +  \frac{R}{R + K_b}m_b^{T} + \frac{R}{{R + K_{\ell}}}m_{\ell}^{T}
\end{equation}
In this first case, it is worth noting that this formulation is independent of the growth rate function. Despite Eq.~\eqref{r tot analysis_M0} being solved numerically, it is possible to find a closed-form solution if the manifolds of $c_i$ (which are Hill functions, according to Eq.~\eqref{manifold c}) are approximated by piecewise linear functions. Without loss of generality, assuming that $K_b < K_\ell$, then the approximated functions are:
\begin{equation}\label{approx functions_M0}
    c_i \approx
		\begin{cases}
	\dfrac{m_i^{T}}{K_i} R & \text{, } R << K_i \\
     \dfrac{m_i^{T}}{2} & \text{, } R=K_i \\	
		m_i^{T} & \text{, } R >> K_i 
		\end{cases}
  \qquad \text{ with } i = \{b,\ell\}
\end{equation}
Due to the mass conservation law, Eq.~\eqref{Rtot_m1_M0} and by exploiting Eq.~\eqref{approx functions_M0}, the approximated function of $R^{T}$ is:
\begin{equation}\label{R_tot sys_M0}
    R^{T}\!=\! 
    \begin{cases}
    \Big(1+ \dfrac{m_b^{T}}{K_b} + \dfrac{m_{\ell}^{T}}{K_{\ell}} \Big)R & \text{, } 0 \leq R \leq K_b \\
       \Big(1+ \dfrac{m_{\ell}^{T}}{K_{\ell}} \Big)R + m_b^{T} & \text{, } K_b < R \leq K_{\ell} \\    
    R + m_b^{T} + m_{\ell}^{T} & \text{, } R > K_{\ell} 
    \end{cases}
\end{equation}
Now it is possible to reverse the equations to obtain $R$ as function of $R^{TOT}$:
\begin{equation}\label{R sys_M0}
    R = 
    \begin{cases}
    \dfrac{1}{\Big(1+ \dfrac{m_b^{T}}{K_b} + \dfrac{m_{\ell}^{T}}{K_{\ell}} \Big)}R^{T}& \text{,} 0 \leq R^{T} \leq \tilde{R}_1^{T} \\
    \dfrac{R^{T} - m_b^{T}}{\Big(1+ \dfrac{m_{\ell}^{T}}{K_{\ell}} \Big)} & \text{,} \tilde{R}_1^{T} < R^{T} \leq  \tilde{R}_2^{T} \\
    R^{TOT} - m_b^{T} - m_{\ell}^{T}& \text{,} R^{T} > \tilde{R}_2^{T} 
    \end{cases}
\end{equation}
where, accordingly to the conditions in Eq.~\eqref{R sys_M0}:
\begin{align}
    \tilde{R}_1^{T} =& \Big(1+ \dfrac{m_b^{T}}{K_b} + \dfrac{m_{\ell}^{T}}{K_{\ell}} \Big)K_b \label{R1 tilde_M0} \\
    \tilde{R}_2^{T} =& \Big(1+ \dfrac{m_{\ell}^{T}}{K_{\ell}} \Big)K_{\ell} + m_b^{T}   \label{R2 tilde_M0}
\end{align}

\paragraph{\textbf{ODE model for M1:}}
The associated kinetic reactions in Eqs.\eqref{react_in}-\eqref{react_fin} 
can be translated into a system of differential equations, as follows (with $i=\{b,\ell, p\}$):
\begin{align}
	\dot{m_i} =& \omega_i D_i - a_i m_i R + (d_i + \beta_i) c_i - \gamma_{R} m_i \label{mb m1} \\
	\dot{c_i} =&  a_i m_i R - (d_i + \beta_i) c_i - \gamma_{R} c_i - \mu c_i \label{cb m1} \\
	\dot{R} =& \sigma_p r P - \sum_j a_j m_j R + \sum_j d_j c_j+  \nonumber\\ 
	        &  +\sum_j \beta_j c_j + \sum_{j}\gamma_{R} c_j- \mu R \quad \text{, } j=\{b,p,\ell\}\label{R m1}\\
	\dot{B} =& \beta_b c_b - \mu B \\
	\dot{L} =& \beta_{\ell} c_{\ell} - \mu L \\
	\dot{r} =& \omega_rD_r -\sigma_p rP - \gamma_{R} r\\
	\dot{P} =& \beta_p c_p - \sigma_p r P- \mu P \label{P_m2}\\
	\dot{R}^{T} =& \sigma_p r P - \mu R^{T} \label{ode_Rtot} \\
    \dot{r}^{T} =& \omega_r D_r - \gamma_{R} r - \mu R^{T} \label{ode_rtot}\\
	\dot{P}^{T} =& \beta_p c_p - \mu P^{T} \label{ode_Ptot}\\
    \dot{m}_i^{T} =& \omega_i D_i - \gamma_{R} m_i^{T} - \mu c_i. \label{mb_tot_ode} 
\end{align}
The same approximation adopted for \textit{M0} can be performed: by considering the relation between $\gamma_{R}$ and $\mu$ ($\gamma_{R} >> \mu$, from~\cite{del_vecchio}) and assuming $m_{i}^{T} \geq c_i$ (from Eq.~\eqref{mb_tot1}), then Eq.~\eqref{mb_tot_ode} can be written in its closed form as in Eq. \eqref{m_tot closed form}.
Hence, the set of Eqs.~\eqref{mb m1}-\eqref{ode_Ptot} can be studied under steady-state conditions using numerical methods as before.
 
Following the same flow described for \textit{M0}, approxiated solutions for Eqs.~\eqref{mb m1}-\eqref{R m1} (and thus for Eq.~\eqref{ode_Rtot}) can be found, with fast dynamics represented by Eq.~\eqref{mb m1} and Eq.~\eqref{cb m1}, while slow dynamics by Eq.~\eqref{mb_tot_ode}. The same final structure of Eqs. \eqref{mtot_in_appendix}-\eqref{manifold m} with $i \in \{ b, \ell, p\} $can be obtained.

In this case, for model \textit{M1}, by replacing Eq.~\eqref{manifold c} in Eq.~\eqref{Rtot_m1}:
\begin{equation}\label{r tot analysis}
    R^{T} = R +  \frac{R}{R + K_b}m_b^{T} + \frac{R}{{R + K_{\ell}}}m_{\ell}^{T} +  \frac{R}{R + K_p}m_p^{T}.
\end{equation}
and again, despite Eq.~\eqref{r tot analysis} being solved numerically, it is possible to find a closed-form solution if the manifolds of $c_i$ (which are Hill functions, according to Eq.~\eqref{manifold c}) are approximated by piecewise linear functions. Without loss of generality, assuming that $K_b < K_p < K_\ell$, then the approximated functions are:
\begin{equation}\label{approx functions}
    c_i \approx
		\begin{cases}
	\dfrac{m_i^{T}}{K_i} R & \text{, } R << K_i \\
     \dfrac{m_i^{T}}{2} & \text{, } R=K_i \\	
		m_i^{T} & \text{, } R >> K_i 
		\end{cases}
  \qquad \text{ with } i = \{b,p,\ell\}
\end{equation}
Because of the mass conservation law in Eq.~\eqref{Rtot_m1} and by exploiting Eq.~\eqref{approx functions}, the approximated function of $R^{T}$ is:
\begin{equation}\label{R_tot sys}
    R^{T}\!=\! 
    \begin{cases}
    \Big(1+ \dfrac{m_b^{T}}{K_b} + \dfrac{m_p^{T}}{K_p} +\dfrac{m_{\ell}^{T}}{K_{\ell}} \Big)R & \text{, } 0 \leq R \leq K_b \\
       \Big(1+ \dfrac{m_p^{T}}{K_p} + \dfrac{m_{\ell}^{T}}{K_{\ell}} \Big)R + m_b^{T} & \text{, } K_b < R \leq K_{p} \\
       \Big(1+ \dfrac{m_{\ell}^{T}}{K_{\ell}} \Big)R + m_b^{T}+ m_p^{T} & \text{, } K_p < R \leq K_{\ell} \\
    R + m_b^{T} + m_p^{T} + m_{\ell}^{T} & \text{, } R > K_{\ell} 
    \end{cases}
\end{equation}
Now it is possible to reverse the equations to obtain $R$ as a function of $R^{T}$:
\begin{equation}\label{R sys}
    R = 
    \begin{cases}
    \dfrac{R^{T}}{\Big(1+ \dfrac{m_b^{T}}{K_b} + \dfrac{m_p^{T}}{K_p} +\dfrac{m_{\ell}^{T}}{K_{\ell}} \Big)} \text{,  } &0 \leq R^{T} \leq \tilde{R}_1^{T} \\
    \dfrac{R^{T} - m_b^{T}}{\Big(1+ \dfrac{m_p^{T}}{K_p} + \dfrac{m_{\ell}^{T}}{K_{\ell}} \Big)} \text{,} &\tilde{R}_1^{T} < R^{T} \leq  \tilde{R}_2^{T} \\
    \dfrac{R^{T} - m_b^{T}-m_p^{T}}{\Big(1+ \dfrac{m_{\ell}^{T}}{K_{\ell}} \Big)}  \text{,} &\tilde{R}_2^{T} < R^{T} \leq  \tilde{R}_3^{T} \\
    R^{T} - m_b^{T} - m_p^{T} - m_{\ell}^{T} \text{,} &R^{T} > \tilde{R}_3^{T} 
    \end{cases}
\end{equation}
where, accordingly to the conditions in Eq.~\eqref{R sys}:
\begin{align}
    \tilde{R}_1^{T} =& \Big(1+ \dfrac{m_b^{T}}{K_b} + \dfrac{m_p^{T}}{K_p} + \dfrac{m_{\ell}^{T}}{K_{\ell}} \Big)K_b \label{R1 tilde} \\
    \tilde{R}_2^{T} =&  \Big(1+ \dfrac{m_p^{T}}{K_p} + \dfrac{m_{\ell}^{T}}{K_{\ell}} \Big)K_p + m_b^{T} \label{R2 tilde} \\
    \tilde{R}_3^{T} =& \Big(1+ \dfrac{m_{\ell}^{T}}{K_{\ell}} \Big)K_{\ell} + m_b^{T} + m_p^{T}   \label{R3 tilde}
\end{align}

\paragraph{\textbf{ODE model for M2:}}
The last model involves a negative feedback mechanism to prevent the endless growth of free ribosomes, thereby limiting the formation of R-protein-associated mRNA. Accordingly, it is reasonable to expect that the feedback assumes remarkable importance when the metabolic load is null or weak. At the same time, it becomes irrelevant when the load is increased (the ribosomes' request increases, as they tend to bind mRNA rather than accumulate it). In this case, all the mass conservation laws remain unchanged except for the correction on $m_p^{T}$, which becomes Eq.~\eqref{mp_tot_m3}.
Starting from \textit{M1}, the ODEs Eq.~\eqref{mb m1} and Eq.~\eqref{mb_tot_ode} for $i=p$, and Eq.~\eqref{P_m2} must also be adjusted:
	\begin{align}
		\dot{m_p}=& \omega_p D_p - a_p m_p R + (d_p  + \beta_p) c_p -  \nonumber\\ 
	 &  \alpha_p m_p P - \gamma_{R} m_p \label{mp_m3}\\
		\dot{m}_p^{T}=& \omega_p D_p - \gamma_{R} m_p - \gamma_{R} Z - \gamma_{R} c_p - \mu c_p-\nonumber\\ 
	 &   \mu Z \label{mp_tot3}\\
     \dot{P}=& \beta_p c_p - \sigma_p r P + \gamma_{R} Z- \alpha_p m_p P - \mu P \label{P_m3}
\end{align}
And the following must be added:
	\begin{equation}\label{P:mp}
		\dot{Z} = \alpha_p  m_p P - \gamma_{R} Z - \mu Z
	\end{equation}
Concerning the study of the equations at steady state, nothing changes compared to model \textit{M1}.
All the parameters used to run simulations have been obtained or derived from the literature, as described in Table~\ref{tab:es1}.
\begin{table}[!h]
\caption{Models parameters \small{(*=computed)}}
\label{tab:es1}
\begin{adjustbox}{max width=\columnwidth}
    \centering
		\begin{tabular} {|c| c| c| c| c|} 
			\hline
			\textbf{Par.} & \textbf{Description} & \textbf{Value} & \textbf{Ref.} & \textbf{Model} \\ 
			\hline
			     $\gamma_{R}$ & Decay of mRNA and rRNA &  13 $h^{-1}$  &~\cite{alon2019introduction} & 0,1,2 \\ 
			$\delta$ &Proteins, Complexes, Ribosomes decay& 0.6 $h^{-1}$ &~\cite{alon2019introduction,decay_proteins} & 0,1,2 \\
			$a_b$ & Association rate of $c_b$ &  $1666$  $(\mu M h)^{-1}$ &~\cite{del_vecchio} & 0,1,2 \\
			$d_b$ & Dissociation rate of $c_b$ &  100 $h^{-1}$ &~\cite{del_vecchio} & 0,1,2 \\
			$a_p$ & Association rate of $c_p$ &   $166$ $(\mu M h)^{-1}$ &~\cite{del_vecchio} & 1,2 \\
			$d_p$ & Dissociation rate of $c_p$ &  100 $h^{-1}$ &~\cite{del_vecchio} & 1,2 \\
			$a_\ell$ & Association rate of $c_\ell$ &   $16$ $(\mu M h)^{-1}$ & $[a]$& 0,1,2 \\
			$d_\ell$ & Dissociation rate of $c_\ell$ &  100 $h^{-1}$ & $[a]$ & 0,1,2 \\
            $a_f$ & Association rate of $c_f$ &   $16$ $(\mu M h)^{-1}$ & $[a]$ & 0,1,2 \\
			$d_f$ & Dissociation rate of $c_f$ &  100 $h^{-1}$ & $[a]$ & 0,1,2 \\
			$\beta_b$ & Translation rate of $B$ & 156 $h^{-1}$ & $[b]$ & 0,1,2 \\
			$\beta_p$ & Translation rate of $P$ &  156 $h^{-1}$ & [b] & 1,2 \\
			$\beta_\ell$ & Translation rate of $L$ &  156 $h^{-1}$ & $[b]$ & 0,1,2 \\
            $\beta_f$ & Translation rate of $F$ &  156 $h^{-1}$ & $[b]$ & 0,1,2 \\
            $\omega_b$ & Transcription rate of $m_b$ & 520 $ h^{-1}$ &~\cite{del_vecchio,GNB} $\circ$ & 0,1,2 \\
            $\omega_{p_1}$ & Transcription rate of $m_p$ of model 1& $\omega_b \cdot 10$ $h^{-1}$ & $\circ[c]$ & 1 \\
            $\omega_{p_2}$ & Transcription rate of $m_p$ of model 0 and 2& $\omega_b \cdot 100$ $h^{-1}$ & \textasciicircum $[c]$ & 0,2 \\
			$\omega_\ell$ & Transcription rate of $m_\ell$ & $\omega_b \cdot 10$ $h^{-1}$ &$[c]$ & 0,1,2 \\
            $\omega_f$ & Transcription rate of $m_f$ (burden monitor) & $\omega_b \cdot 100$ $h^{-1}$ &$[c]$& 1,2 \\
			$\sigma_p$ & Association rate between $r$ and $P$ &  $1666$ $(\mu M h)^{-1}$ & * & 1,2 \\
			$\alpha_p$ & Association rate between $m_p$ and $P$ &  $0.1 \cdot \sigma_p$  $(\mu M h)^{-1}$ & $\dagger$& 2 \\
			$\omega_r$ & Transcription rate of $r$ &  $\omega_b \cdot 100$ $h^{-1}$ &$[c]$& 1,2 \\
            $\beta_r$ & Synthesis rate of $R$ in \textit{M0} & 156 $h^{-1}$ & $\circ$ & 0 \\
			$D_b$ & Basal \textit{DNA} &  1 $\mu M$ &\cite{dna_of_e_coli} & 0,1,2 \\
			$D_p$ & R-Protein \textit{DNA} &  0.01 $\mu M$ &\cite{dna_of_e_coli} & 1,2 \\
			$D_\ell$ & Load \textit{DNA} &  0.1 $\mu M$ &\cite{dna_of_e_coli} & 0,1,2 \\
			$D_r$ & rRNA &  0.01 $\mu M$ &\cite{dna_of_e_coli} & 0,1,2 \\
            $D_f$ & Constitutive reporter gene (burden monitor)&  0.01 $\mu M$ &\cite{dna_of_e_coli} & 0,1,2 \\
   \hline

   \multicolumn{5}{p{251pt}}{$*$ Fast binding kinetics \cite{bunner2010kinetic}, same order of magnitude of $a_b$.}\\
   \multicolumn{5}{p{251pt}}{$\dagger$ Set to be smaller than association rate between $r$ and $P$.}\\
      \multicolumn{5}{p{251pt}}{$\circ$ Set to achieve the desired maximum growth rate. }\\
      \multicolumn{5}{p{251pt}}{\textasciicircum  Set to compensate the negative feedback. }\\
      \multicolumn{5}{p{251pt}}{$[a]$ Set in order to satisfy the relation $K_b< K_p< K_{\ell,f}$}.\\
      \multicolumn{5}{p{251pt}}{$[b]$ All translation rates are assumed equal.}\\
      \multicolumn{5}{p{251pt}}{$[c]$ Transcription rates $\omega_p$ and $\omega_r$ are much higher than $\omega_b$ as stated in \cite{mueller1977capacity, dennis2009varying,maeda2015strength} while $\omega_{\ell,f}$ can be properly designed.}\\
		\end{tabular}
    \end{adjustbox}
\end{table}

\subsection{Growth Law}\label{literature}
Given the designed ribosome synthesis and gene expression models, an underlying growth rate function is needed to link ribosomal allocation to cell growth. The \textit{Bacterial growth and components dilution} sidebar provides details on bacterial growth, component dilution, and their mathematical representation. A literature review on different options for mapping ribosomal levels to bacterial growth rate is reported in the \textit{Growth laws overview} sidebar. As already explained in the introduction, equations that express growth rate as a function of the total amount of ribosomes (e.g.,~\cite{growth_rate_of_Ecoli,A_numbers_game_ribosome_density}) could not be directly applied, as they generally lack a description of metabolic load. In fact, in conditions of burdensome heterologous expression, the distribution of ribosomes is affected, and these models are are unable to capture the resulting growth rate reduction. The growth functions of other models that explicitly include heterologous genes generally relate cell growth with the total protein synthesis rate, given by the intracellular amount of mRNA-ribosome complexes (e.g.,~\cite{proof_load_effect,weisse}). However, these growth functions also include energy (e.g., nutrients, ATP)-dependent translation rates and a mechanistic composition of the cellular protein mass. For this reason, such functions cannot be adopted in the minimal modeling framework used in this work, as it lacks the needed mechanistic detail.

As contradictions or inaccuracies may arise between model simulations and biological evidence when these functions are derived without accounting for metabolic burden, the chosen growth function was inspired by studies that explicitly tested cell load conditions. A linear function of the active ribosomes ($R^{a}$) of the endogenous proteins was adopted:

\begin{equation}
\lambda(R^{a})=\alpha \cdot R^{a} \label{growthlaw}\\
\end{equation}
with $R^{a} = \sum_{i}c_i$ and $\alpha=0.045 \, h^{-1}\mu M^{-1}$
fitted from the data in~\cite{growth_rate_of_Ecoli}, adequately adapted. Specifically, the fitted line was the one passing for the origin and for $R^{a} = \frac{2}{3} R^{T}_{max}$, according to \cite{del_vecchio}, given $R_{max}^T\approx60 \mu M$ for a maximum growth rate of $\lambda\approx 2 h^{-1}$.
The $R^{a}$ complex considers $c_b$ in the \textit{M0} model, as it is the only active complex in the absence of load. Conversely, $R^{a}$ is the sum of $c_b$ and $c_p$ in \textit{M1} and \textit{M2} as they include R-protein translation explicitly. In the chosen function, only the translationally active complexes of the proteins contribute to cell growth, consistent with \cite{santosnavarro21}. The inclusion of complexes related to heterologous proteins in $R^{a}$, reported in previous cell models (e.g., \cite{sechkar24,weisse,proof_load_effect}) described in the \textit{Growth laws overview} sidebar, was not considered because they could lead to a biologically implausible increase in growth rate for specific parameter values. Their inclusion would have required the addition of cell load dependent translation rate or intracellular protein composition, to obtain a mechanistically complete equation.

\begin{sidebar}{Bacterial growth and components dilution}
\setcounter{equation}{40}
\renewcommand{\thesequation}{S\arabic{sequation}}
\renewcommand{\thestable}{S\arabic{stable}}
\renewcommand{\thesfigure}{S\arabic{sfigure}}
\label{growth definition}
\sdbarinitial{T}he \textit{Escherichia coli} bacterium, a unicellular organism usually containing just one chromosome in the form of a circular \textit{DNA} molecule, reproduces by cell division, and this process is called \textit{Binary Fission}. 

This process begins with the replication of \textit{DNA}, which is responsible for the genetic pool of the cell. Because of that, the replication of \textit{DNA} is a fundamental process, as it must ensure that the new cell contains all the proteins needed for life. Although the entire process of cell duplication will not be debated, it is interesting to analyze the required time and conditions for cell growth, as well as the effects on the system. 

For example, it is known that the binary fission process of a cell of \textit{E. coli}, at $ \ang{37} C$, takes about 40 minutes. Still, it can be reduced to 20 minutes \cite{growth_rate_of_Ecoli,hans_bremer} under particular nutrient conditions. The needed time for the duplication of a single cell is called \textit{Doubling Time} $\tau$.

 Cell division has a significant influence on the cell system. Indeed, it is worth recalling that this process involves dividing the intracellular content equally between the newborn sister cells. Consequently, the loss of some genetic material due to duplication must be accounted for in the model, since it concerns a single cell rather than the entire population. The components can be assumed to be lost due to the actual dilution and transfer to a newborn cell. Therefore,	this can be considered to be equal to a molecule loss reaction (Sidebar \textit{Biological scenario}), namely:
	\begin{equation}\label{reaction lambda}
	\ce{A ->[$\lambda$] 0}
	\end{equation} 
	which indicates that the generic component \textit{A} of the cell is dissipated. The rate $\lambda$, which is called \textit{Growth Rate}, indicates the velocity of the growth, and it is related to time $t$ needed for the growth; specifically, it is related to the $\tau$ parameter, as it will be shown below. 
	
The equation that rules the growth dynamics of a cellular population is:
	\begin{equation}\label{exp growth}
		x(t) = x_0 e^{t \lambda} 
	\end{equation}
where $x$ is the number of cells in the temporal instant $t$, and $x_0$ is the initial condition of the population.
To calculate the doubling time of a cell population, Eq.~\eqref{exp growth} can be rewritten as:
	\begin{equation}
		2x_0 = x_0 e^{\tau \lambda}
	\end{equation}
	where $\tau$ is the doubling time. Consequently:
	\begin{equation}\label{mu gr}
		\lambda = {\dfrac{\ln 2}{ \tau }}
	\end{equation}
Even though Eq.~\eqref{mu gr} has been derived from an equation that expresses the dynamics of a population of cells, it can be easily adapted to the single-cell case. Indeed, all the elements in the cell grow linearly with the cell itself, so it is sufficient to consider $x$ as the amount of a specific component, rather than the number of cells. This parameter is applied to the single-cell model, but it is also used to express the growth velocity of a cellular population.\\
Finally, it is essential not to confuse the growth rate $\lambda$ with the spontaneous decay, even though they are represented by the loss reaction in Eq.~\eqref{degradation}, where $\psi_A$ can be either the growth rate or the spontaneous decay rate. Indeed, while the first concerns cell growth, the second follows from the assumption that all components cannot last forever but are subject to a natural decline. Consequently, the pool of chemical reactions will contain two different loss reactions for each element. Thus, they can be summed up in a unique response:
\begin{equation}
		\ce{A ->[$\mu_A$] 0}
\end{equation}
where $\mu_A$ is called \textit{Loss rate} and it is the summation between the growth and the spontaneous decay rate.
    
In most of the experiments and articles from the literature, the growth rate has been considered constant, namely, all the calculations have been made based on the assumption that the cell (or the population) is always in its maximal growth condition.\\
Instead, this dissertation aims to perturb the system using a metabolic load to appreciate the effect on the growth. Because of that, in this work the growth rate cannot be assumed constant; it is instead studied as a function of specific components of the cell.
\end{sidebar}

\begin{sidebar}{\continuesidebar}
\renewcommand{\thesequation}{S\arabic{sequation}}
\renewcommand{\thestable}{S\arabic{stable}}
\setcounter{sfigure}{3}
\renewcommand{\thesfigure}{S\arabic{sfigure}}
\section{Growth laws overview}
The main functions proposed in the literature to relate ribosomal content with cell growth rate are discussed, along with their form (linear or nonlinear), key variables (e.g., a given ribosomal species), and suitability for models aimed at capturing cell load conditions.

In \textit{Marr}~\cite{growth_rate_of_Ecoli}, the proposed model for the cell growth is a linear function 
of the total number of ribosomes $R^{T}$, considered as the summation of free ribosomes $R$ and the active ribosomal complexes $c_i$, which are obtained upon ribosome and \textit{mRNA} binding (where $i$ stands for the $i-th$ species). Cell load has not been taken into account. The critical aspect of this work comes from the assumption that the growth rate depends on $R^{T}$. This implies that the metabolic burden will not impact the growth rate of the cell. In other words, the addition of the load will rearrange the inner proportions of the compartment $R^{T}$. Hence, it should be possible to observe the same growth in both configurations (with and without load). As cell load is known to affect growth rate, the growth function of Marr et al. is not consistent with biological evidence. Even though some biological experiments show that adding and/or increasing the concentration of the load does not lead to an instant alteration in the growth rate, it is not reasonable to expect that the rate will be maintained invariant compared to the unloaded case.

\textit{Levin et al.} also considers total ribosomes $R^{T}$ as the variable affecting growth rate~\cite{A_numbers_game_ribosome_density}. Different from the previous study, the proposed formulation of growth rate is a Michaelis-Menten function instead of a linear relation:
\begin{equation}
	\lambda = (\lambda_{MAX} - \lambda_{MIN}){\dfrac{R^{T}}{R^{T} + K}} + \lambda_{MIN}
\end{equation}	
where $\lambda_{MAX}$ ($>0$) is the maximum growth rate, $\lambda_{MIN}$ ($<0$) is the death rate and $K$ is the Hill (shape) parameter.
The existence of minimum and maximum values for the growth suggests that there should exist an equivalent for $R^{T}$. In other words, this model states that the cell cannot grow if $R^{T} < R^{T}_{min}$. Furthermore, it has been assumed that there also exists a threshold for the maximum amount of ribosomal content that can be produced by the cell, namely $R^{T}_{max}$. Again, this model has the same issue as the first one discussed (\cite{growth_rate_of_Ecoli}), namely, its dependence on $R^{T}$ is not suitable for modeling bacterial growth rate under cell-load conditions.

In \textit{Wu et al.}~\cite{Cellular_perception_of_growth_rate}, growth rate is a linear function of a ribosomal species.
However, instead of relating growth rate with the total amount of ribosomes, the growth rate function now depends on the difference between total active ($N_R$) and inactive number ($N_R^{inact}$) of ribosomes and on the total protein mass ($M_p$) in a cell:
\begin{equation}\label{gr article 2}	\lambda M_p = \epsilon \cdot (N_R - N_R^{inact})
\end{equation}
Although the proposed formulation does not clearly show critical points, the study focused on nutrient contributions without any heterologous expression load. Indeed, even though the addition of a metabolic load will influence both $N_R$ and $N_R^{inact}$ (generally increasing $N_R$ and decreasing $N_R^{inact}$) and consequently modify the growth rate, the equation was not compared with any cell-load condition.

A mechanistic model relating growth rate and ribosomal species was adopted in different works~\cite{weisse,proof_load_effect,Resource_aware_whole_cell_model}:
\begin{equation}
	\lambda = {\dfrac{\gamma(a)}{M}} \sum_{x} c_x
\end{equation}
Where $\gamma(a)$ is the rate of translation of all the proteins dependent on the available energy ($a$), and $M$ is the total proteome content of the cell. Although this relation appears to be a linear function of all ribosome complexes, i.e., ribosomes bound to \textit{mRNA}, it also includes complex relationships with energy and proteome composition. In fact, cell load conditions caused by heterologous expression would affect not only ribosome complexes through resource allocation but also the distribution of protein species and the ability of cells to import and metabolize nutrients into energy (e.g., ATP). For these reasons, this growth law is compatible with a cell load description but is only consistent with detailed modeling of cell processes that explicitly include proteome composition, energy, and endogenous genes involved in nutrient processing (uptake and metabolism), such as membrane proteins and enzymes.

Finally, in \textit{Del Vecchio and Murray}~\cite{del_vecchio}, the growth rate function is not clearly defined. Instead, the authors have provided the ratio of free ribosomes, which is $30 \%$ over the total number of ribosomes $R^{T} \approx 34 \mu M$ (in the exponential phase). This proportion has been used in this work to initialize the models. According to this information, the growth rate has been analytically determined by fitting a Hill function to the points obtained from the available information.
\end{sidebar}
\setcounter{equation}{85}
\setcounter{figure}{3}

\subsection{Choice of Parameter Values.}
To set the initial state of the systems (i.e., the bacterial ribosomal pool before applying the burden), we assume maximal growth to estimate the distribution of ribosomal species in conditions of high resource availability; however, this point also varies depending on the chosen literature and growth rate function. The complexity in setting proper parameter values is accentuated by the limited availability of streamlined and real-time approaches for measuring ribosomes, although recent works proposed experimental and computational strategies to indirectly measure their abundance in single cells under dynamic conditions~\cite{LP_Pavlou2025}. Nonetheless, the most accessible techniques rely on measuring growth rate and the expression of reporter proteins~\cite{LP_Borkowski2018,LP_Shopera2017}. The recently proposed techniques are expected to pave the way for a deeper understanding of the interplay between growth conditions and stress in bacterial populations and single cells.

\paragraph{\textbf{Initialization in M0:}} 
By assumption, to be conservative, the initial state proposed in~\cite{del_vecchio} was pursued, leading to the following partition of the ribosomes under maximal growth conditions:
\begin{equation}
	c_b = \frac{2}{3} R^{T}_{max}\text{,}\quad R = \frac{1}{3} R^{T}_{max} \label{cb max gr m1}\\
\end{equation}
From Eq.~\eqref{r_tot m1} at the steady state, the value of $\beta_r D_r$ can be derived simply by substituting $R^{T}_{max}$ and $\mu_{max}$.
\begin{equation}
   \beta_r D_r = \mu_{max} R^{T}_{max}
\end{equation}
To obtain $\omega_b D_b$, $m_b^{T}$ in Eq.~\eqref{mb_tot_ode_M0} at the steady state must be replaced with its value derived from Eq.~\eqref{manifold c}:
\begin{equation}\label{sys max gr}
    0 = \omega_b D_b - \gamma_{R} \frac{R + K_b}{R} c_b - \mu c_b
\end{equation}
By replacing $c_b$ and $R$ in Eq.~\eqref{sys max gr} with their values in Eq.~\eqref{cb max gr m1}, the value of $\omega_b D_b$ for maximal growth condition becomes:
\begin{align}
    \omega_b D_b&=\frac{2}{3} R^{TOT}_{max} (\gamma_{R} + \mu_{max})+ 2 \gamma_{R} K_b \nonumber\\ 
	 &\approx 2 \gamma_{R} \Big(\frac{R^{TOT}_{max}}{3} + K_b\Big)
\end{align}
The basal genes' mRNA is expected to bind strongly with ribosomes due to the importance of basal proteins, committed to the fundamental processes of the cell. Therefore, the dissociation constant $K_b$ (i.e., the ratio between $d_b$ and $a_b$) is assumed to be much lower than the fraction of $R^{T}_{max}$ in Eq.~\eqref{sys max gr}, and can be neglected. Therefore, a further approximation is possible, namely:
\begin{equation} \label{eq_omegab}
    \omega_b D_b \approx 2 \gamma_{R} \Big(\frac{R^{T}_{max}}{3} + K_b\Big) \approx \frac{2}{3} R^{T}_{max} \gamma_{R}
\end{equation}

\paragraph{\textbf{Initialization in M1:}} The definition of the parameters at maximal growth presents some issues. The main difficulty arises from the unavailability of experimental data needed to discriminate between basal and R-protein species due to technological complexity. Additionally, information on the amounts of rRNA and R-proteins is reported only in one paper, as relative quantities~\cite{hans_bremer}; this makes the model non-identifiable \textit{a priori}. Thus, the already estimated parameters in~\cite{GNB}, namely the transcription rate $\omega_b$ for the basal species and the translation rate $\beta_r$ of the process of ribosome formation, were also used in this model. However, the ribosome formation now is decoupled (ribosomal rRNA $r$ and R-proteins $P$) and the values of $\omega_p$, $\sigma_p$ and $\beta_p$ are still missing.\\
For the transcription rates, it has been assumed that those of R-protein $\omega_p$ and of rRNA $\omega_r$ species are much higher than the basal transcription rate $\omega_b$, as stated in \cite{mueller1977capacity, dennis2009varying,maeda2015strength}. $\sigma_p$, the association rate between $r$ and $P$, is assumed to be fast, in agreement with reported kinetics in \cite{bunner2010kinetic} and has been parametrized to be of the same order of magnitude as typical protein–RNA interactions. Therefore, it has been set to be comparable to the binding rates of ribosome and basal mRNA interactions. For modeling purposes, all the translation rates have been assumed to be of the same magnitude and equal to the synthesis rate $\beta_r$ of the free ribosomes computed in the first model.
In accordance with all the stated assumptions, it is reasonable to expect that these parameters do not yield an effective condition for maximal growth, regardless of the adopted growth rate function.

\paragraph{\textbf{Initialization in M2:}}
As for model \textit{M1}, the estimation of parameters under maximal growth conditions is not possible. For this reason, all the assumptions made with \textit{M1} have been preserved, and the parameters $\omega_p$, $\sigma_p$, and $\beta_p$ are assumed to be unvaried with respect to \textit{M1}. The only additional parameter to be set was the association rate $\alpha_p$ between the R-protein and its mRNA, which can be interpreted as strength of the feedback effect. This parameter value must be in the interval \mbox{$0 < \alpha_p < \sigma_p$}; otherwise, the free ribosomes formation process (given by the binding between $r$ and $P$, occurring at rate $\sigma_p$) would be highly limited by the inhibition, which is inconsistent with the aim of the feedback. Previous works confirm that R-proteins have a higher affinity for rRNA than for their mRNA targets~\cite{LP_Burgos2017}. However, since it is not possible to derive its precise value from the literature nor from data, it has been arbitrarily set as $\alpha_p=0.1 \, \sigma_p$. This choice reflects the assumption that the feedback should have a limited influence and strength compared to the synthesis of ribosomes.
To assess model sensitivity to this parameter and evaluate the robustness of its behavior under variations of $\alpha_p$, the impact of feedback on the system was simulated by altering the parameter by one order of magnitude relative to its nominal value. The results are presented in the \textit{Robustness Analysis} section, along with an analysis of the effects induced by these parameter variations. 

Finally, to compensate for the effect of the negative feedback on $P$, the transcription rate $\omega_p$ in model \textit{M2} is increased with respect to model \textit{M1}.

\section{RESULTS}
\subsection{Analysis of Total Ribosomes Variation}

Since the total number of ribosomes $R^{T}$ is limited in the cell, mRNAs must compete for obtaining the needed ribosomes to form their translational complexes $c_i$.
A preliminary analysis of the distribution of the total number of ribosomes ($R^{T}$) among different ribosomal species was performed to identify which parameters govern this competition.

\begin{figure}[t]
    \centering{\includegraphics[width =1\columnwidth]{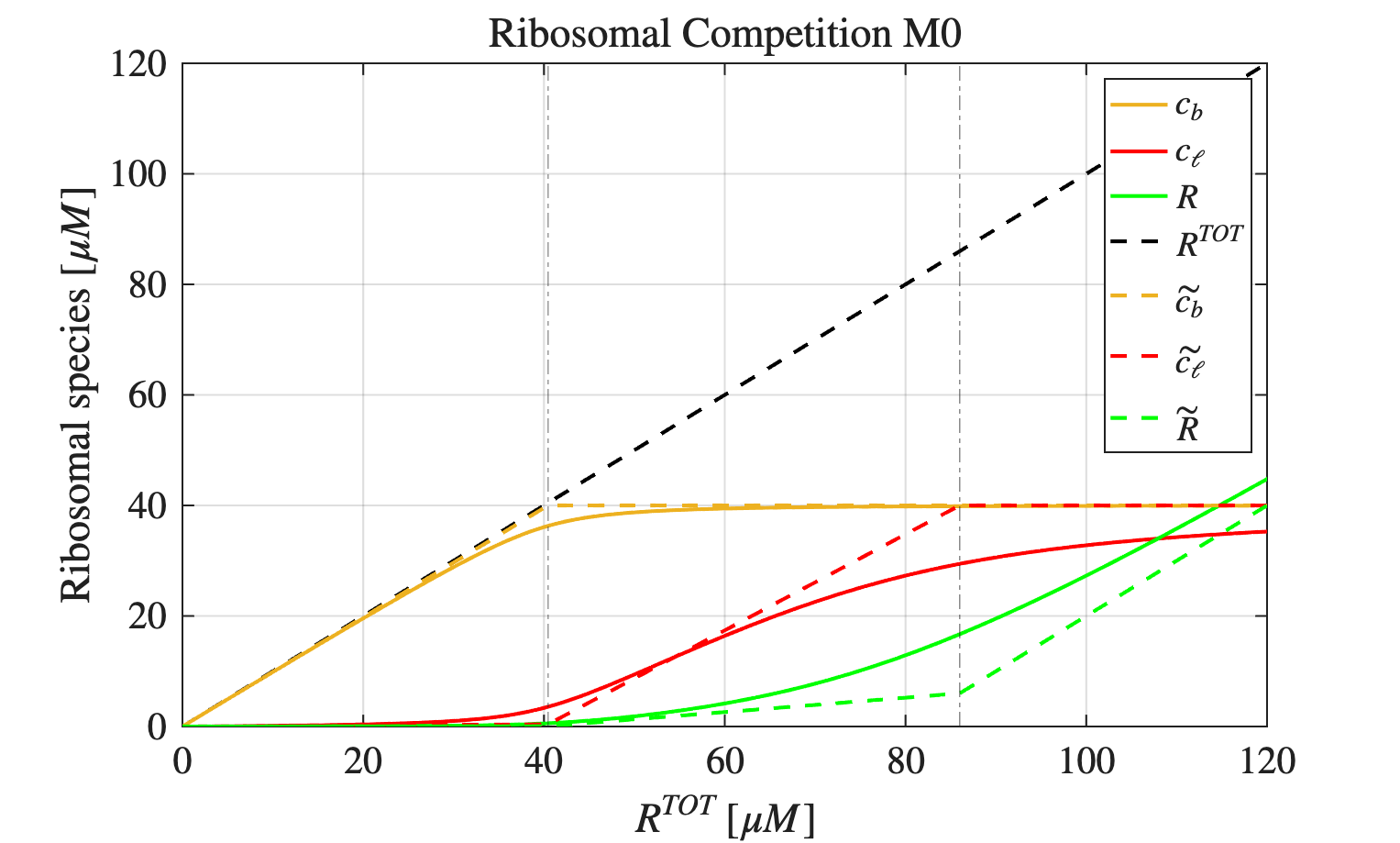}}\caption{\textbf{Competition between ribosomal species in model \textit{M0}.} Competition was studied with $K_b << K_{\ell}$ and $m_b^{T} = m_{\ell}^{T}$. All the ribosomal species described in M0 are shown, namely the free ribosomes, and the active complexes between ribosomes and the mRNAs of basal and load genes. Total ribosomes are also reported as a reference. Dashed lines represent the approximated piece-wise solutions from Eqs. \eqref{R sys_M0}.\label{fig:ribomial competition M0}}
\end{figure}

Starting from model \textit{M0}, which involves the species $R$, $c_b$, and $c_{\ell}$, the evolution of different components as a function of $R^{T}$ is reported in Fig.~\ref{fig:ribomial competition M0}. 
Approximate functions (computed in Eqs.~\eqref{approx functions_M0}-\eqref{R sys_M0}) are used to predict the behavior of ribosomal components. Indeed, the study of the break points provides information regarding the increase in complex formation. From Eq.~\eqref{approx functions_M0}, it is straightforward that the first complex to reach its maximum value has the smallest $K$. Thus, ribosomal sequestration priority among different species is only regulated by dissociation constants. This is consistent with biological knowledge: at the beginning, the amount of available ribosomes is limited; when the species start to compete, only the mRNAs with the strongest association rate (which implies the smallest $K$) can generate complexes; when the first complex reaches mRNAs saturation, which means that it does not need other ribosomes, the species with the second smallest dissociation constant can increase. Moreover, Eq.~\eqref{R1 tilde_M0} and Eq.~\eqref{R2 tilde_M0} show that the break points depend on both the dissociation constants and the total amount of mRNA. This is reasonable because, as already explained, the species $j$ starts increasing when the previous one $i$ (with $K_i < K_j$) reaches its maximal value $m_i^{T}$.

\begin{figure}[!ht]
    \centering
    \includegraphics[width =0.9\columnwidth]{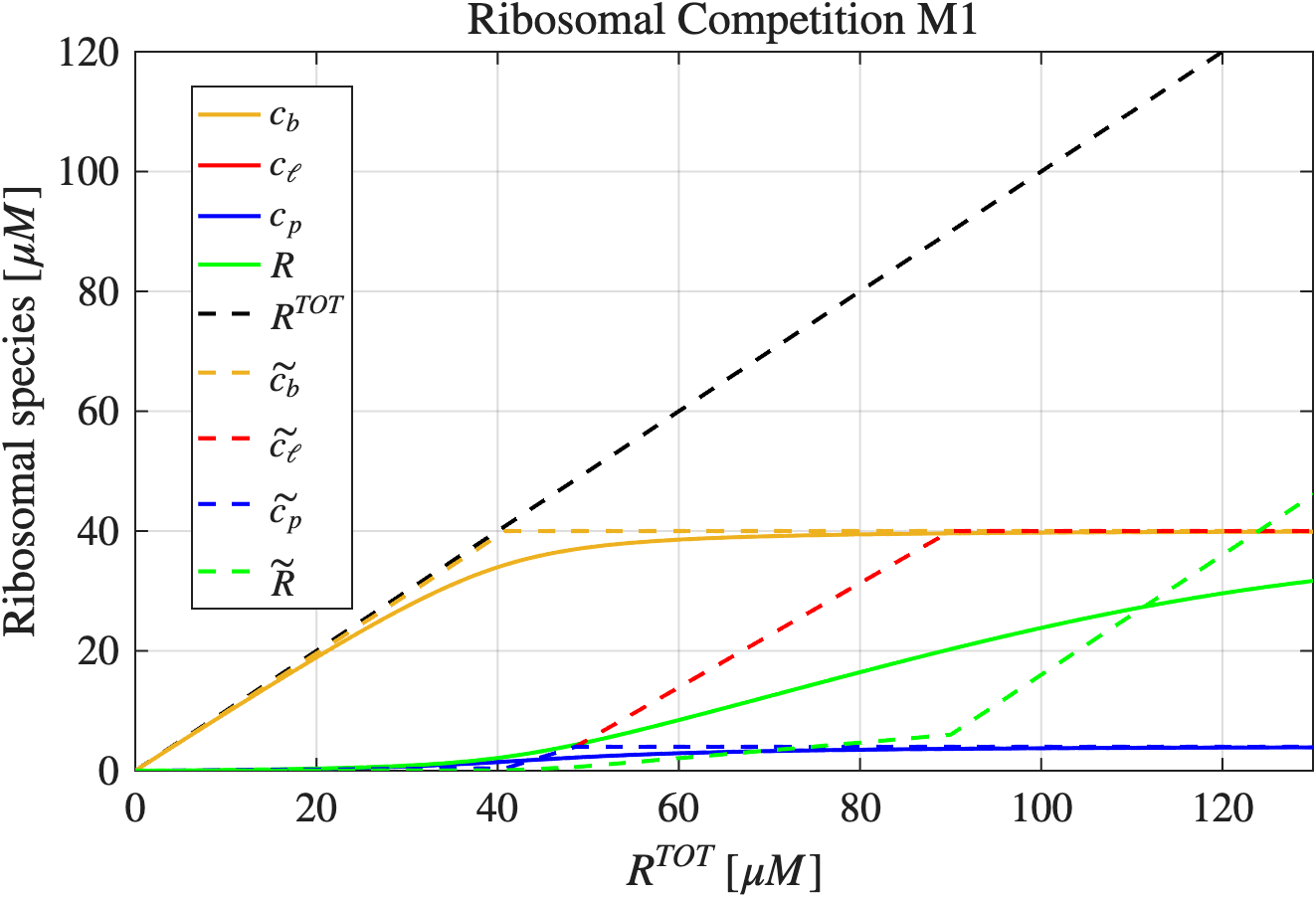}
    \caption{\label{fig:ribomial competition}\textbf{Competition between ribosomal species in model \textit{M1}.} Competition was studied with $K_b << K_{p}<< K_{\ell}$ and $m_b^{T} = m_{\ell}^{T}$. All the ribosomal species described in M1 are shown, namely the free ribosomes, and the active complexes between ribosomes and the mRNAs of basal, ribosomal, and load genes. Total ribosomes are also reported as a reference. Dashed lines represent the approximated piece-wise solutions from Eqs.~\eqref{approx functions}-\eqref{R sys}.}
\end{figure}

In model \textit{M1}, $R^{T}$ is distributed among four ribosomal species $c_b$, $c_p$, $c_{\ell}$ and $R$. The evolution of different components as a function of $R^{T}$ is reported in Fig.~\ref{fig:ribomial competition}. 
As before, approximate functions (computed in Eqs.~\eqref{approx functions}-\eqref{R sys}) are used to predict the behavior of ribosomal components and similar observations can be drawn: when the first complex reaches mRNAs saturation, the species with the second smallest dissociation constant can increase, and so on; also in this case, Eqs.~\eqref{R1 tilde}-\eqref{R3 tilde} show that the break points depend on both the dissociation constants and the total amount of mRNA.

Therefore, $R^{T}$ does not depend on the chosen growth rate function but uniquely on the manifolds:
	\begin{equation}\label{manifolds m2}
		c_i = \frac{R}{R + K_i}m_i^{T}\quad \text{with }i=\{b,\ell,p\}
	\end{equation}
and the mass conservation law of $R^{T}$:
	\begin{equation}\label{Rtot m2 analysis}
		R^{T} = R + c_b + c_p + c_{\ell}
	\end{equation}
where $c_b$, $c_p$ and $c_{\ell}$ must be substituted with their manifolds. 

In model \textit{M2}, differently from the previous cases, the analysis of the distribution of $R^{T}$ among the ribosomal species is no longer available, due to the alternative formulation of the mass conservation law of $m_p^{T}$ (Eq.~\eqref{mp_tot_m3}), which has been made necessary after the addition of the feedback. Indeed, it involves the quantities $m_p$ and $c_p$, as well as $Z$, which does not allow us to derive the manifolds. In particular, being:
	\begin{equation}
		m_p^{T} = \frac{R}{K_p} m_p + m_p + Z
	\end{equation}
	This leads to:
	\begin{equation}\label{manifold m3}
		c_p = \frac{R}{R + K_p}(m_p^{T} - Z)
	\end{equation}
	\begin{equation}
  m_p = \frac{K_p}{R + K_p}(m_p^{T} - Z)
	\end{equation}
and these equations depend not only on $R$, but also on $Z$. Notably, despite the approximation for $m_p^{T}$ derived in Eq.~\eqref{miTOT_approx} remains valid also for \textit{M2}, an analytical study of the evolution of Eq.~\eqref{manifold m3} should also include $Z$ (Eq.~\eqref{P:mp}), which in turn should consider $P$ (Eq.~\eqref{P_m3}). Consequently, a closed-form analysis of the ribosomal competition is no longer feasible; therefore, we have primarily limited our analysis to simulations.

\subsection{Analysis of Load effect}
To analyze the load effects on gene expression and cell growth, starting from the dynamics described by Eqs.~\eqref{mb m1_M0}-\eqref{mb_tot_ode_M0},~\eqref{mb m1}-\eqref{mb_tot_ode}, and Eqs.~\eqref{mp_m3}-\eqref{P:mp} for models \textit{M0}, \textit{M1}, and \textit{M2}, respectively, equilibria were studied as a function of the load. The latter is characterized by four parameters: the dissociation constant $K_{\ell}$, the transcription rate $\omega_{\ell}$, $D_\ell$ (which denotes the concentration of the load gene copies available for transcription), and the translation rate $\beta_{\ell}$. For the sake of simplicity, only $\omega_{\ell} D_\ell$ was varied, mimicking transcriptional tuning, e.g., by an inducible activation of gene expression, and reported on the x-axis of all the following figures. Additionally, as a clearer proxy for resource availability, a further weak constant load $F$ (and thus its mRNA $m_f$ and complex $c_f$) has been added to the system, with kinetic reactions

\begin{align}
    \ce{D_f& ->[$\omega$_f] D_f + m_f} \label{react_in_MF}\\
	\ce{m_f + R& <=>[a_f][d_f] c_f ->[$\beta$_f] m_f + R + F}\\
   \ce{m_f& ->[$\gamma_{R}$]0}\\
    \ce{c_f& ->[$\gamma_{R}$]R}\\
    \ce{F& ->[$\mu$]0}
\end{align}    
and the corresponding parameters reported in Tab.\ref{tab:es1}. The $F$ species is not expected to significantly increase cell load, and its introduction is not essential to capture ribosome allocation in the minimal model defined in this work; nonetheless, it represents a convenient species that corresponds to the resource competition monitors usually adopted in the literature \cite{LP_Ceroni2015,LP_CarbonellBallestero2016,LP_Shopera2017,Pasotti2017} and represented by the constitutive GFP protein in the experimental data of Fig. \ref{fig:spaso}.

It is worth noting that equilibria can change depending on the cell growth phase (i.e., lag, exponential, and stationary phases); besides this, such variations can be of interest in several applications. The inclusion of such dependencies implies a significantly more complex model granularity, which is beyond the scope of this study. Therefore, the system under study is assumed to be observed during its optimal exponential growth phase, as is often the case in synthetic biology applications (e.g., sampling during exponential growth or adopting turbidostat mode in bioreactors~\cite{LP_MiliasArgeitis2016}).

In particular, the evolution of $c_b$, $c_p$, $c_{\ell}$, $c_f$, $R$ and $R^{T}$ is considered. Assuming cell growth is uniquely dependent on ribosomal components, mRNA and protein evolution (except for the $P$ protein) are not investigated in this analysis. On the other hand, it is important to recall that a growth rate $\mu$-dependent decrease in molecule concentration is also involved in the equations and even though the evolution of the complexes can still be derived from the manifold equations, Eq.~\eqref{ode_Rtot} at the steady state depends on $\mu$, with results strongly dependent on the adopted growth rate function (see the \textit{Growth Law} section for details on the growth rate functions).

\begin{figure}[t]
    \centering
    
    \begin{subfigure}{.8\columnwidth}
        \centering
        \includegraphics[width=\linewidth]{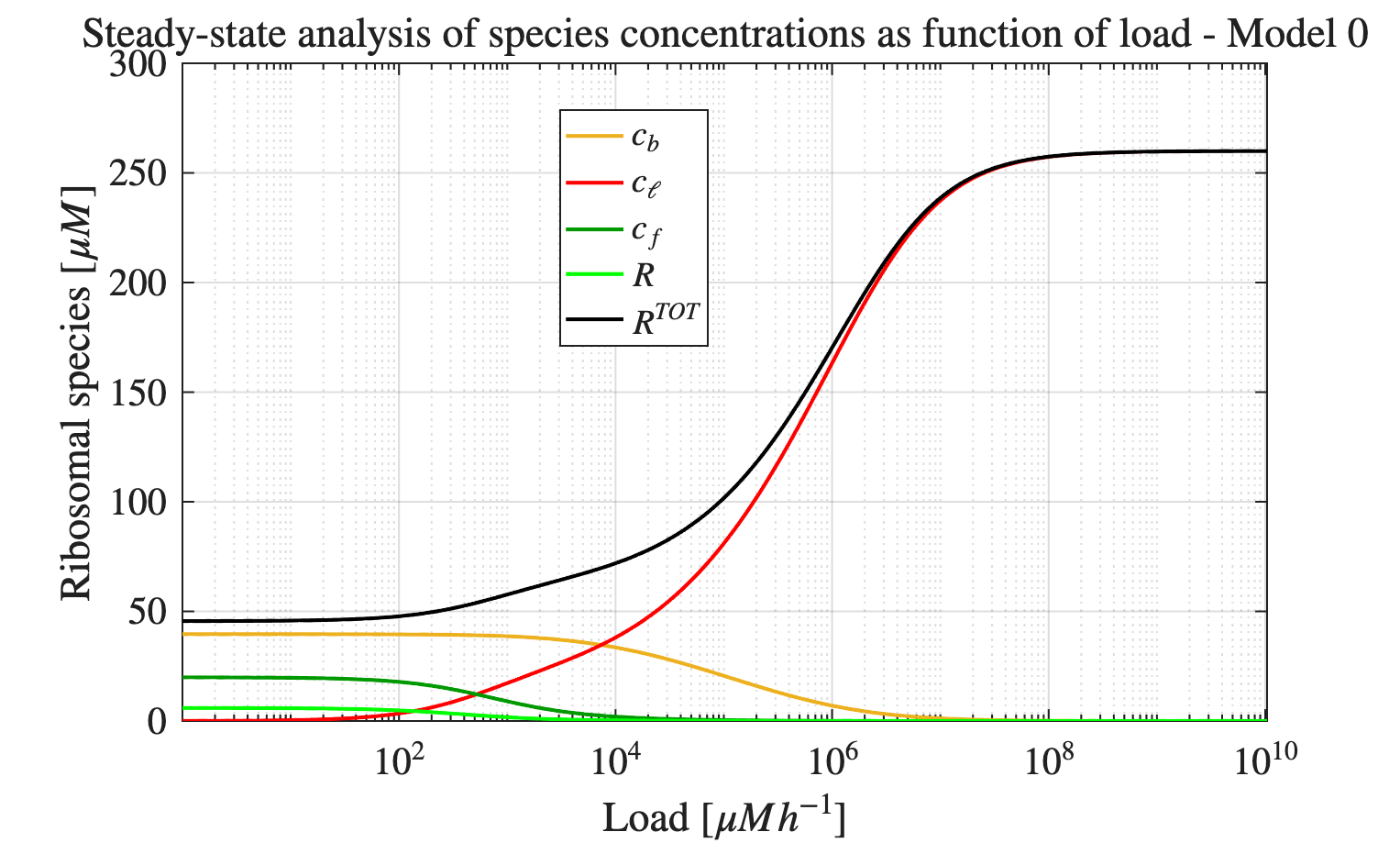}
        \caption{}
        \label{fig:species_m1}
    \end{subfigure}
    
    \vspace{0.5em}
    
    \begin{subfigure}{.8\columnwidth}
        \centering
        \includegraphics[width=\linewidth]{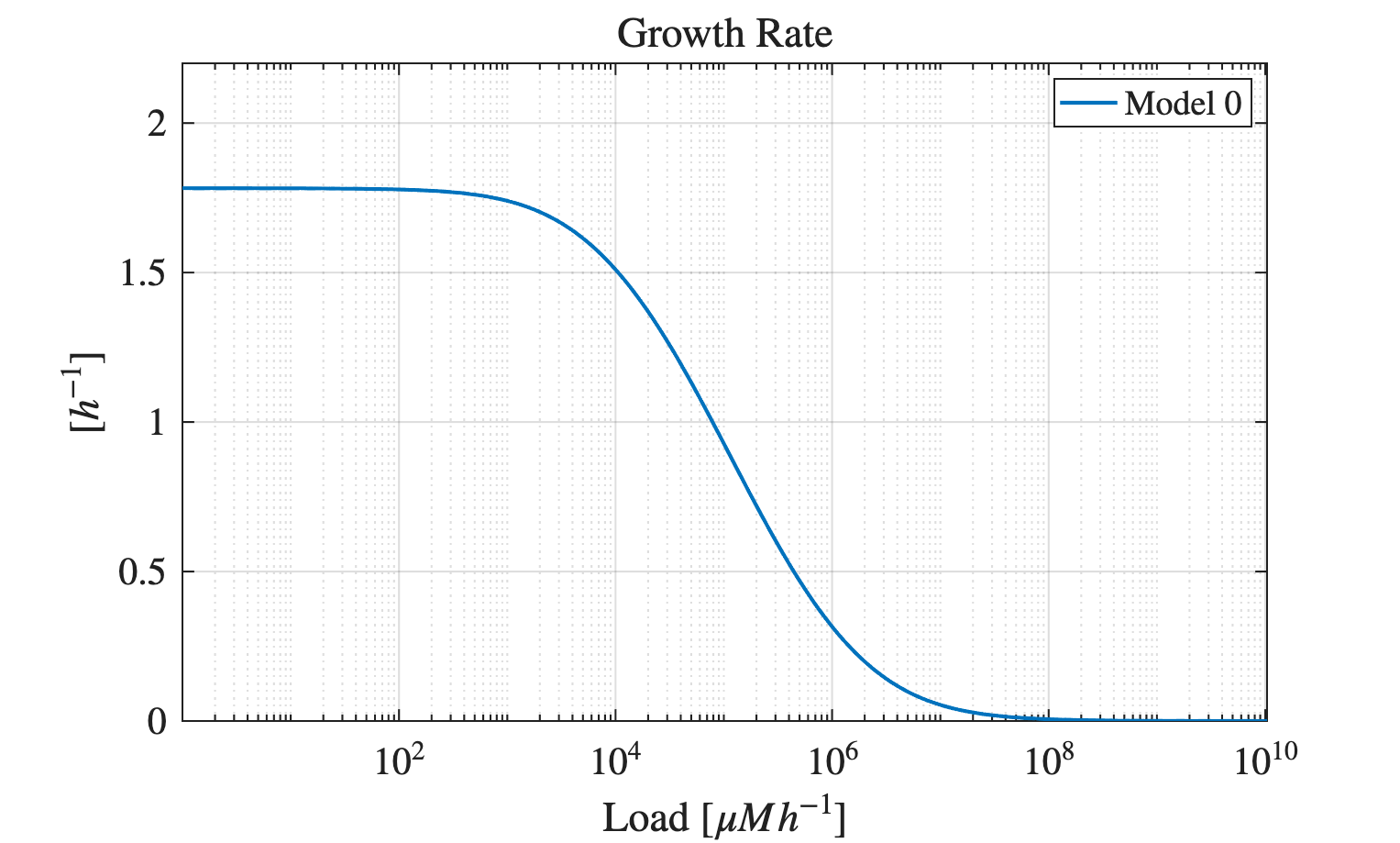}
        \caption{}
        \label{fig:growth_m1}
    \end{subfigure}
    \caption{\textbf{Analysis of load effects in model \textit{M0}.} (a) Steady state analysis of species concentrations as function of metabolic load $\omega_l D_l$ for model \textit{M0}. (b) Steady state analysis of the growth rate as a function of the metabolic load $\omega_l D_l$ for model \textit{M0}.}
    \label{fig:M1}
\end{figure}

In model \textit{M0}, which does not involve ribosome autocatalysis through R-protein $P$, a first insight into the validity of this model only for reduced load can be observed in Fig. \ref{fig:M1}. Only for very low load synthesis rates $x < 10^2 \mu Mh^{-1}$, the system is not affected by metabolic load, with no appreciable effects on either $c_f$ or growth rate. For load synthesis rates up to $x\sim 10^3 \mu Mh^{-1}$, the system is starting to show burden effects, as the monitor $c_f$ is decreasing while the tunable load $c_{\ell}$ rises. As the load is increased further, growth starts to decrease, and the total concentration of ribosomes starts to accumulate, likely due to slowed cell division, which reduced the dilution rate of molecules. At a load level of $x\sim10^4 \mu Mh^{-1}$, the system has no more free ribosomes, and the basal protein complexes $c_b$ start to drastically decrease, along with the growth rate. For a load greater than $x\sim 10^6 \mu Mh^{-1}$, the system collapses. Still, as shown in Fig. \ref{fig:species_m1}, the number of load complexes $c_{\ell}$, along with the total number of ribosomes, continues to rise until it reaches a biologically nonsensical plateau for unrealistically high load synthesis rate (x-axis reaches $10^{10}\mu Mh^{-1}$) at which all the ribosomes are used to produce the load. Due to the lack of dependence on R-proteins (no $P$-related autocatalysis feedback assumption), the synthesis of ribosomes can reach an equilibrium (i.e., $R$ production rate over spontaneous degradation) as well as all the mRNAs and thus complexes, which is something that cannot happen upon cell death.

\begin{figure}[!h]
    \centering
    
    \begin{subfigure}{.8\columnwidth}
        \centering
        \includegraphics[width=\linewidth]{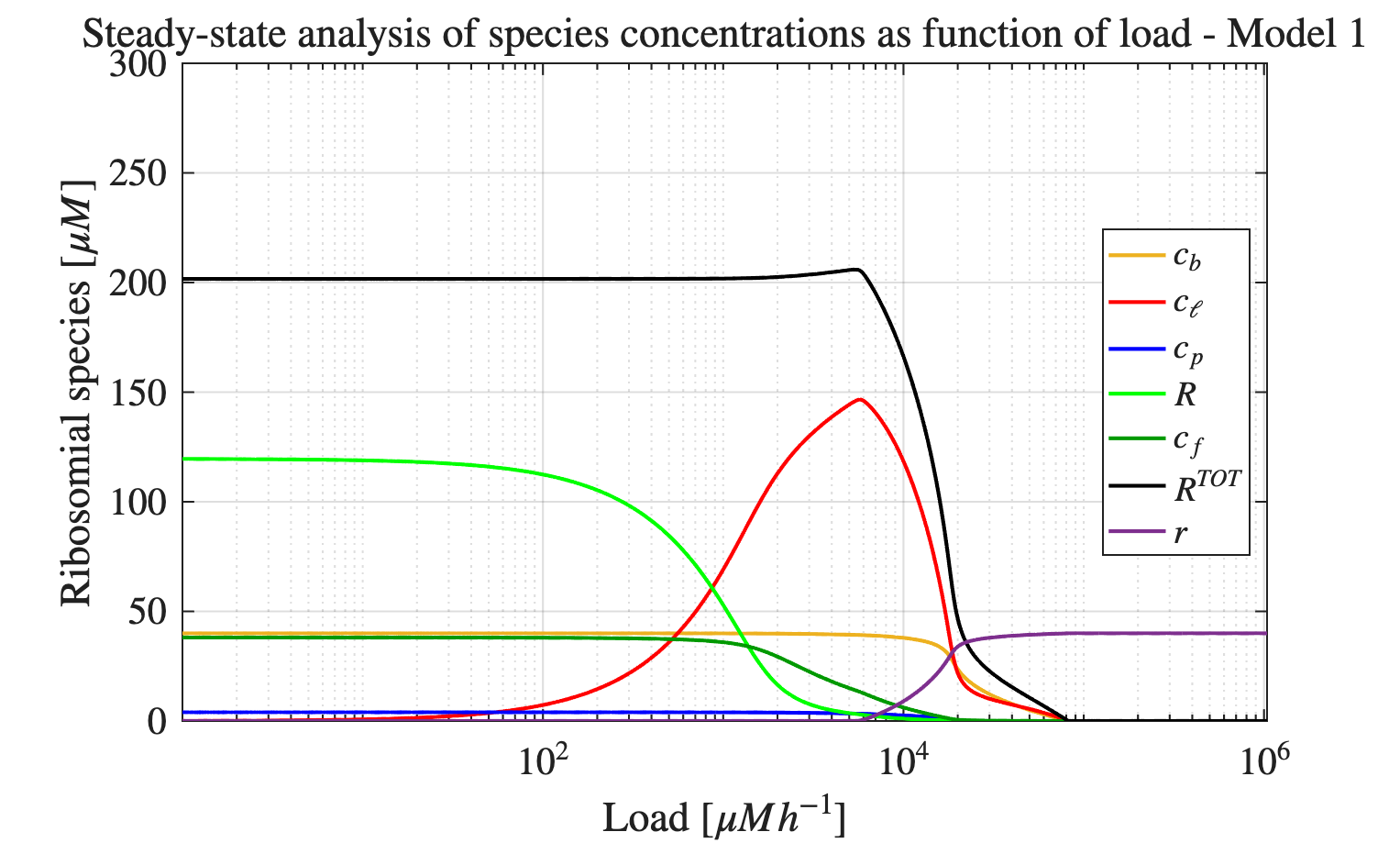}
        \caption{}
        \label{fig:species_m2}
    \end{subfigure}
    
    \begin{subfigure}{.8\columnwidth}
        \centering
        \includegraphics[width=\linewidth]{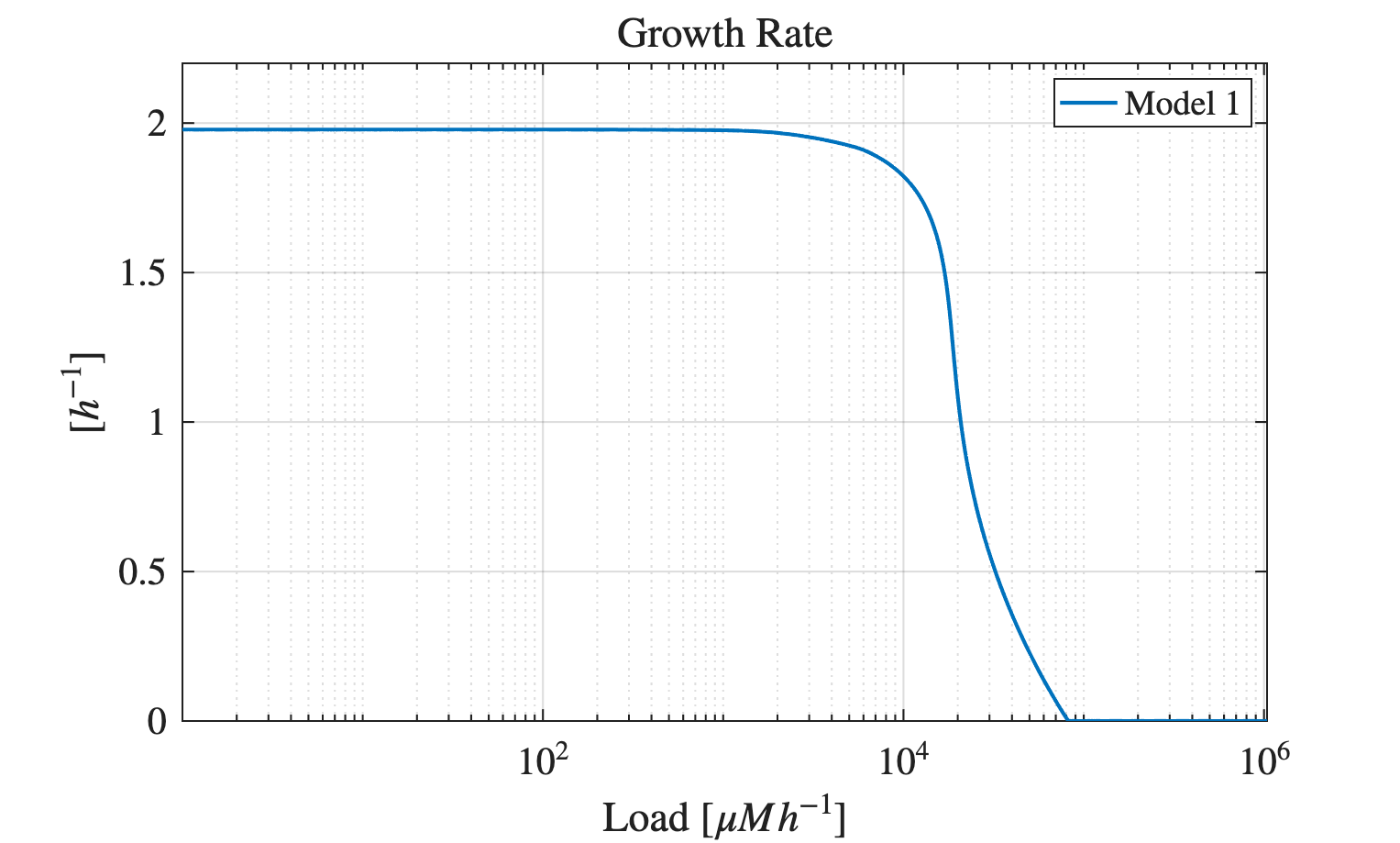}
        \caption{}
        \label{fig:growth_m2}
    \end{subfigure}
    \caption{\textbf{Analysis of load effects in model \textit{M1}.} (a) Steady state analysis of species concentrations as a function of metabolic load $\omega_l D_l$ for Model \textit{M1}. (b) Steady state analysis of the growth rate as a function of the metabolic load $\omega_l D_l$ for Model \textit{M1}.}
    \label{fig:M2}
\end{figure}

In model \textit{M1}, for an initial region of applied load ($x\lesssim 5\times10^2 \mu Mh^{-1}$ regions in Fig.~\ref{fig:M2}), the system is not affected by metabolic load, with stable growth rate and only free ribosomes $R$ decreasing due to the $c_{\ell}$ formation. For a greater load synthesis rate, up to a $x\lesssim 6\times10^3 \mu Mh^{-1}$, the burden can be observed through the decrease in the monitor $c_f$, with only a reduced effect on growth rate. Interestingly, during the transition to the next "overloaded" region - corresponding to a Load between $1\times10^3\mu Mh^{-1}$ and $6\times10^3 \mu Mh^{-1}$ - a small increase in the total number of ribosomes is observed; this is likely due to reduced dilution from cell division, corresponding to the onset of the decreased growth rate. Finally, for higher values, the load complex $c_{\ell}$ collapses first, followed by the basal complexes $c_b$, along with the total number of ribosomes and the growth rate, leading to cell death. Interestingly, while cell death is now described, another biologically nonsensical effect can be observed when looking at the amount of free rRNA $r$: as all complexes decrease, it increases due to the lack of R-proteins to bind to. This behavior arises because RNA polymerase synthesis is not explicitly modeled and polymerase levels are instead assumed to remain constant. However, since this assumption does not affect the predicted metabolic burden or growth rate, we consider it a reasonable simplification and do not investigate it further.

\begin{figure}[!h]
    \centering
    
    \begin{subfigure}{.8\columnwidth}
        \centering
        \includegraphics[width=\columnwidth]{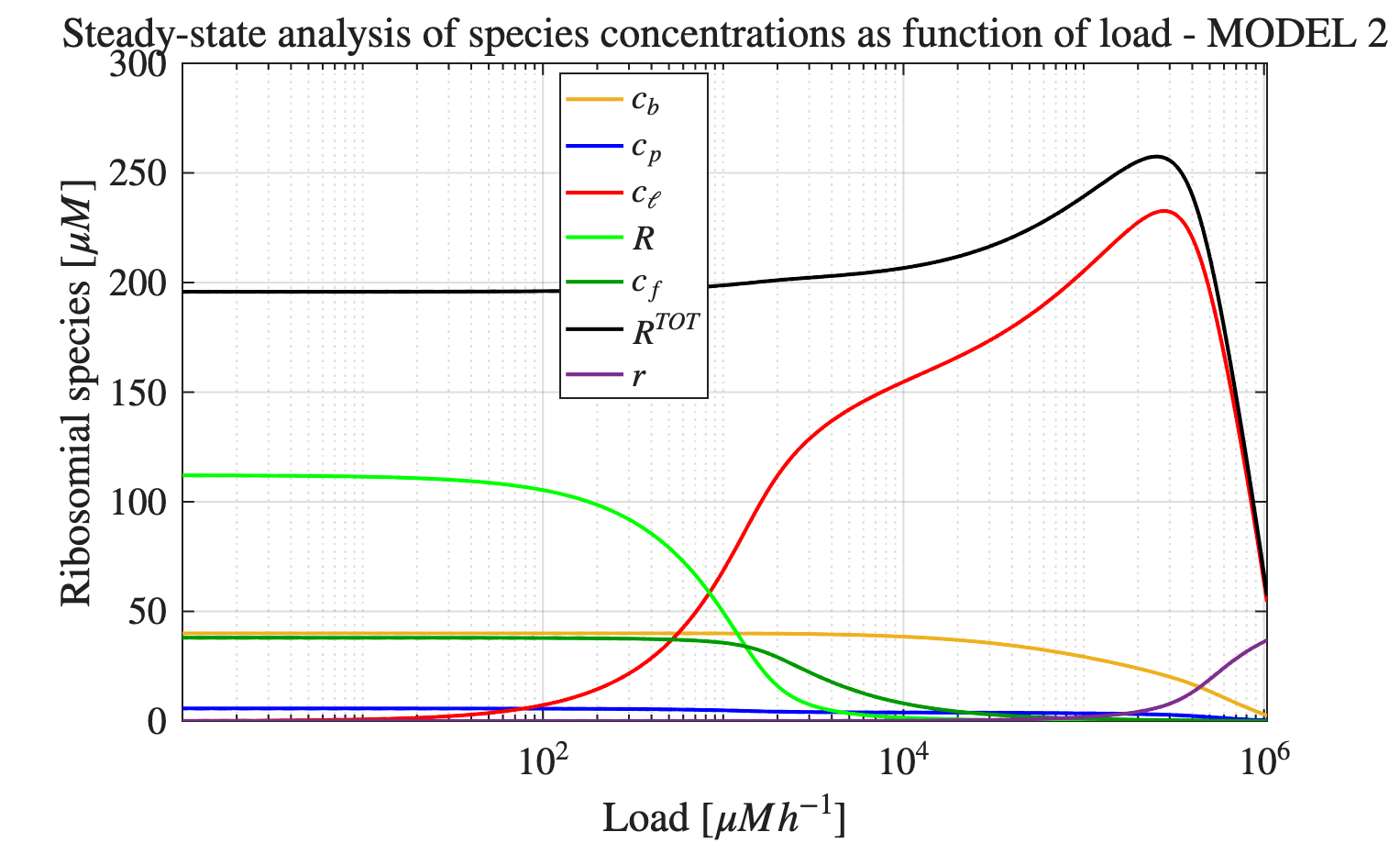}
        \caption{}
        \label{fig:species_m3}
    \end{subfigure}
    
    \begin{subfigure}{.8\columnwidth}
        \centering
        \includegraphics[width=\columnwidth]{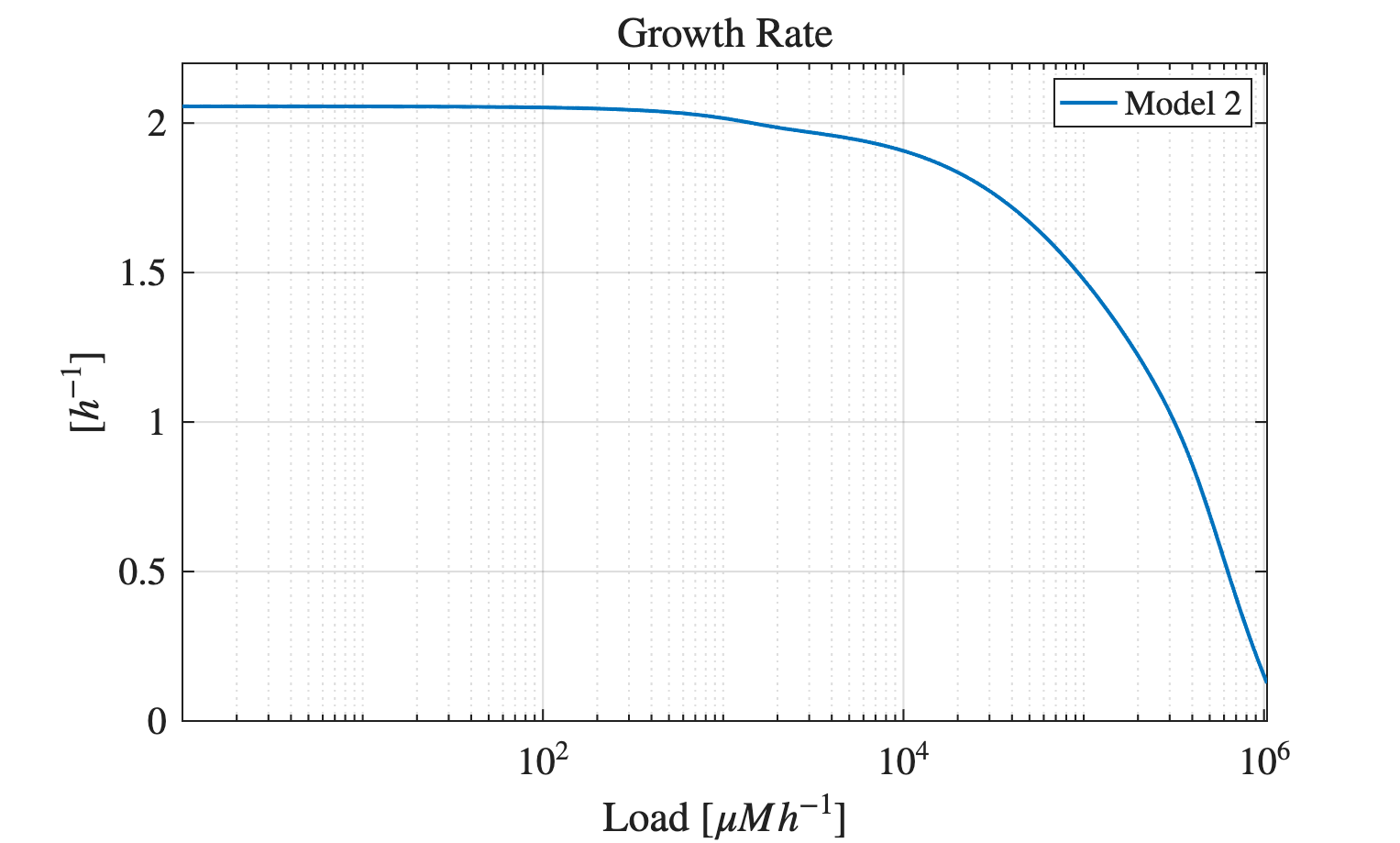}
        \caption{}
        \label{fig:growth_m3}
    \end{subfigure}

    \begin{subfigure}{.8\columnwidth}
        \centering
        \includegraphics[width=\columnwidth]{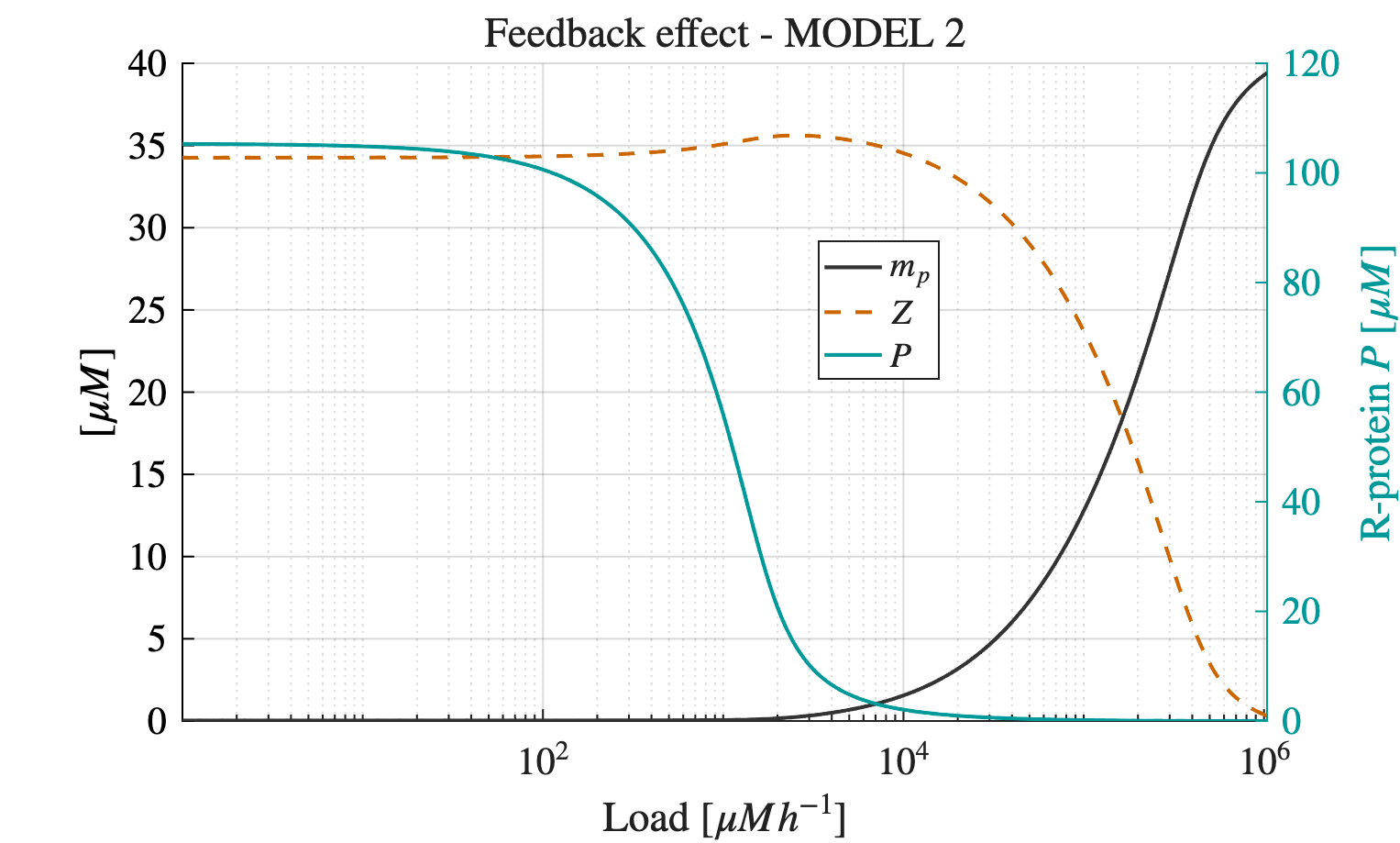}
        \caption{}
        \label{fig:feedback_m3}
    \end{subfigure}
    \caption{\textbf{Analysis of load effects in model \textit{M2}.} (a) Steady state analysis of species concentrations as a function of metabolic load $\omega_l D_l$ for Model \textit{M2}. (b) Steady state analysis of the growth rate as a function of the metabolic load $\omega_l D_l$ for Model \textit{M2}. (c) Steady state analysis of feedback effect as a function of the metabolic load $\omega_l D_l$ for Model \textit{M2}.}
    \label{fig:M3}
\end{figure}

Adopting model \textit{M2}, with plotted results in Fig.~\ref{fig:M3}, a behavior similar to what was shown for model \textit{M1} can be observed, but with a wider range of tolerance to increasing load and smoother system collapse to cell death. Moreover, the load complex collapses after the growth rate starts to decrease, which is biologically more plausible than what is observed for \textit{M1}.

Considering the components directly involved in the negative feedback between R-proteins and their mRNA, namely $m_p$, $Z$ and $P$ in Fig.~\ref{fig:feedback_m3}, the concentration of the complex $Z$ can be regarded as an indicator of feedback strength. Since $Z$ is formed when free R-proteins bind their own mRNA, a larger $Z$ concentration corresponds to a stronger inhibition of further R-protein production, as follows from the kinetic scheme in Fig.~\ref{fig:es1}.

The feedback is effective only within a specific range of load values: 
\begin{itemize}
    \item When $Z > m_p$, most of the available $m_p$ is sequestered into the inhibited complex, meaning that the feedback is strongly limiting the production of new R-proteins. In this regime, mRNA molecules that could otherwise be translated into $P$ are instead trapped in $Z$ and can also degrade because mRNAs in complexes are assumed not to be protected from degradation.
    \item When $Z < m_p$, the feedback is relatively weak. In this case, the available R-proteins are preferentially used for ribosome assembly rather than for feedback inhibition. Indeed, $P$ is required both for the formation of $Z$ and for the formation of ribosomes $R$, so these two processes compete for the same pool of free R-proteins. Since the affinity of R-proteins for rRNA is higher than for their own mRNA, ribosome assembly is favored over feedback complex formation whenever $P$ becomes limiting.
\end{itemize}
Although it may seem counterintuitive that the feedback is more active at lower burden values, this behavior can be explained by the cell's metabolic requirements. At low load, the demand for new ribosomes is smaller, allowing R-proteins to accumulate and activate the feedback loop. As the load increases, a larger fraction of R-proteins is consumed for ribosome production, reducing the amount available for feedback inhibition.\\

\begin{figure}[!htp]
    \centering
    
    \begin{subfigure}{.8\columnwidth}
        \centering
        \includegraphics[width=\columnwidth]{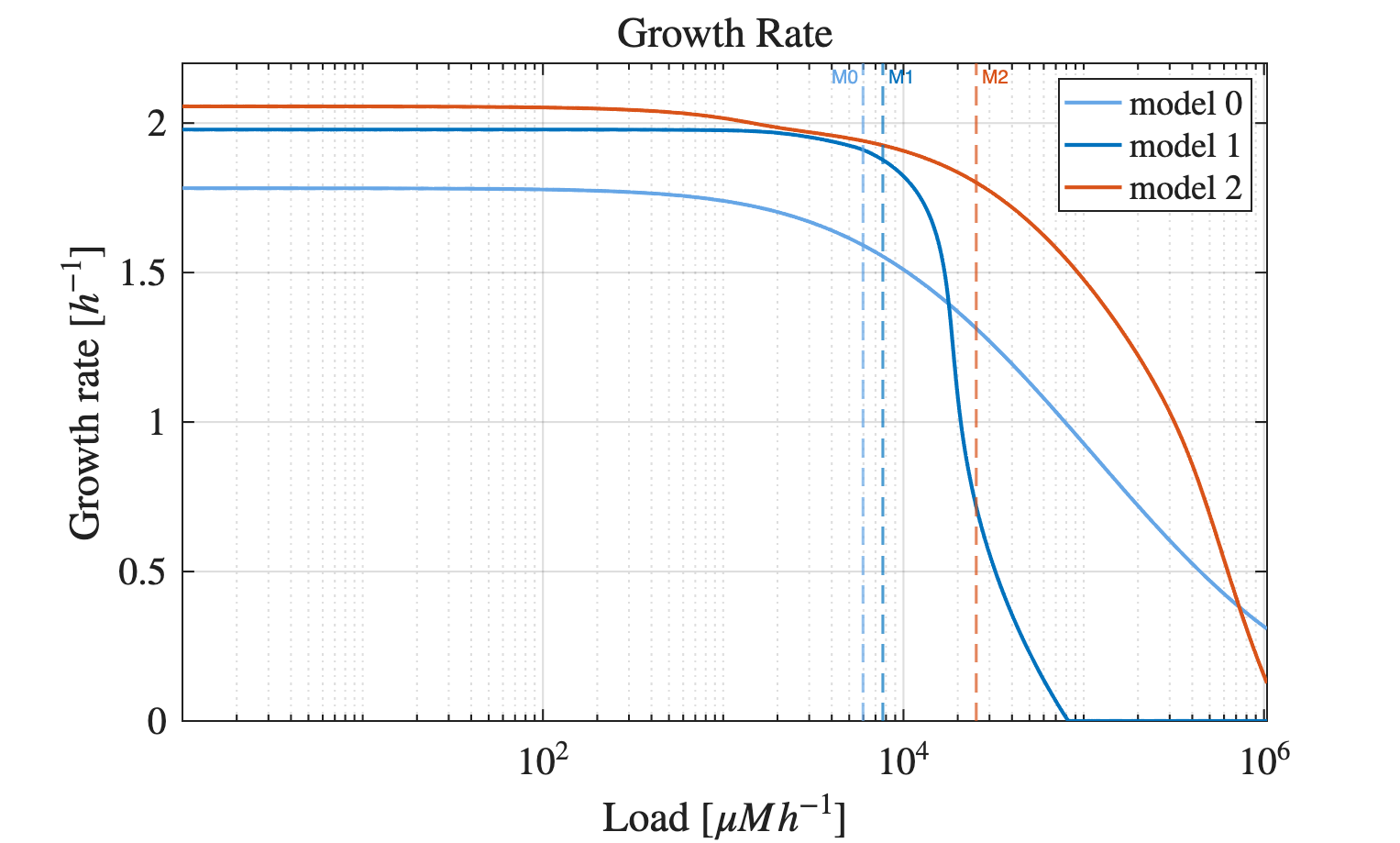}
        \caption{}
        \label{fig:growth_m2_m3}
    \end{subfigure}
    
    \begin{subfigure}{.8\columnwidth}
        \centering
        \includegraphics[width=\columnwidth]{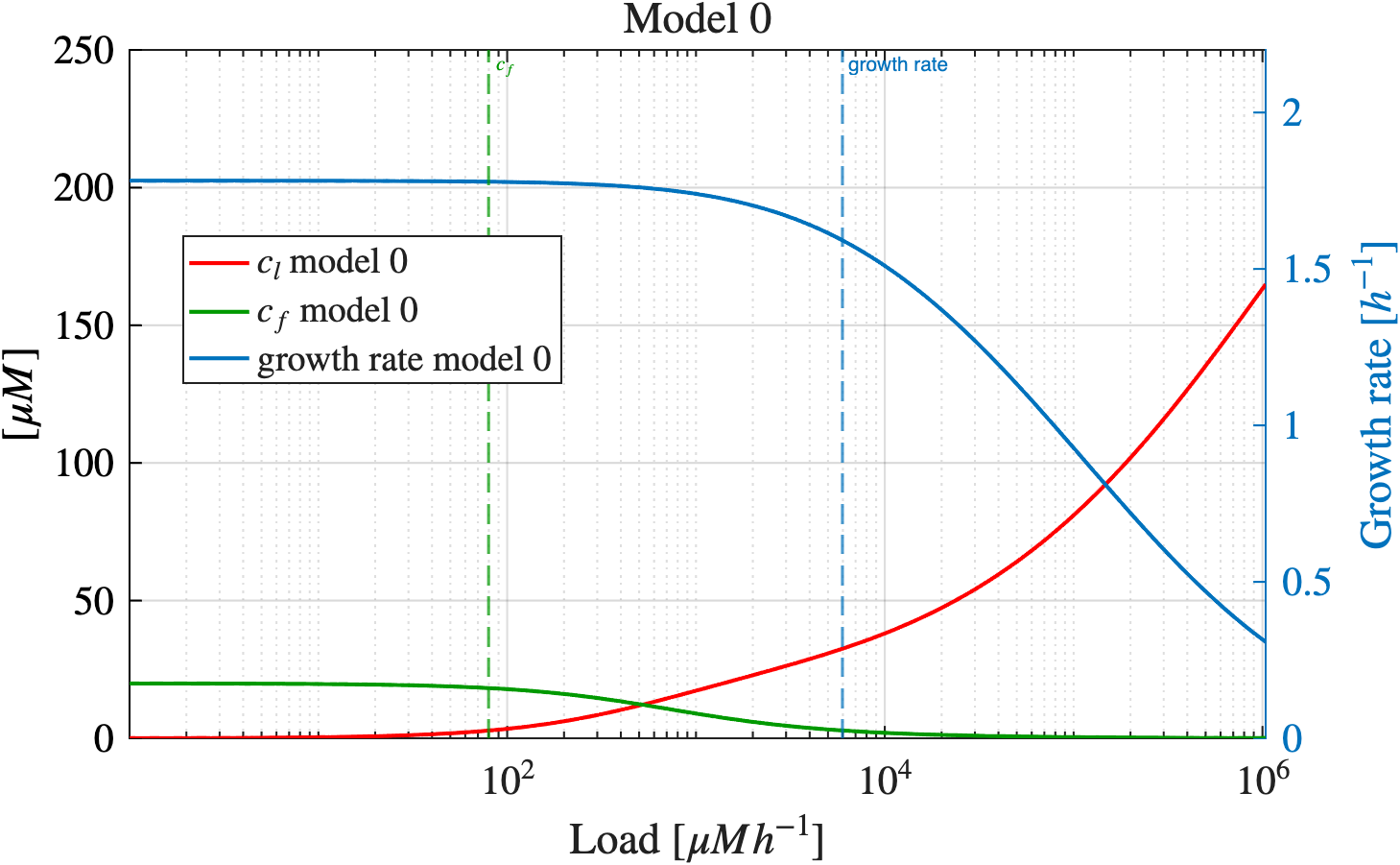}
        \caption{}
        \label{fig:final_m1}
    \end{subfigure}
    
    \begin{subfigure}{.8\columnwidth}
        \centering
        \includegraphics[width=\columnwidth]{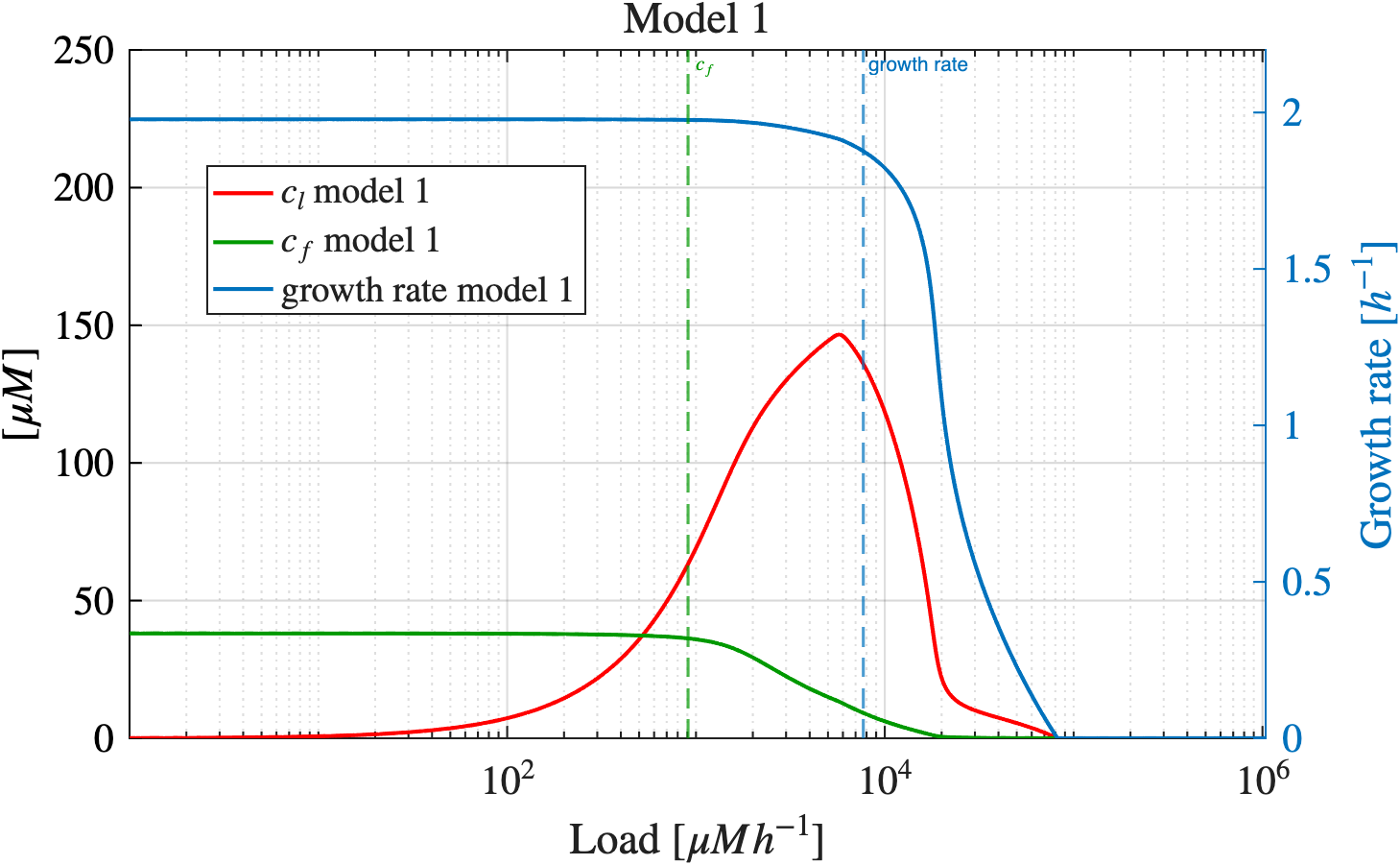}
        \caption{}
        \label{fig:final_m2}
    \end{subfigure}

    \begin{subfigure}{.8\columnwidth}
        \centering
        \includegraphics[width=\columnwidth]{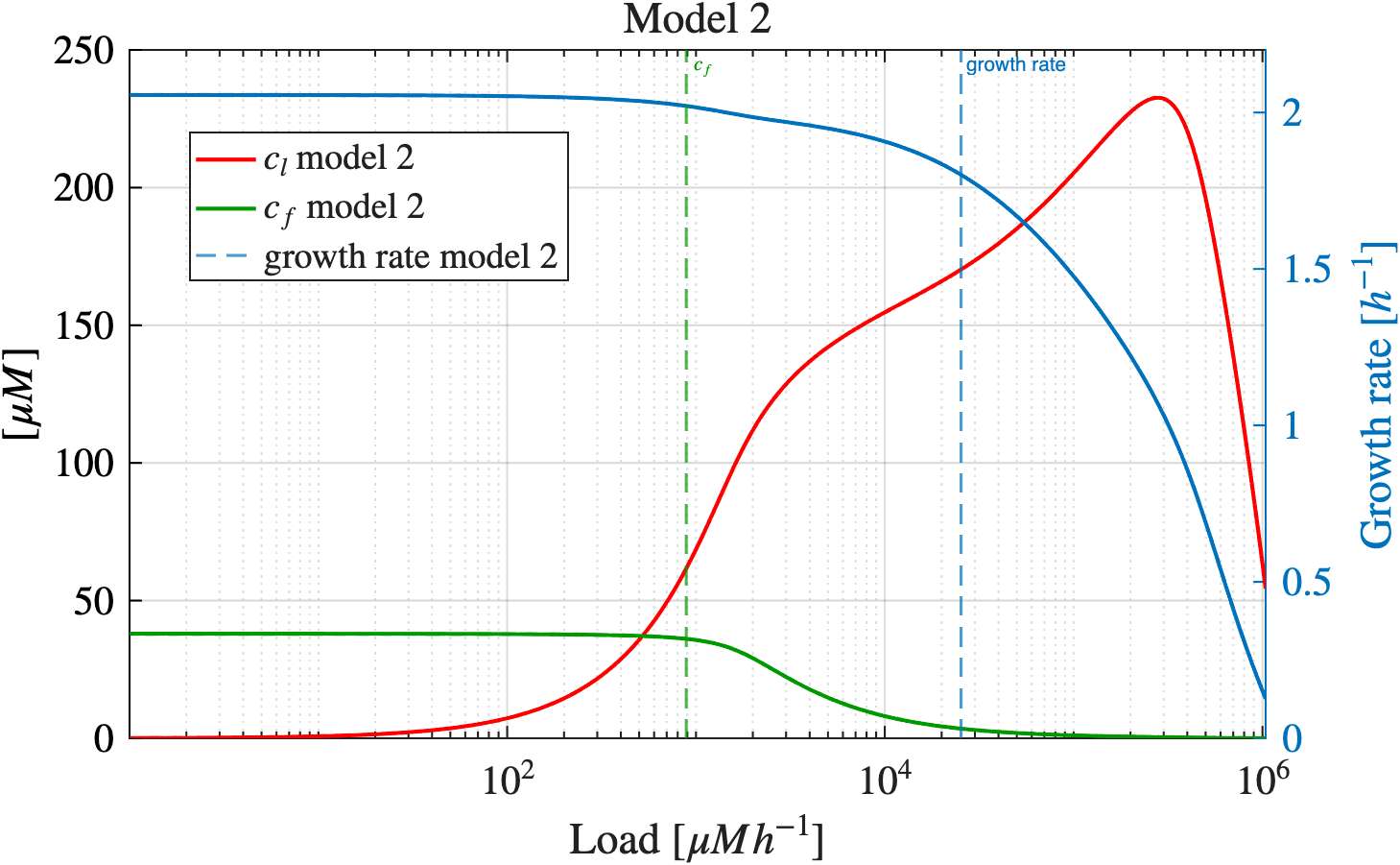}
        \caption{}
        \label{fig:final_m3}
    \end{subfigure}
    \caption{\textbf{Comparison of models M0, M1 and M2. }(a) Comparison of growth rates of models \textit{M0}, \textit{M1} and \textit{M2} as a function of metabolic load $\omega_l D_l$. Dashed lines indicate the load at which the growth rate decreases with a slope of 20\%. (b-c-d) Steady state analysis of $c_f$, $c_l$ and growth rate as a function of metabolic load $\omega_l D_l$ for models \textit{M0}, \textit{M1} and \textit{M2}, respectively. Dashed lines indicate the load at which growth rate (blue) and $c_f$ (green, complex of the burden monitor) decrease with a slope of 20\%.}
    \label{fig:comparison}
\end{figure}

For a more direct and quantitative comparison of the three models, Fig.\ref{fig:comparison} highlights the thresholds for the decrease in burden monitor $c_f$ and the growth rate. In particular, in Figs. \ref{fig:comparison}b-d, the three regions observed in the experimental data and described in \textit{Experimental evidence of metabolic burden} Section are highlighted. 

As shown in Fig.\ref{fig:comparison}a, the load corresponding to a growth rate decrease with a slope of 20\% shifts from a value of $\sim6\times10^3 \mu Mh^{-1}$ in \textit{M0} and $\sim7.5\times10^4 \mu Mh^{-1}$ in \textit{M1}, to $\sim2.5\times10^4 \mu Mh^{-1}$ in \textit{M2}. This underlines that the addition of R-proteins implementing a positive feedback for the ribosome autocatalysis in \textit{M1} does not significantly contribute to maintaining the growth rate at its initial level as the load increases. Conversely, the contribution of the negative feedback on $P$ synthesis from \textit{M2} helps the system to be stable for higher recombinant gene expression in terms of growth rate perturbation. Looking at the other three panels, it can be further observed that the burden monitor remains more stable by introducing the R-protein description of \textit{M1} and \textit{M2}, with a decrease that starts for a load of $8\times10^{1} \mu Mh^{-1}$ for \textit{M0} to a value of $9\times10^3 \mu Mh^{-1}$ for \textit{M1} and \textit{M2}.

\begin{figure}[!ht]
    \centering
    \includegraphics[width=1\columnwidth]{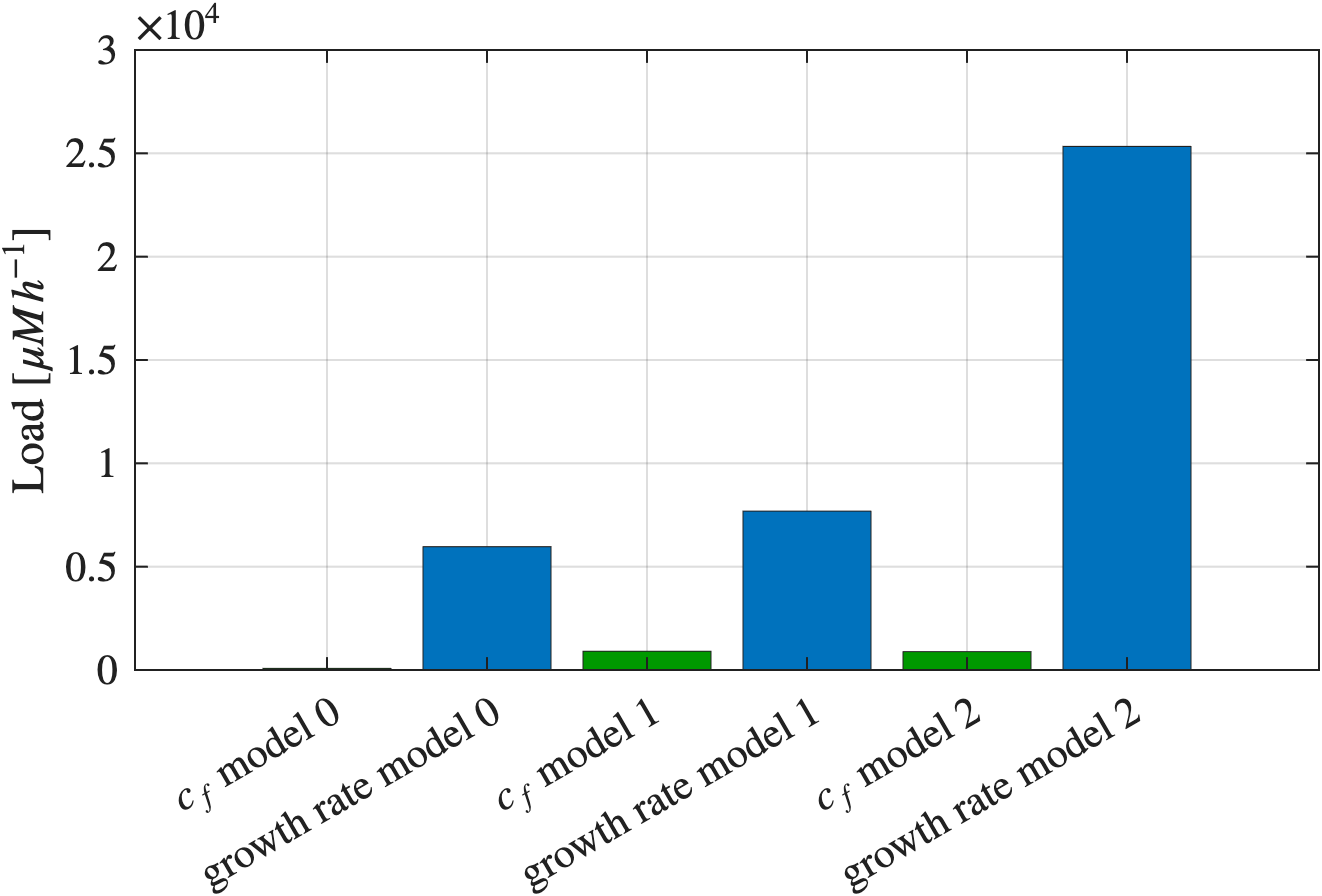}
    \caption{\textbf{Comparison of the metabolic load values at which the burden monitor $c_f$ and the growth rate start showing a decrease for all the models.} As reported in the text, the decrease values correspond to a 20\% slope reduction from their initial values, for models \textit{M0}, \textit{M1}, and \textit{M2}.}
    \label{fig:bar_plot}
\end{figure}

Focusing on the two primary manifestations of metabolic load — namely, the growth rate and the burden monitor $c_f$ decrease — the bar plot in Fig.~\ref{fig:bar_plot} summarizes the corresponding metabolic load at which they start to visibly decrease (with a 20\% reduction slope). The incorporation of R-protein in model \textit{M1} and the additional negative feedback in model \textit{M2} yield qualitatively similar, albeit quantitatively modest, shifts in the load threshold at which the burden monitor becomes responsive. However, the strongest effect is observed in \textit{M2}, where a visibly higher load is necessary to have a decrease in growth rate. A further insight into the functional role of ribosomal feedback is that such regulation enables a more sustained recombinant protein production at higher load levels, with load expression maintained even after the onset of the growth rate decline, as shown in Fig.~\ref{fig:M3}.

\subsection{Robustness analysis}
A significant limitation in the modeling of biological systems lies in the uncertainty - or complete absence - of reliable parameter values. Many kinetic constants, such as binding or association rates, could be highly context-dependent or even experimentally inaccessible. Furthermore, even if the parameter values can be estimated from experimental measurements, biological variability - both within individual cells and across various conditions - can lead to substantial deviations in parameter values and system behavior. In fact, synthetic circuit parameters can change upon variations in the circuit composition (e.g., when biological parts are reused in different networks), host context (e.g., due to the bacterial strain selected as a chassis for the circuit), and environmental context (e.g., temperature, nutrient levels, oxygen availability) \cite{arkin_cardinale}.
As a result, it is critical to assess whether a model maintains its qualitative predictions despite these uncertainties. This robustness is a hallmark of well-constructed biological models, supporting their applicability to real-world scenarios where parameters are variable.

For these reasons, sensitivity analyses are reported to evaluate how model output changes as a function of their parameter values. The strength parameter of the ribosomal feedback in model \textit{M2}, and the parameters known to change for different nutrient availability conditions were selected as key parameters for this analysis. Their variation reflects parameter uncertainty and changes in the environmental context, respectively.

\paragraph{\textbf{Effects of ribosomal feedback strength}}

In model \textit{M2}, particular attention is given to the association rate $\alpha_p$ between the R-protein and its mRNA, a parameter that modulates the strength of the negative feedback loop. As this parameter was not directly available from the literature, a plausible value was assigned within a physiologically meaningful range.
To assess the robustness of the system dynamics to uncertainties in $\alpha_p$, we simulated the model by increasing and decreasing this parameter by one order of magnitude relative to its nominal value ($\alpha_p = 0.1\, \sigma_p$).

\begin{figure*}[t]
    \centering
    \includegraphics[width=\linewidth]{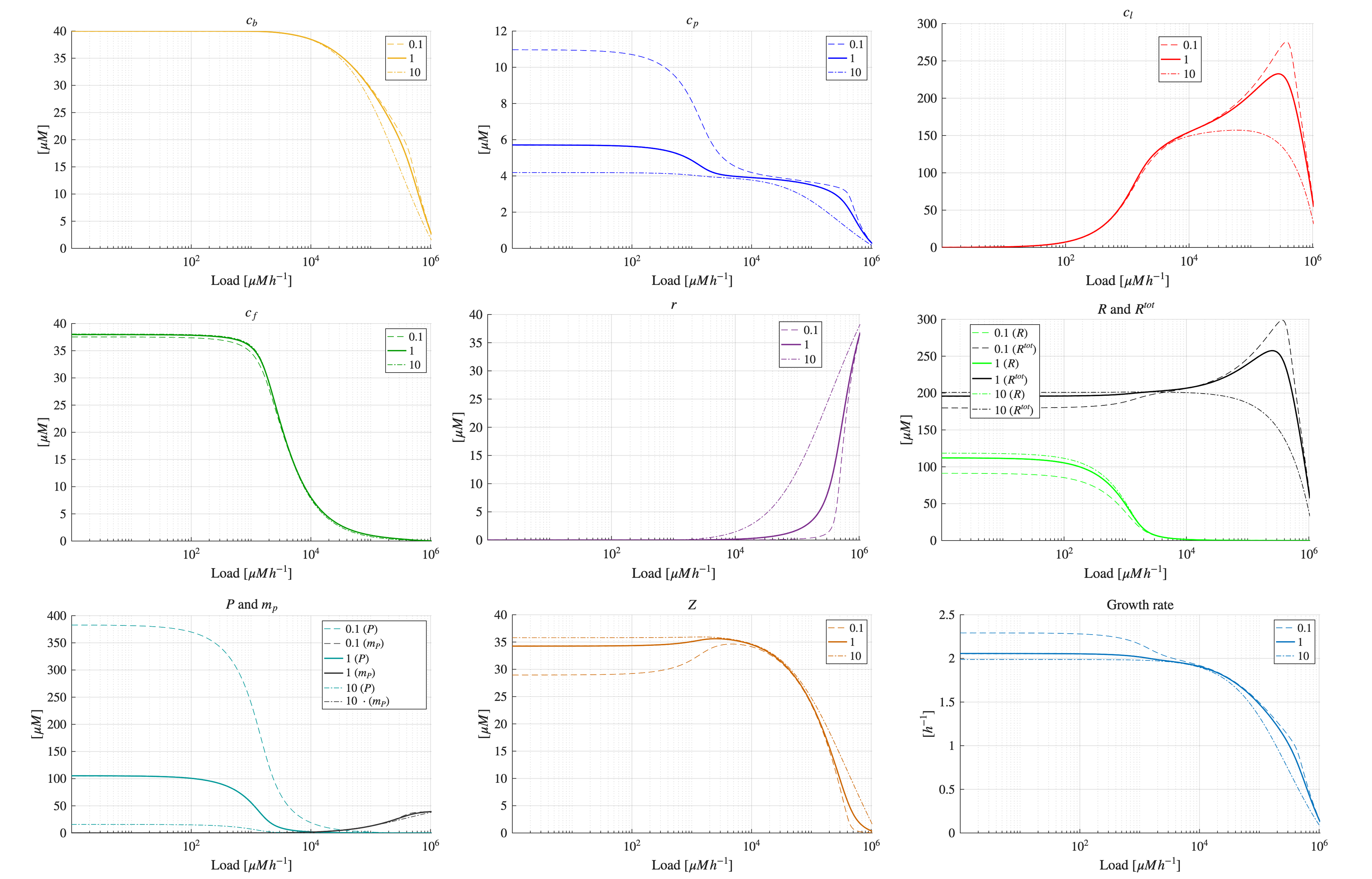}
    \caption{\textbf{Sensitivity analysis of model \textit{M2} as a function of ribosomal feedback strength $\alpha_p$ parameter.} The $\alpha_p$ parameter has a biologically plausible range $0<\alpha_p < \sigma_p$, with nominal value $\alpha_p = 0.1 \sigma_p$. A null value corresponds to no ribosomal feedback, while a value equal to $\sigma_p$ corresponds to a binding constant rate between R-proteins and their own mRNAs that is equal to the binding constant rate between R-proteins and rRNAs. As the affinity between R-proteins and rRNA is known to be higher $\alpha_p = \sigma_p$ is an upper bound. In these simulations, the $\alpha_p$ parameter is decreased to \textbf{0.1} times its nominal value ($\alpha_p = 0.01 \sigma_p$) and increased to \textbf{10} times its nominal value ($\alpha_p = \sigma_p$).}
    \label{fig:S_m3_alpha_p}
\end{figure*}

The simulations, reported in Fig. \ref{fig:S_m3_alpha_p}, show that variations in $\alpha_p$ primarily affect the species involved in the negative feedback loop, namely $P$, $Z$, and $c_p$. Specifically, sensitivity to $\alpha_p$ is most pronounced in the low metabolic load regime, corresponding to the region where the negative feedback is active.
In contrast, the behavior of the other species remains largely unchanged, even in the extreme case where $\alpha_p = \sigma_p$. This indicates that the overall system dynamics are not overly sensitive to this parameter.
These results support the robustness and reliability of the model, especially in biological contexts where precise parameter values may be unavailable, uncertain, or host-dependent.

\paragraph{\textbf{Effects of suboptimal culture conditions}}

Although this study aimed to identify a minimal model that could recapitulate the effects of metabolic load on both the expression of other genes and growth, it is worth noting that these results are valid under optimal culture conditions. Indeed, it is well known that the introduction of additional variables and contexts (such as reduced availability of carbon or nitrogen sources for protein synthesis and metabolic processes, or variations in temperature and oxygenation of bacterial cultures) can severely influence the cellular state. Some studies have attempted to model these aspects~\cite{towbin2017optimality, zhu2019growth}.
Besides extending beyond (if not counter to) the objectives of this study, it is possible to gain a qualitative insight into the system's response to such perturbations through a robustness analysis on Models \textit{M1} and \textit{M2}.

Specifically, in microbial cultures, environmental conditions such as temperature, nutrient availability, and oxygenation play a critical role in shaping cellular behavior. These factors directly influence the metabolic state of the cell, gene expression patterns, and consequently the rate of growth and protein synthesis. It is therefore essential to assess whether model predictions remain reliable under variations that reflect real-world fluctuations. For example, temperature changes can alter enzymatic reaction rates, affecting transcription and translation dynamics; nutrient limitation can lead to stress responses and shifts in resource allocation; and oxygen levels can modulate energy production pathways, especially in facultative anaerobes. To test whether our model captures the core dynamics of the system robustly, we performed a sensitivity analysis by varying the parameters most likely to be influenced by such environmental changes in terms of nutrient availability - namely, the protein translation rates $\beta_i$, where $i \in \{b,p,\ell,f\}$, and $\alpha$, the scaling factor linking active ribosomes to cellular growth in Eq.\eqref{growthlaw}. These parameters have been varied together, as they are all related to the translational efficiency that is reported to be affected by the energy taken from nutrients \cite{weisse,proof_load_effect}. In fact, the $\alpha$ parameter in the growth rate function used in this work is proportional to the translation rate according to the mechanistic basis of this growth law (explained in the \textit{Growth laws overview} sidebar).

The nominal parameter values adopted in this study correspond to maximal growth conditions. Accordingly, in the sensitivity analysis we systematically reduced $\beta_i$ and $\alpha$ to 0.5 and 0.1 times their nominal values, corresponding to moderate and one-order-of-magnitude perturbations, respectively. This approach allows us to evaluate whether the qualitative behavior of the model is preserved and whether its predictions remain biologically plausible across a physiologically relevant range of conditions. A robust model should not only reproduce the expected dynamics under ideal conditions but also be able to tolerate moderate perturbations, reflecting the inherent adaptability of living systems.

\begin{figure*}[!h]
    \centering
    \includegraphics[width=\linewidth]{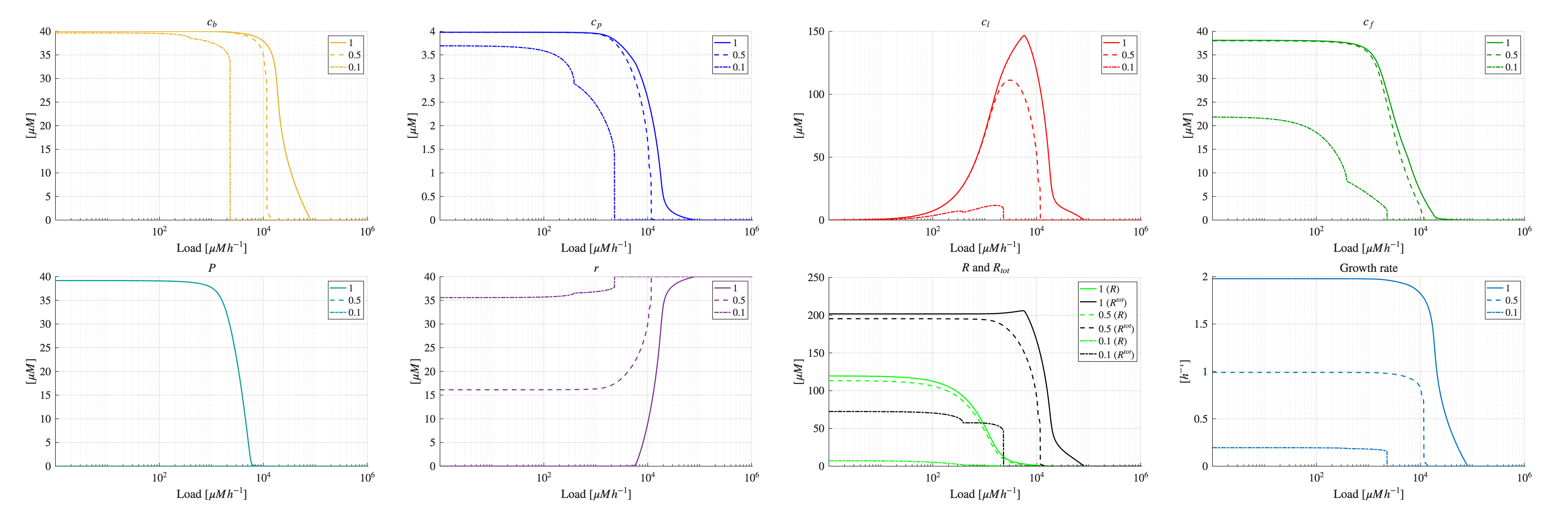}
    \caption{\textbf{Sensitivity analysis of model \textit{M1} as a function of translation rates $\beta_i$ with $i\in \{b,p,l,f\}$ and the scaling factor $\alpha$}. Simulations of suboptimal culture conditions reducing $\beta_i$ (translation rates) and $\alpha$ (scaling factor linking active ribosomes to cellular growth) to 0.5 and 0.1 times their nominal values.}
    \label{fig:S_m2_beta}
\end{figure*}

\begin{figure*}[!h]
    \centering
    \includegraphics[width=\linewidth]{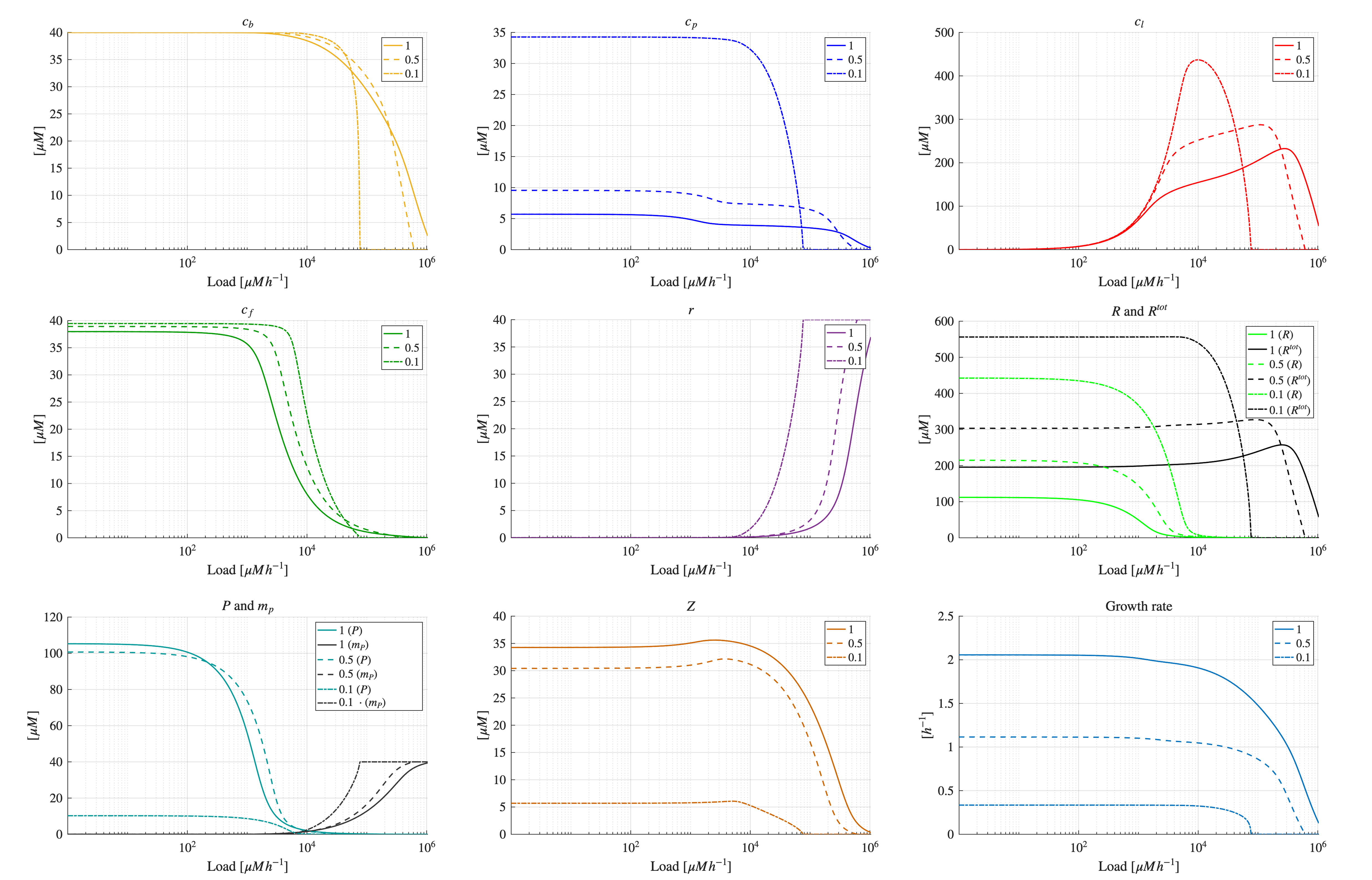}
    \caption{\textbf{Sensitivity analysis of model \textit{M2} as a function of translation rates $\beta_i$ with $i\in \{b,p,l,f\}$ and the scaling factor $\alpha$}. Simulations of suboptimal culture conditions reducing $\beta_i$ (translation rates) and $\alpha$ (scaling factor linking active ribosomes to cellular growth) to 0.5 and 0.1 times their nominal values.}
    \label{fig:S_m3_beta}
\end{figure*}

The results are presented in the Fig. \ref{fig:S_m2_beta} and \ref{fig:S_m3_beta} for Models \textit{M1} and \textit{M2}, respectively. 
As indicated by the simulation results, the main qualitative behavior of the system is preserved. A reduction of these parameters generally leads to an earlier collapse of the system in the burdensome protein and growth rate, with model \textit{M1} showing an earlier collapse than model \textit{M2}. In the presence of ribosomal feedback, the sequestered R-protein mRNAs $Z$ decrease earlier for the lowest $\beta_i$ and $\alpha$ parameter values, i.e., in the most severe nutrient-poor condition tested (Fig. \ref{fig:S_m3_beta}), providing indication on the load ranges in which the negative feedback becomes released due to R-protein demand.


\section{Conclusions}
This work investigated the role of ribosomal feedback in bacterial cell growth and metabolic load through a systems biology lens, proposing a minimal mathematical model that explicitly incorporates the post-transcriptional negative feedback of R-proteins on their own mRNA synthesis, a mechanism typically neglected in cell models of resource allocation.

Three models of increasing complexity were developed and compared:
\begin{itemize}
\item\textit{M0}, a simplified baseline lacking ribosome autocatalysis;
\item\textit{M1}, introducing the decoupled synthesis of rRNA and R-proteins;
\item\textit{M2}, augmenting \textit{M1} with the negative feedback loop on R-protein translation.
\end{itemize}

Despite its simplicity and the fact that it neglects several important intracellular mechanisms, model \textit{M0} remains a valid approximation under low metabolic burden conditions, which explains why it is still widely used in the literature for modeling purposes. However, the analysis shows that its validity degrades at higher loads, where it can lead to unrealistic behavior.

The results demonstrate that the minimal model structure requires the ribosomal negative feedback (\textit{M2}) to reproduce three biologically plausible regimes of cellular burden : (i) a tolerance region with no appreciable effect on growth or gene expression, (ii) an intermediate burden regime where protein synthesis is impaired while growth rate remains relatively stable, and (iii) a severe overload condition leading to simultaneous collapse of both protein production and growth rate. Critically, \textit{M2} showed that the negative feedback enables sustained recombinant protein production at higher load levels, with load expression maintained even after the onset of growth rate decline, a behavior not captured by \textit{M0} or \textit{M1}.

Sensitivity analysis confirmed the robustness of these qualitative predictions under parameter variations, including uncertainty in the feedback strength parameter $\alpha_p$ and suboptimal culture conditions. Notably, \textit{M2} proved more resilient than \textit{M1} to one-order-of-magnitude reductions in translation rates and growth-coupling parameters, underscoring the stabilizing role of the negative feedback loop. Moreover, the explicit description of R-proteins and the complex they form with their own sequestered mRNAs enables the study of parameter ranges in which the ribosomal feedback mechanism is working, or when the ribosomal demand exceeds the capacity of this mechanism.

Additionally, the minimal model defined in this work investigates only part of the biological complexity of translation regulation. The pool of intracellular molecules (e.g., amino acids and ATP) that provide energy to biosynthetic processes, not explicitly represented in the present model, can affect some key reactions such as transcription and translation rates; cell load can increase the demand of energy, e.g., by recruiting more tRNAs for protein synthesis, negatively affecting global protein expression including proteins devoted to nutrient uptake or enzymatic transformations that provide energy. This positive feedback loop, in which energy is a limiting resource, is a viable alternative to introducing the autocatalytic production of ribosomes, which is another positive feedback, and is expected to effectively capture load behaviors due to resource limitation, like model \textit{M1} defined in this work. However, the use of an energy-based model rather than a ribosome-based one limits the incorporation of other specific feedback mechanisms operating at the ribosomal level, such as the post-transcriptional negative regulation investigated in this work.

Other feedback mechanisms involving ribosomes that have not been specifically considered in our minimal model include the negative regulation of rRNA and R-protein transcription, which occurs under cell load conditions in bacteria with a functional stringent response, mediated by alarmone molecules (e.g., ppGpp). This feedback decreases ribosome production and leads to a reduction in growth rate. This counterintuitive effect in the presence of cell load aims to reallocate the resources towards essential biosynthetic processes, related to cell survival, by decreasing fast growth. The inclusion of this mechanism in the model proposed in this work would require the explicit introduction of energy (e.g., ATP, produced by an endogenous proteome fraction like the $B$ fraction) and alarmone (ppGpp, decreasing function of energy); the energy-based loop described above should be included, with ATP affecting transcription and translation rates; on the other hand, ppGpp should repress the transcription rates of rRNA and R-proteins ($\omega_r$ and $\omega_p$, while activating the transcription of basal genes ($\omega_b$) to mimic an effective resource allocation. As a consequence, the growth law should be updated to introduce energy dependence on the $\alpha$ scale parameter, as described in \cite{weisse,proof_load_effect,santosnavarro21}. As these updates would have significantly increased model complexity, we did not include any energy or ppGpp species, thereby keeping the modeling framework minimal. 

Although the contribution of the mentioned energy- and ppGpp-dependent mechanisms is considered central in computational systems and synthetic biology models, the post-transcriptional R-protein feedback may also play a key role in resource allocation. The post-transcriptional feedback loop could be predominant in bacteria with compromised alarmone biosynthesis (e.g., \textit{relA1} mutants, such as the widely used $DH10B$ laboratory strain of \textit{E. coli}), further confirming the importance of studying the individual effects of each regulatory mechanism.
As described in this work, the post-transcriptional feedback can explain important changes in relevant cellular outputs, such as protein expression and growth. Moreover, recent works highlighted the importance of heterogeneity in rRNA and R-protein species and inter-cell variability in the expression of ribosomal components~\cite{LP_Shen2025,LP_Pavlou2025}. Analysis of these features may elucidate ribosomal function and is expected to improve our understanding of translational regulation. Adoption of synthetic biology solutions such as reporter genes fused to ribosomes or systematic adoption of burden monitors in engineered systems \cite{LP_Pavlou2025,LP_Ceroni2015} could support the refinement of these aspects in a quantitative manner, although intensive experimental work would be required to collect data.

Nonetheless, in this work, we focused on a single, often neglected, mechanism to define a minimal model that ultimately links burdensome protein expression and cell load, thereby elucidating biologically plausible load-dependent trends.

From a synthetic biology perspective, these findings carry important practical implications. The explicit modeling of ribosomal feedback provides a more accurate predictive framework for the rational design of synthetic genetic circuits, enabling engineers to better anticipate and mitigate the detrimental effects of metabolic burden on both cellular fitness and circuit performance. The minimal nature of the proposed model makes it computationally accessible and readily integrable into broader circuit design workflows, without requiring the extensive parameterization typical of more detailed whole-cell models.

This work highlights the importance of molecular feedback loops, which are ubiquitous in naturally occurring biological systems~\cite{kam_anti,LP_Harris2015,LP_DDVcontrol2018,LP_Chevalier2019,LP_Pasotti2019,LP_Zand2024,LP_Hsiao2018,LP_Hu2022} and, analogously, play a crucial role in engineering-inspired synthetic biological circuits.

\section{ACKNOWLEDGMENT}

This work was partially funded by: Fondazione Cariparo grant “Bando Ricerca Scientifica di Eccellenza 2021 n 59576”; Department of Information Engineering - type B Senior research grant, call 2022; PhD Grant PON-RI 2014-20 (CCI2014IT16M2OP005) on FSE REACT-EU funds.

\section{Author Information}

\begin{IEEEbiography}{{C}hiara Cimolato}(chiara.cimolato@studenti.unipd.it) received the M.Sc. degree (cum laude) in control engineering in 2021 and the Ph.D. in control engineering in 2025, both from the University of Padova, Italy.
She is currently a Postdoctoral Researcher in the Department of Information Engineering at the University of Padova. Her research focuses on applying control theory and systems biology approaches to synthetic biology to address the challenge of antibiotic resistance. During her Ph.D., she undertook a research visit with Prof. Mustafa Khammash’s laboratory at ETH Zurich and spent an additional visiting period at the COSBI Center in Rovereto. She is a Graduate Student Member of IEEE. 
\end{IEEEbiography}

\begin{IEEEbiography}{Elisa Gaetan}(elisa.gaetan@unimore.it) received the B.Sc. degree in information engineering and the M.Sc. degree in control engineering from the University of Padova, Italy. She is currently pursuing a Ph.D. degree in Autonomous Systems (DAUSY) with the Department of Engineering "Enzo Ferrari" at the University of Modena and Reggio Emilia, Italy. In 2024-2025, she was a visiting researcher at UC Berkeley, in the Berkeley Artificial Intelligence Research (BAIR) Lab, under the supervision of Prof. Alexandre M. Bayen.
Her research interests include modeling and control with applications to multi-agent systems, traffic control, and human-machine interactions.
\end{IEEEbiography}

\begin{IEEEbiography}{Lorenzo Pasotti}(lorenzo.pasotti@unipv.it) received the M.Sc. (cum laude) degree in biomedical engineering in 2008 and the Ph.D. in bioengineering and bioinformatics in 2012 from the University of  Pavia, Italy. In 2011, he was a visiting researcher at the School of Biological Sciences, University of Edinburgh. From 2012 to 2015, he was a Postdoctoral Researcher with the Center for Tissue Engineering at the University of Pavia. From 2015 to 2023, he served as an Assistant Professor in the Department of Electrical, Computer, and Biomedical Engineering, University of Pavia, where he became an Associate Professor in 2023. Since 2020, he has been a visiting scientist at the Department of Computational Biology, Institut Pasteur, Paris. His interests include computational biology, synthetic biology, and metabolic engineering.
\end{IEEEbiography}

\begin{IEEEbiography}{Luca Schenato}- IEEE Fellow, (l.schenato@unipd.it) received the Dr. Eng. degree in electrical engineering from the University of Padova in 1999, and the Ph.D. degree in electrical engineering and computer sciences from UC Berkeley in 2003, US. He held a postdoctoral position in 2004 and served as a Visiting Professor at the University of California, Berkeley from 2013 to 2014. He is currently a full professor in the Department of Information Engineering at the University of Padova. His interests include networked control systems, multi-agent systems, wireless sensor networks, distributed optimization, and synthetic biology. He has been awarded the 2004 Researchers Mobility Fellowship by the Italian Ministry of Education, University and Research (MIUR), the 2006 Eli Jury Award at U.C. Berkeley, and the EUCA European Control Award in 2014. He served as an Associate Editor for IEEE TRANSACTIONS ON AUTOMATIC CONTROL from 2010 to 2014. He is currently a Senior Editor of IEEE TRANSACTIONS ON CONTROL OF NETWORK SYSTEMS and an Associate Editor of Automatica.
\end{IEEEbiography}

\begin{IEEEbiography}{Massimo Bellato}- IEEE Member, (massimo.bellato@unipd.it) obtained his M.Sc. degree in bioengineering in 2015 and his Ph.D. in bioengineering and bioinformatics in 2019 from the University of Pavia, Italy. He currently serves as an Assistant Professor of Bioengineering within the Department of Molecular Medicine and the Department of Information Engineering at the University of Padova, Italy. His research concentrates on synthetic biology and mathematical modeling, with a specific focus on the rational design of synthetic genetic circuits, including a research period at the MIT Synthetic Biology Center. His work in Synthetic Biology has a clinical microbiology orientation at the University of Padova, emphasizing microbial communities, CRISPR interference, and phage engineering for the treatment of multidrug-resistant bacterial infections. He initiated the iGEM Teams at the University of Padova and oversees research initiatives in Synthetic Biology. Additionally, he is a member of IEEE and serves on the IEEE-CSS Systems and Synthetic Biology Technical Committee.
\end{IEEEbiography}
\printbibliography 
\end{document}